\documentclass[11pt]{article}
\pdfoutput=1 
\usepackage{jheppub}
\usepackage{bbm,bm,mathtools,slashed}
\usepackage[T1]{fontenc}
\usepackage[nottoc]{tocbibind}
\usepackage[toc,page]{appendix}
\usepackage{youngtab}
\usepackage{url}
\usepackage{ulem}
\usepackage{physics}

\graphicspath{{./figures/}}

\newcommand{\ri}{\mathrm{i}}
\newcommand{\ER}{{R}}
\newcommand{\EL}{{L}}

\newcommand{\be}{\begin{equation}}
\newcommand{\ee}{\end{equation}}
\newcommand{\bea}{\begin{eqnarray}}
\newcommand{\eea}{\end{eqnarray}}

\newcommand{\hop}{{w}}

\newlength{\fighskip} \fighskip=2pt
\newlength{\figvskip} \figvskip=3pt

\preprint{KEK-TH-2540, J-PARC-TH-0293, RIKEN-iTHEMS-Report-23}
\title{Dense $\textrm{QCD}_2$ with matrix product states}

\author[a,b]{Tomoya Hayata,}
\author[b,c,d,e,f]{Yoshimasa Hidaka}
\author[c,g]{and Kentaro Nishimura}

\affiliation[a]{Departments of Physics, Keio University, 4-1-1 Hiyoshi, Kanagawa 223-8521, Japan}
\affiliation[b]{RIKEN iTHEMS, RIKEN, Wako 351-0198, Japan}
\affiliation[c]{KEK Theory Center, 1-1 Oho, Tsukuba, Ibaraki 305-0801, Japan}
\affiliation[d]{Graduate Institute for Advanced Studies, SOKENDAI, 1-1 Oho, Tsukuba, Ibaraki 305-0801, Japan}
\affiliation[e]{International Center for Quantum-field Measurement Systems for Studies of the Universe
and Particles (QUP), KEK, 1-1 Oho, Tsukuba, Ibaraki 305-0801, Japan}
\affiliation[f]{Department of Physics, Faculty of Science, University of Tokyo, 7-3-1 Hongo Bunkyo-ku Tokyo 113-0033, Japan}
\affiliation[g]{Research and Education Center for Natural Sciences, Keio University, 4-1-1 Hiyoshi, Yokohama, Kanagawa 223-8521, Japan}

\emailAdd{hayata@keio.jp}
\emailAdd{yoshimasa.hidaka@yukawa.kyoto-u.ac.jp}
\emailAdd{nishiken@hiroshima-u.ac.jp}

\abstract{
We study one-flavor $\mathrm{SU}(2)$ and $\mathrm{SU}(3)$ lattice QCD in ($1+1$) dimensions at zero temperature and finite density using matrix product states and the density matrix renormalization group.
We compute physical observables such as the equation of state, chiral condensate, and quark distribution function as functions of the baryon number density. 
As a physical implication, we discuss the inhomogeneous phase at nonzero baryon density, where the chiral condensate is inhomogeneous, and baryons form a crystal.
We also discuss how the dynamical degrees of freedom change from hadrons to quarks through the formation of quark Fermi seas.
}

\begin{document}
\maketitle

\section{Introduction}
The study of the phase structure of cold dense QCD is of great importance to understand the physics inside neutron stars as well as to deepen our understanding of QCD as a part of the fundamental theory of our universe~\cite{Baym:2017whm,Rezzolla:2018jee}.
In dense QCD matter, the physics of confinement and colored Fermi seas of quarks play an essential role in determining the phase structures. 
Indeed, various interesting phases have been conjectured so far such as the quarkyonic matter~\cite{McLerran:2007qj,Kojo:2009ha,Kojo:2011cn}, and color superconductivity~\cite{Alford:2007xm} (see e.g., refs~\cite{Fukushima:2010bq,Fukushima:2013rx,Fischer:2018sdj} for reviews of the QCD phase diagram).
However, the direct study of cold dense QCD is very difficult except at large chemical potential, where interactions between quarks become weak due to asymptotic freedom, and perturbative QCD is applicable. 
In the presence of the baryon chemical potential, which is not high enough to accommodate color superconductivity, the ab-initio study of QCD is demanded, but conventional Monte Carlo simulations based on lattice QCD are no longer valid due to the notorious sign problem; the fermion determinant in the path integral becomes complex at nonzero baryon chemical potential, so that the importance sampling breaks down~\cite{Nagata:2021ugx}.
Several methods have been proposed to circumvent the sign problem in the path integral formulation~\cite{Aarts:2015tyj,Berger:2019odf}.

Recently, the Hamiltonian lattice gauge theory is reconsidered with the help of tensor networks and quantum computing to attack the problems suffering from the sign problem such as finding groundstate at finite-density or in the presence of the $\theta$ term, and real-time problems~\cite{Dalmonte:2016alw,Banuls:2019bmf}.
In particular, numerical simulations based on a tensor network known as matrix product state (MPS)~\cite{perez2006matrix} are found to be effective to solve (1+1)-dimensional QED, which is also known as the Schwinger model~\cite{Banuls:2013jaa,Banuls:2015sta,Banuls:2016gid,Buyens:2013yza,Buyens:2015tea,Buyens:2016ecr,Kuhn:2014rha,Pichler:2015yqa,Funcke:2019zna} (see e.g., refs.~\cite{Shimizu:2017onf,Butt:2019uul} for applications of the tensor network in the path integral formulation of the Schwinger model).
Variational computations based on MPSs have also been adopted to ($1+1$)-dimensional QCD, which we refer to as $\textrm{QCD}_2$ for $\mathrm{SU}(2)$~\cite{Kuhn:2015zqa,Silvi:2016cas,Banuls:2017ena,Sala:2018dui}, and $\mathrm{SU}(3)$~\cite{Silvi:2019wnf,Rigobello:2023ype}.
However, to our knowledge, comprehensive study of $\textrm{QCD}_2$ at finite density has not been done yet.

In this paper, we study one-flavor $\mathrm{SU}(2)$ and $\mathrm{SU}(3)$ $\textrm{QCD}_2$ at zero-temperature and finite-density based on Hamiltonian lattice QCD and matrix product states~\cite{perez2006matrix}. 
We adopt the formulation developed in ref.~\cite{Sala:2018dui} to decouple nonabelian gauge fields explicitly in open boundary conditions.
We use a variational approximation for the wave function of the ground state of quark many-body systems interacting with color coulomb force based on density matrix renormalization group (DMRG)~\cite{PhysRevLett.69.2863,SCHOLLWOCK201196},
and compute physical observables such as the equation of state, chiral condensate, and quark distribution function as a function of the baryon number density, where the conventional lattice QCD simulation suffers from the sign problem. 
As a physical implication, we discuss the inhomogeneous phase at nonzero baryon density, where the chiral condensate is inhomogeneous, and baryons form a crystal.
We also discuss how dynamical degrees of freedom change from hadrons to quarks through the formation of quark Fermi seas.
We find that the study of dense $\textrm{QCD}_2$ is a good playground to test theoretical tools developed for ab-initio study of cold-dense states in QCD or QCD-like theory.

The remainder of this paper is organized as follows.
In section~\ref{sec:formulation}, we review the Hamiltonian formulation of $\mathrm{SU}(N_\mathrm{c})$ lattice QCD in $(1+1)$ dimensions. We eliminate the gauge field from the Hamiltonian by a unitary transformation under open boundary conditions in section~\ref{sec:rotated_Hamiltonian}.
We map our system to the spin system through the Wigner-Jordan transformation~\cite{Jordan:1928wi}  in section~\ref{sec:map_to_spin}.
In section~\ref{sec:DMRG}, we introduce a matrix product state and matrix product operator, and explain the DMRG technique briefly.
In section~\ref{sec:Numerical_results}, we show the numerical results.
We used iTensor~\cite{itensor} for numerical calculations.
Section~\ref{sec:summary} is devoted to the summary.
In Appendix~\ref{sec:Free_theory}, we show several results of the free theory with open boundary conditions to compare with $\textrm{QCD}_2$.

\section{\texorpdfstring{$\textrm{QCD}_2$ on a lattice}{QCD2 on a lattice}}\label{sec:formulation}
\subsection{Hamiltonian formulation}\label{sec:KS_hamiltonian_formulation}

We consider $\mathrm{SU}(N_\mathrm{c})$ gauge theory with $N_{\mathrm{f}}$ flavors in $(1+1)$ dimensions, although our numerical calculations mainly deal with the cases of $N_\mathrm{c}=2, 3$ and $N_\mathrm{f}=1$.
 We employ the Kogut-Susskind Hamiltonian with staggered fermions~\cite{Kogut:1974ag}, which is composed of three parts: electric $H_\mathrm{E}$, hopping $H_{\mathrm{hop}}$, and mass $H_\mathrm{m}$
 parts: $ H = H_\mathrm{E} + H_\mathrm{hop} +H_\mathrm{m}$, where
\begin{align}
  H_\mathrm{E}&=\frac{ag_0^2}{2}\sum_{n=1}^{N-1}E_i^2(n), \\
  H_\mathrm{hop}&=\frac{1}{2a}\sum_{n=1}^{N-1}\qty(\chi^\dag(n+1) U(n) \chi(n)+\chi^\dag(n) U^\dag(n) \chi(n+1)), \\
  H_\mathrm{m} &=m_0\sum_{n=1}^N(-1)^{n} \chi^\dag(n) \chi(n) .
\end{align}
Here, $g_0$, $a$, $m_0$, and $N$ are the bare gauge coupling, lattice spacing, bare mass, and number of sites, respectively.
$\chi$ and $\chi^\dag$ are annihilation and creation operators of $N_{\mathrm{c}}\times N_{\mathrm{f}}$ component fermions, $U$ is the link variable with the fundamental representation that is a $N_\mathrm{c}\times N_\mathrm{c}$ matrix, and $E_i^2(n)$ is the square of the electric field. The index ``$i$'' denotes the color of adjoint representation ($i=1,\cdots N_\mathrm{c}^2-1$), and repeated indices are summed unless otherwise noted.
For convenience in later numerical calculations, we introduce adimensional Hamiltonians defined by
\begin{align}
  H_\mathrm{E}/g_0&=J\sum_{n=1}^{N-1}E_i^2(n), \\
  H_\mathrm{hop}/g_0&=\hop\sum_{n=1}^{N-1}\qty(\chi^\dag(n+1) U(n) \chi(n)+\chi^\dag(n) U^\dag(n) \chi(n+1)), \\
  H_\mathrm{m}/g_0 &=m\sum_{n=1}^N(-1)^{n} \chi^\dag(n) \chi(n) .
\end{align}
Here, $J=ag_0/2$, $\hop=1/(2ag_0)$, and $m=m_0/g_0$ are the dimensionless couplings.
Hereafter, we use the unit system with $g_0=1$ and omit $g_0$ where it does not lead to confusion.

The fermion fields satisfy the anticommutation relation:
\begin{equation}
  \{\chi_{f,c}(n),\chi^{f',c'\dag}(n')\}\coloneqq
  \chi_{f,c}(n)\chi^{f',c'\dag}(n')+\chi^{f',c'\dag}(n')\chi_{f,c}(n)
  =\delta_{n,n'}\delta_{c}^{c'}\delta_{f}^{f'},
  \label{eq:anticommutationrelation}
\end{equation}
where $c$ and $f$ represent the color and flavor indices, respectively.
In the Hamiltonian formulation on a lattice, there are two types of electric fields, $\ER_i(n)$ and $\EL_i(n)$, which generate the gauge transformation of $U$ from the right and left, respectively. Their commutation relations are as follows:
\begin{align}
  [\ER_i(n),U(n')] &=U(n) T_i\delta_{n,n'}, \label{eq:[E_s,U]}\\
  [\EL_i(n),U(n')] &=T_iU(n)\delta_{n,n'} , \label{eq:[E_t,U]}
  \\
[\ER_i(n),\ER_j(n')] &=\ri f^{k}_{~ij}\ER_{k}(n)\delta_{n,n'},    \label{eq:[E_s,E_s]} \\
[\EL_i(n),\EL_j(n')] &=-\ri f^k_{~ij}\EL_{k}(n)\delta_{n,n'}    \label{eq:[E_t,E_t]}.
\end{align}
Here, $T_i$ are generators of fundamental representation that satisfy the commutation relation,
\begin{equation}
  [T_i,T_j]=\ri f^k_{~ij} T_k
\end{equation}
with the structure constant $f^k_{~ij}$.
$\ER_i(n)$ and $\EL_i(n)$ are not independent but related by
a parallel transport:
\begin{equation}
  \ER_i(n) =\EL_j(n)[U_{\mathrm{adj}}(n)]_i^j,\quad
  \mathrm{or} \quad
  \EL_i(n) = \ER_j(n)[U^\dag_{\mathrm{adj}}(n)]_i^j,
  \label{eq:E_relation}
\end{equation}
where $[U_{\mathrm{adj}}(n)]_i^j$ is the link variable with the adjoint representation defined by
\begin{equation}
  \begin{split}
    U T_i U^\dag &=T_j[U_{\mathrm{adj}}]_i^j,
    \label{eq:adjoint}
  \end{split}
\end{equation}
which satisfies $[U^\dag_{\mathrm{adj}}(n)]_i^j=[U_{\mathrm{adj}}(n)]^i_j$.
From eq.~\eqref{eq:E_relation}, the square of $\ER_i(n)$ and $\EL_i(n)$ are identical,
$(\ER_i(n))^2=(\EL_i(n))^2\eqqcolon E_i^2(n)$. Therefore, $H_\mathrm{E}$ is independent of the choice of electric fields.

A physical state $\ket{\Psi}$ is gauge invariant, which needs to satisfy the Gauss-law constraint:
\begin{equation}
  G_i(n)\ket{\Psi}=0,
\end{equation}
where $G_i(n)$ is defined by
\begin{equation}
    G_i(n)\coloneqq\ER_i(n)-\EL_i(n-1)+Q_i(n)
\end{equation}
with the color-charge density operator,
\begin{equation}
  Q_i(n)\coloneqq\chi^\dag(n)T_i \chi(n).
\end{equation}
$Q_i(n)$ satisfy the following commutation relation as do electric fields:
\begin{equation}
  [Q_i(n),Q_j(n')]= \ri f^k_{~ij}Q_k(n) \delta_{n,n'}.
\end{equation}
Note that the Hamiltonian is gauge invariant, so that the Hamiltonian commutes with $G_i(n)$, $[H, G_i(n)]=0$.

\subsection{Hamiltonian in a rotated basis}\label{sec:rotated_Hamiltonian}
In $(1+1)$ dimensions, the gauge fields are non-dynamical and can be eliminated from the Hamiltonian in open boundary conditions.
Following refs.~\cite{Sala:2018dui,Atas:2021ext},
we remove link variables $U(n)$ from the Hopping term $H_{\mathrm{hop}}$ by a unitary transformation, i.e., 
\begin{equation}
  \chi^\dag(n+1) U(n) \chi(n)\to
  \Theta\chi^\dag(n+1) U(n) \chi(n) \Theta^\dag
  =\chi^\dag(n+1)\chi(n).
  \label{eq:elmination_of_U}
\end{equation}
This can be accomplished if the unitary operator $\Theta$ satisfies
\begin{align}
  \Theta\chi(n) \Theta^\dag
  &= U(n-1)U(n-2)\cdots U(1)\chi(n)
  \label{eq:unitary_trans_1+1},\\
  \Theta U(n) \Theta^\dag &=U(n).
  \label{eq:unitary_trans_1+1-2}
\end{align}
The construction of $\Theta$ is as follows:
We can express the link variable $U(n)$ by introducing a gauge field $A^i(n)$ as
\begin{equation}
  U(n)=\exp(\ri T_iA^i(n)).
\end{equation}
Using this gauge field $A^i(n)$, we introduce the following operator:
\begin{equation}
  V(n) =\exp\qty(-\ri A^i(n)\sum_{m=n+1}^NQ_i(m)),
\end{equation}
which satisfies
\begin{equation}
  V(k)\chi(n)V^\dag(k)= 
  \begin{cases}
    U(k)\chi(n)& \mathrm{if}\quad n>k\\
    \chi(n)& \mathrm{else}\\ 
  \end{cases}.
\end{equation}
The unitary operator $\Theta$ satisfying eqs.~\eqref{eq:unitary_trans_1+1} and \eqref{eq:unitary_trans_1+1-2} is constructed from $V$ as
\begin{equation}
  \Theta
  =V(1)V(2)V(3)\cdots V(N-1).
\end{equation}
Using $\Theta$, we eliminate the link variables from the Hopping term $H_{\mathrm{hop}}$,
\begin{equation}
  \tilde{H}_{\mathrm{hop}}\coloneqq\Theta H_{\mathrm{hop}}\Theta^\dag
  =\hop\sum_{n=1}^{N-1}\qty(\chi^\dag(n+1)\chi(n)+\chi^\dag(n)  \chi(n+1)).
\end{equation}
The consequence of this transformation is that the electric part of the Hamiltonian, $H_\mathrm{E}$, becomes nonlocal and more complex. However, the Gauss-law constraint becomes simpler and can be easily solved under open boundary conditions.

The right-electric field $R_i(n)$ and the color-charge density $Q_i(n)$ transform under $V$ as
\begin{equation}
  \begin{split}
    V(n) \ER_i(n)V^\dag(n) &= V(n) V^\dag(n) \ER_i(n)
    +V(n)[\ER_i(n),V^\dag(n)]\\
    &=\ER_i(n)+ \sum_{m=n+1}^NQ_i(m),
  \end{split}
\end{equation}
and
\begin{equation}
  V(k)Q_i(n) V^\dag(k)=
  \begin{cases}
    Q_j(n)[U_{\mathrm{adj}}^\dag(k)]^j_i & \mathrm{if}\quad n>k\\
    Q_i(n) & \text{else}
  \end{cases},
\end{equation}
which lead to 
\begin{equation}
  \begin{split}
    \Theta \ER_i(n)\Theta^\dag&=
    V(1)V(2)\cdots V(n-1)
    \qty(
      \ER_i(n)+ \sum_{m=n+1}^NQ_i(m)
    )
    V^\dag(n-1)\cdots V^\dag(2) V^\dag(1)\\
      &=
      \ER_i(n)+ \sum_{m=n+1}^N
      Q_j(m)[U_{\mathrm{adj}}^\dag(1)U_{\mathrm{adj}}^\dag(2)
      \cdots U_{\mathrm{adj}}^\dag(n-1)]^j_i.
      \label{eq:Theta_Es}
  \end{split}
\end{equation}
The transformation of the left-electric field $\EL_i(n)$ can be obtained from eqs.~\eqref{eq:Theta_Es} and \eqref{eq:E_relation} as
\begin{equation}
  \Theta \EL_i(n)\Theta^\dag
    =
    \EL_i(n)+ \sum_{m=n+1}^N
    Q_j(m)[U_{\mathrm{adj}}^\dag(1)U_{\mathrm{adj}}^\dag(2)
    \cdots U_{\mathrm{adj}}^\dag(n)]^j_i.
\end{equation}
Using these transformations,
the Gauss-law constraint $G_i(n)$ for $n>1$ reduces to
\begin{equation}
  \begin{split}
    \Theta  G_i(n)\Theta^\dag
    &=\Theta \ER_i(n)\Theta^\dag
    -\Theta \EL_i(n-1)\Theta^\dag
    +\Theta Q_i(n)\Theta^\dag\\
    &=
    \ER_i(n)+ \sum_{m=n+1}^N
    Q_j(m)[U_{\mathrm{adj}}^\dag(1)U_{\mathrm{adj}}^\dag(2)
    \cdots U_{\mathrm{adj}}^\dag(n-1)]^j_i\\
    &\quad-\EL_i(n-1) - \sum_{m=n}^N Q_j(m)[U_{\mathrm{adj}}^\dag(1)U_{\mathrm{adj}}^\dag(2)
    \cdots U_{\mathrm{adj}}^\dag(n-2)
    U_{\mathrm{adj}}^\dag(n-1)]^j_i
    \\
    &\quad+Q_j(n)[U_{\mathrm{adj}}^\dag(1)U_{\mathrm{adj}}^\dag(2)\cdots U_{\mathrm{adj}}^\dag(n-1)]_i^j\\
    &=\ER_i(n)-\EL_i(n-1).
  \end{split}
\end{equation}
On the other hand, for $n=1$, it is
\begin{equation}
  \begin{split}
    \Theta  G_i(1)\Theta^\dag
    &= \ER_i(1)+ \sum_{m=2}^NQ_i(m)
    -\EL_i(0)
    +Q_i(1)\\
    &=\ER_i(1)+ \sum_{m= 1}^NQ_i(m)
    -\EL_i(0).
    \end{split}
\end{equation}
Here, $\EL_i(0)$ represents the incoming flux to the system.
In the following, we impose no incoming flux condition, $\EL_i(0)=0$;
then we have
\begin{equation}
  \ER_i(1)=- \sum_{m=1}^NQ_i(m).
  \label{eq:initial_ER}
\end{equation}
The Gauss-law constraint is now a simple recurrence relation with the initial value~\eqref{eq:initial_ER}, which can be analytically solved:
\begin{equation}
  \begin{split}
    \ER_i(n)&=\EL_i(n-1)\\
    &=\ER_j(n-1)[U^\dag_{\mathrm{adj}}(n-1)]_i^j\\
    &=\ER_j(1)[U^\dag_{\mathrm{adj}}(1)U^\dag_{\mathrm{adj}}(2)\cdots U^\dag_{\mathrm{adj}}(n-1)]^j_i\\
    &=-\sum_{m= 1}^NQ_j(m)[U^\dag_{\mathrm{adj}}(1)U^\dag_{\mathrm{adj}}(2)\cdots U^\dag_{\mathrm{adj}}(n-1)]^j_i.
    \label{eq:Es}
  \end{split}
\end{equation}
Substituting eq.~\eqref{eq:Es}
into eq.~\eqref{eq:Theta_Es}, we obtain
\begin{equation}
  \begin{split}
    \Theta \ER_i(n)\Theta^\dag
    &=   \qty(\sum_{m=n+1}^{N_\mathrm{c}}-\sum_{m=1}^{N_\mathrm{c}})
    Q_j(m)[U_{\mathrm{adj}}^\dag(1)U_{\mathrm{adj}}^\dag(2)
    \cdots U_{\mathrm{adj}}^\dag(n-1)]^j_i\\
    &= -\sum_{m=1}^n
    Q_j(m)[U_{\mathrm{adj}}^\dag(1)U_{\mathrm{adj}}^\dag(2)
    \cdots U_{\mathrm{adj}}^\dag(n-1)]^j_i.
  \end{split}
\end{equation}
Therefore, the electric part of the Hamiltonian in the rotated basis becomes
\begin{equation}
  \begin{split}
    \tilde{H}_\mathrm{E}\coloneqq\Theta H_\mathrm{E}  \Theta^\dag &= J\sum_{n=1}^{N-1}\qty(\sum_{m= 1}^nQ_i(m))^2\\
    &= J\qty(
      \sum_{n=1}^{N-1}(N-n)Q^2_i(n)  
    +2\sum_{n=1}^{N-2}\sum_{m= n+1}^{N-1}(N-m)Q_i(n)Q_i(m)
    ),
  \end{split}
\end{equation}
which is expressed only by the fermion fields.
Note that transformation of the mass term is trivial: $\tilde{H}_\mathrm{m} =H_\mathrm{m}$.
Consequently, the Hamiltonian in the rotated basis becomes
\begin{equation}
  \tilde{H}=\Theta H\Theta^\dag = \tilde{H}_\mathrm{E} + \tilde{H}_\mathrm{hop} +\tilde{H}_\mathrm{m}
\end{equation}
with
\begin{align}
  \tilde{H}_\mathrm{E}&=J\qty(
    \sum_{n=1}^{N-1}(N-n)Q^2_i(n)  
  +2\sum_{n=1}^{N-2}\sum_{m= n+1}^{N-1}(N-m)Q_i(n)Q_i(m)
  ), \label{eq:tilde_HE}\\
  \tilde{H}_\mathrm{hop}&=\hop\sum_{n=1}^{N-1}(\chi^\dag(n+1) \chi(n)+\chi^\dag(n)\chi(n+1)), \label{eq:tilde_Hhop}\\
  \tilde{H}_\mathrm{m} &=m\sum_{n=1}^N(-1)^{n} \chi^\dag(n) \chi(n). \label{eq:tilde_HM} 
\end{align}
Thus far, the outgoing flux $\ER_i(N)$ is not restricted, although we imposed the no incoming flux condition $\EL_i(0)=0$.
If we impose no outgoing flux condition, $\ER_i(N)=0$, we may add a penalty term,
\begin{equation}
  H_\lambda  = \lambda (\ER_i(N))^2 = \lambda\qty(\sum_{n=1}^N Q_i(n))^2,
\end{equation}
into the Hamiltonian, $\tilde{H}+ H_\lambda$. We will consider this setup in the numerical calculations.

\subsection{Observables}\label{sec:observables}
We are interested in local observables, correlation functions, and thermodynamic quantities. Here, we summarize these observables represented on a lattice.
\subsubsection{Local observables}
The two-component fermions $\psi=(\psi_\mathrm{L},\psi_\mathrm{S})$ in the Dirac representation correspond to 
\begin{align}
  \psi_\mathrm{L}(n)&=(\ri)^{2n}{\hop^{\frac{1}{2}}}\chi(2n),\\
  \psi_\mathrm{S}(n)&=(\ri)^{2n-1}{\hop^{\frac{1}{2}}}\chi(2n-1),
\end{align}
and gamma matrices are $\gamma^0=\sigma^z$, $\gamma^1=\ri\sigma^y$, $\gamma^5=\gamma^0\gamma^1=\sigma^x$ (See Appendix \ref{sec:Free_theory} for the detailed correspondence in the case of the free theory).
The physical coordinate $n$ runs from $1$ to $L\coloneqq N/2$.
Bilinear local operators, the quark-number density, the current, scalar, and pseudo-scalar operators in the rotated basis,
are 
\begin{align}
   \bar{\psi}\gamma^0\psi(n) &=\hop\chi^\dag(2n-1)\chi(2n-1)+\hop\chi^\dag(2n)\chi(2n), \label{eq:j0}\\
   \bar{\psi}\gamma^1\psi(n) &= -\ri\hop\chi^\dag(2n)\chi(2n-1)+\ri\hop\chi^\dag(2n-1)\chi(2n),\label{eq:j1}\\
   \bar{\psi}\psi(n) &=\hop\chi^\dag(2n)\chi(2n)-\hop\chi^\dag(2n-1)\chi(2n-1),\label{eq:scalar}\\
   \bar{\psi}\ri\gamma_5\psi(n) &= \hop\chi^\dag(2n)\chi(2n-1)+\hop\chi^\dag(2n-1)\chi(2n), 
   \label{eq:pseudoscalar}
\end{align}
respectively. The expectation value of an operator $O$ is defined as $\expval{O}=\bra{\Omega} O\ket{\Omega}$, where $ \ket{\Omega}$ is the ground state wave function.
\subsubsection{Correlation and quark distribution functions}\label{sec:distribution}
Another observable in the rotated basis is a two-point function,
\begin{equation}
  \begin{split}
    S^<_{s,s'}(n,n')&\coloneqq\expval{\psi_{s'
    }^\dag(n')\psi_{s}(n)}\\
    & =
    \hop(-1)^{n'-n}
    \begin{pmatrix}
      \expval{\chi^\dag(2n')\chi(2n)} &  \ri\expval{\chi^\dag(2n'-1)\chi(2n)}\\
      -\ri\expval{\chi^\dag(2n')\chi(2n-1)} & \expval{\chi^\dag(2n'-1)\chi(2n-1)}
    \end{pmatrix},
    \label{eq:two-point_function}
  \end{split}
\end{equation}
where $s,s'\in\{\mathrm{L},\mathrm{S}\}$.
We note that eq.~\eqref{eq:two-point_function} is gauge invariant. This is because $\expval{\chi^\dag(n')\chi(n)}$ in the rotated basis corresponds to $\expval{\chi^\dag(n)U(n-1)U(n-2)\cdots U(n'+1)U(n')\chi(n')}$ in the original basis, which is manifestly gauge invariant.

We define the Wigner transform of $\sum_{s} S_{s,s}^<(n,n')$ by
\begin{equation}
  W(x,p) = \frac{1}{\hop}\sum_{\ell=-K}^{K} e^{\ri \frac{p\ell}{\hop} }\sum_{s=\mathrm{L},\mathrm{S}}S_{s,s}^<\qty(\Bigl\lfloor \hop x+\frac{1}{2}+\frac{\ell}{2}\Bigr\rfloor,\Bigl\lfloor \hop x+\frac{1}{2}-\frac{\ell}{2}\Bigr\rfloor),
  \label{eq:W(x,p)}
\end{equation}
which represents the distribution function of quarks.
Here, $\lfloor\cdot\rfloor$ is the floor function,
$ x$ is the position in physical units
with $\hop x=1,2,\cdots,L-1$, and $p=2\hop\pi k/(2K+1)$ is the momentum with $k=-K,-K+1,\cdots, K$ ($K=2\min(\hop x,L-\hop x)-1$).
Because there is a boundary, the possible momentum is restricted, unlike the infinite system.
For numerical calculations, we will use the center position, $x=L/(2\hop)$, which allows the maximum number of points for the momentum.
We have introduced the floor function to restrict $\hop x+1/2 \pm \ell/2$ to an integer.
The $1/2$ in the argument is a shift term introduced to ensure the distance between two points changes by one when $\ell$ is increased by one.
One could use the ceiling function instead of the floor function, 
whose difference disappears in the large volume limit $L\to\infty$.
Using eq.~\eqref{eq:W(x,p)}, we define one particle distribution function $n(p)$ by 
\begin{equation}\label{eq:n(p)}
  n(p) \coloneqq \left.\frac{1}{N_\mathrm{c}N_\mathrm{f}}W(x,p)\right|_{x=L/(2\hop)}-1.
\end{equation}
The vacuum contribution is subtracted so that the number density vanishes at the vacuum.
In the case of the free theory at zero temperature and finite density in the continuum $a\to0$ and large volume $L\to\infty$ limits, it corresponds to
\begin{equation}
  n(p)=\theta(\mu-E_p), \label{eq:def_of_Wigner_function}
\end{equation}
where $\theta(z)$ is the Heaviside step function, and $E_p=\sqrt{p^2+m^2}$ is the energy, $\mu$ is the quark chemical potential.
There is the Fermi surface at $p=\sqrt{\mu^2-m^2}$. If one turns on an attractive interaction, the Fermi surface will disappear, and the distribution function $n(p)$ will become a smooth function. We will see this is the case for both two and three colors in section~\ref{sec:Numerical_results}.

\subsubsection{Thermodynamic quantities}
We are interested in a finite-density system at zero temperature.
To this end, we minimize 
\begin{equation}\label{eq:H-muN}
  H-\mu_{\mathrm{B}} N_\mathrm{B}= H-\mu N_\mathrm{q},
\end{equation}
instead of the Hamiltonian,
where $N_\mathrm{B}= N_\mathrm{q}/N_\mathrm{c}$ is the baryon number operator, and $N_\mathrm{q}$ is the quark number operator,
\begin{equation}
  N_\mathrm{q}={\frac{1}{\hop}}\sum_{n=1}^L\bar{\psi}(n)\gamma^0\psi(n)-L N_\mathrm{c}N_\mathrm{f}=
  \sum_{n=1}^N\chi^\dag(n)\chi(n)-L N_\mathrm{c}N_\mathrm{f}.
\end{equation}
Here, we subtract the constant $L N_\mathrm{c}N_\mathrm{f}$ to ensure the baryon number vanishes at the vacuum.
Similarly, the Hamiltonian is used with the vacuum energy implicitly subtracted.
$\mu_\mathrm{B}$ is the baryon chemical potential that relates the quark chemical potential $\mu$ through $\mu_\mathrm{B}=N_\mathrm{c}\mu$.

The pressure $P$ is given by
\begin{equation}\label{eq:pressure}
  P = \mu_\mathrm{B} n_\mathrm{B} - \varepsilon,
\end{equation}
where we defined the baryon number $n_\mathrm{B}$ and energy density $\varepsilon$ as
\begin{align}
  n_\mathrm{B} &= \frac{1}{V}\expval{N_\mathrm{B}},\label{eq:baryon_number}\\
  \varepsilon &= \frac{1}{V}\expval{H},
\end{align}
with $V = L/\hop$ being the physical volume.

\subsection{Mapping to spin system}\label{sec:map_to_spin}
For numerical calculations, it is more convenient to use spin degrees of freedom than fermionic ones.
Using the Wigner-Jordan transformation~\cite{Jordan:1928wi}, we can map our fermionic system to a $(NN_\mathrm{f}N_\mathrm{c})$-sites spin system:
\begin{align}
  \chi_{f,c}(n) &=\qty[\prod_{k<\ell(n,f,c)}(-\sigma^z_k)]\sigma^-_{\ell(n,f,c)}, \label{eq:chi_spin}\\
  \chi^\dag_{f,c}(n) &=\qty[\prod_{k<\ell(n,f,c)}(-\sigma^z_k)]\sigma^+_{\ell(n,f,c)},
  \label{eq:chidag_spin}
\end{align}
where $\sigma^{\pm}_{l}=(\sigma^{x}_{l}\pm\ri\sigma^{y}_{l})/2$ with the Pauli matrices $\sigma^{i}$ ($i=x,y,z$), and we introduced the mapping of sites from the original system $(n,f,c)$ to the spin system, $\ell(n,f,c)=(n-1)N_\mathrm{f}N_\mathrm{c}+N_\mathrm{c}(f-1)+c-1$.
By construction, $\chi_{f,c}(n)$ and $\chi^\dag_{f,c}(n)$ in eqs.~\eqref{eq:chi_spin} and \eqref{eq:chidag_spin} satisfy the canonical anti-commutation relation~\eqref{eq:anticommutationrelation} thanks to $\sigma^z\sigma^{\pm}=-\sigma^{\pm}\sigma^z$.

Using the spin degrees of freedom, the hopping term~\eqref{eq:tilde_Hhop} is expressed as
\begin{equation}
  \tilde{H}_\mathrm{hop}=\hop\sum_{n=1}^{N-1}
  \sum_{f=1}^{N_\mathrm{f}}\sum_{c=1}^{N_\mathrm{c}}\qty(
  \sigma_{\ell(n+1,f,c)}^+ \Sigma_{f,c}(n)\sigma_{\ell(n,f,c)}^-
  +\sigma_{\ell(n,f,c)}^+\Sigma^\dag_{f,c}(n) \sigma^-_{\ell(n+1,f,c)}
  ),
\end{equation}
 where $ \Sigma_{f,c}(n)$ is defined as
 \begin{equation}
  \begin{split}
  \Sigma_{f,c}(n)
  &=\prod_{k=\ell(n,f,c)}^{\ell(n+1,f,c)-1}(-\sigma_{k}^z).
  \end{split}
\end{equation}
Similarly, the mass term is
\begin{equation}
  \tilde{H}_\mathrm{m} =m\sum_{n=1}^N(-1)^{n}\sum_{f=1}^{N_\mathrm{f}}\sum_{c=1}^{N_\mathrm{c}}\frac{1}{2}(\sigma^z_{\ell(n,f,c)}+1).
\end{equation}
For the electric part of the Hamiltonian~\eqref{eq:tilde_HE}, we need to express the color-charge densities by spins.
For this purpose, we introduce the following basis of generators:
\begin{equation}
  [T_{(ij)}]_{kl}= \frac{1}{\sqrt{2}}\delta_{ik} \delta_{jl},
\end{equation}
for $i\neq j$,
and 
\begin{equation}
  [T_{(ii)}]_{kl}=\frac{1}{\sqrt{2i(i-1)}}\times \begin{cases}
    \delta_{kl} &  k<i\\
    (1-i)\delta_{kl} & k=i\\
    0 & k>i
  \end{cases}
\end{equation}
for $i>1$. These generators satisfy
\begin{equation}
  \tr T^\dag_{(ij)}T_{(kl)} = \frac{1}{2}\delta_{ik}\delta_{jl}.
\end{equation}
Since $T_{(ji)}=T_{(ij)}^\dag$, we can restrict the indices of $T_{(ij)}$ to $i>j$ and use $T_{(ij)}^\dag$ for $j>i$.
Using this basis, $Q_i(n)Q_i(m)$ in the electric part~\eqref{eq:tilde_HE} is expressed as
\begin{equation}
  Q_i(n)Q_i(m)  = \sum_{i=2}^{N_\mathrm{c}}\sum_{j=1}^{i-1}\qty(
    Q^\dag_{(ij)}(n)Q_{(ij)}(m)+Q_{(ij)}(n)Q^\dag_{(ij)}(m))
  + \sum_{i=2}^{N_\mathrm{c}} Q_{(ii)}(n)Q_{(ii)}(m).
  \label{eq:Qsqure_ij}
\end{equation}

Let us now map $Q_{(ij)}(n)$ to the spin degree of freedom:
\begin{equation}
  Q_{(ij)}(n)
  =\chi^\dag(n)T_{(ij)}\chi(n)
  = \sum_{f=1}^{N_\mathrm{f}}\frac{1}{\sqrt{2}}\sigma_{\ell(n,f,i)}^+\Lambda_{f(ij)}(n)\sigma_{\ell(n,f,j)}^-,
  \label{eq:Qij_spin}
\end{equation}
for $i> j$,
and
\begin{equation}
  Q_{(ii)}(n) 
  =\sum_{f=1}^{N_\mathrm{f}}\frac{1}{2\sqrt{2i(i-1)}}\sum_{c=1}^{i-1}\qty(\sigma^z_{\ell(n,f,c)}-\sigma^z_{\ell(n,f,i)}),
  \label{eq:Qii_spin}
\end{equation}
for the diagonal parts,
where
\begin{equation}
  \Lambda_{f(ij)}(n)=\prod_{k=\ell(n,f,j)}^{\ell(n,f,i)-1}(-\sigma_k^z).
\end{equation}
Substituting eqs.~\eqref{eq:Qij_spin} and \eqref{eq:Qii_spin} into \eqref{eq:tilde_HE}
combined with eq.~\eqref{eq:Qsqure_ij},
we obtain the electric part represented by the spin degrees of freedom.

Finally, let us show the observables introduced in section~\ref{sec:observables}.
Bilinear local operators~\eqref{eq:j0}-\eqref{eq:pseudoscalar} are expressed as
\begin{align}
  \bar{\psi}\gamma^0\psi(n)
  &=\hop\sum_{f=1}^{N_\mathrm{f}}\sum_{c=1}^{N_\mathrm{c}}\frac{1}{2}\qty(\sigma_{\ell(2n,f,c)}^z+\sigma_{\ell(2n-1,f,c)}^z)+\hop N_\mathrm{c}N_\mathrm{f},\\
  \bar{\psi}\gamma^1\psi(n)
  &=\hop\sum_{f=1}^{N_\mathrm{f}}\sum_{c=1}^{N_\mathrm{c}}\Bigl(
    -\ri\sigma_{\ell(2n,f,c)}^+ \Sigma_{f,c}(2n-1)\sigma_{\ell(2n-1,f,c)}^-
    +\ri\sigma_{\ell(2n-1,f,c)}^+\Sigma^\dag_{f,c}(2n-1) \sigma^-_{\ell(2n,f,c)}\Bigr),\\
  \bar{\psi}\psi(n)
    &=\hop\sum_{f=1}^{N_\mathrm{f}}\sum_{c=1}^{N_\mathrm{c}}\frac{1}{2}\qty(\sigma_{\ell(2n,f,c)}^z-\sigma_{\ell(2n-1,f,c)}^z),\\
  \bar{\psi}\ri\gamma_5\psi(n) 
  &=\hop\sum_{f=1}^{N_\mathrm{f}}\sum_{c=1}^{N_\mathrm{c}}\Bigl(
    \sigma_{\ell(2n,f,c)}^+ \Sigma_{f,c}(2n-1)\sigma_{\ell(2n-1,f,c)}^-
    +\sigma_{\ell(2n-1,f,c)}^+\Sigma^\dag_{f,c}(2n-1) \sigma^-_{\ell(2n,f,c)}\Bigr).
\end{align}
Similarly, the two-point function is
\begin{equation}
  \begin{split}
    S^<(n,m)& =
    \hop\sum_{f=1}^{N_{\mathrm{f}}}\sum_{c=1}^{N_\mathrm{c}}
    (-1)^{m-n}
    \begin{pmatrix}
      D_{\ell(2m,f,c),\ell(2n,f,c)}
       &  \ri D_{\ell(2m-1,f,c),\ell(2n,f,c)}\\
      -\ri D_{\ell(2m,f,c),\ell(2n-1,f,c)} &
      D_{\ell(2m-1,f,c),\ell(2n-1,f,c)}
    \end{pmatrix},
    \label{eq:two-point_function2}
  \end{split}
\end{equation}
where we defined
\begin{equation}
  D_{n,m} = \expval{\sigma_{m}^+\qty[\prod_{k=\min(m,n)}^{\max(m,n)-1}(-\sigma^z_k)]\sigma^-_n}.
\end{equation}

\section{Matrix Product state and Density matrix renormalization group}\label{sec:DMRG}
We utilize the MPS under open boundary conditions as a variational ansatz, which is given explicitly as 
\begin{equation}
  \ket{\psi} =\sum_{\alpha_1,\cdots, \alpha_N}\sum_{i_1,i_2,\cdots, i_N}^2 A_{\alpha_1}^{i_1}A_{\alpha_1\alpha_2}^{i_2}\cdots A_{\alpha_N}^{i_N}\ket{i_1,i_2,\cdots i_N} ,
  \label{eq:wave_function}
\end{equation}
where $\ket{i_1,i_2,\cdots i_N}$ form the $2^N$-dimensional Hilbert space of the $N$-site spin chain, and $A^{i=2,\cdots, N-1}$ [$A^{i=1,N}$] are $D\times D$ complex matrices [$D$-dimensional complex vectors], respectively.
$\alpha_i$ are called bond indices, and $D$ is the bond dimension.
MPS is graphically expressed e.g., for $N=8$ as
\begin{equation}
\sum_{\alpha_1,\cdots \alpha_N}A_{\alpha_1}^{i_1}A_{\alpha_1\alpha_2}^{i_2}\cdots A_{\alpha_N}^{i_N}=
  \parbox[c]{15em}{
  \centering
  \includegraphics[scale=0.5]{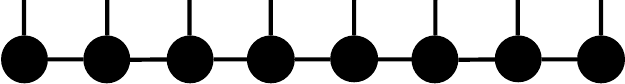} \ ,
  }
 \end{equation}
where shapes and lines represent tensors and their legs, respectively. The connected lines imply contractions of connected indices of tensors.
Similarly, we express the spin Hamiltonian $\tilde{H}$ in the matrix product operator (MPO) form:
\begin{equation}
  \tilde{H} =\sum_{\alpha_1,\cdots \alpha_N}\sum_{i_1,i_2,\cdots i_N}^2\sum_{j_1,j_2,\cdots j_N}^2 C_{j_1}^{i_1\alpha_1}C_{\alpha_1 j_2}^{i_2\alpha_2}\cdots C_{\alpha_N j_N}^{i_N}\ket{i_1,i_2,\cdots i_N}\bra{j_1,j_2,\cdots j_N} .
\end{equation}
Graphically, it is expressed as
\begin{equation}
  \centering
  \parbox[c]{17em}{
  \includegraphics[scale=0.5]{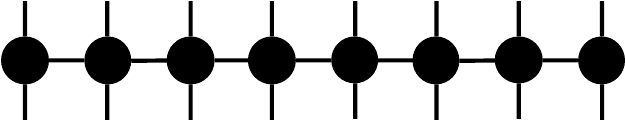}
  }.
 \end{equation}
Our purpose is to find the wave function~\eqref{eq:wave_function} that minimizes the expectation value of the Hamiltonian.
To achieve this, we adopt the MPO version of the DMRG (two-site update)~\cite{PhysRevLett.69.2863,SCHOLLWOCK201196}.

The procedure of the algorithm is as follows:
Using the singular value decomposition, we first prepare an initial MPS that is expressed into an orthogonal form (canonical form),
\begin{equation}
  \parbox[c]{18em}{
  \centering
  \includegraphics[scale=0.5]{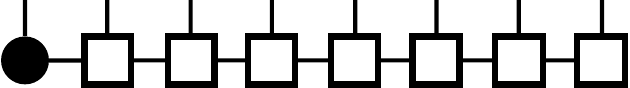}
  } ,
 \end{equation}
where the orthogonality center is set to the site $1$, and the tensors represented by open squares at $i=2,\cdots,N$ are right-orthogonal, i.e.,
 \begin{equation}
  \parbox[c]{5em}{
  \centering
  \includegraphics[scale=0.5]{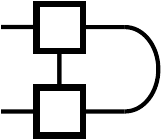}
  }
=
  \parbox[c]{4em}{
  \centering
  \includegraphics[scale=0.5]{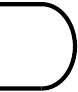}
  }.
 \end{equation}
As an initial step, we update the tensor on the left two sites,
\begin{equation}
  \parbox[c]{10em}{
  \centering
  \includegraphics[scale=0.5]{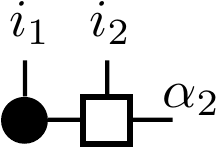}
  }
  .
  \label{eq:MPS12}
\end{equation}
To achieve this, we consider an effective Hamiltonian obtained by contracting the remaining tensor with physical indices $i_3,i_4,\cdots i_N$ and bond index $\alpha_2$.
Due to the orthogonal property (partial isometry), MPS tensors such as
 \begin{equation}
  \parbox[c]{15em}{
  \centering
  \includegraphics[scale=0.5]{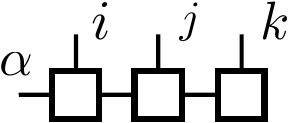}
  }
  \end{equation}
can be understood as the basis transformation from the physical indices ($i$, $j$, $k$) to the virtual bond indices $\alpha$.
Thus, we obtain the effective Hamiltonian by transforming $\tilde{H}$ into the ($i_1$, $i_2$, $\alpha_2$) basis as explicitly given as
\begin{equation}
  \parbox[c]{18em}{
  \centering
  \includegraphics[scale=0.5]{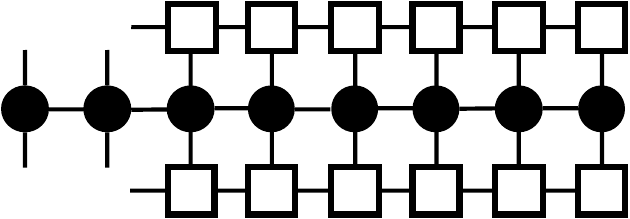}
  }.
  \end{equation}
Since the Hamiltonian is given by MPO, the contractions of tensors can be done sequentially and efficiently by defining and computing $R_j$ tensors as
\begin{equation}
\begin{split}
  &\parbox[c]{15em}{
  \centering
  \includegraphics[scale=0.5]{figures/Hamiltonian_reduced.pdf}
  }
  \\
  &=\ \parbox[c]{15em}{
  \centering
  \includegraphics[scale=0.5]{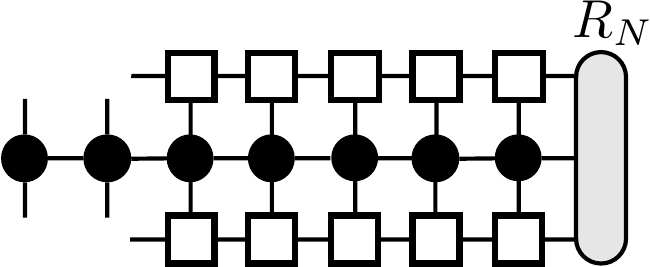}
  }
  \\
  &=\cdots
  \\
  &=\parbox[c]{8em}{
  \centering
  \includegraphics[scale=0.5]{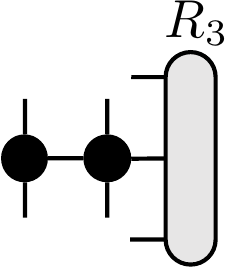} 
  }.
  \end{split}
\label{eq:effective_hamiltonian}
 \end{equation}
We note that the $R_j$ tensors can be recycled during the sweep of DMRG.
In order to update the tensor~\eqref{eq:MPS12}, our task is now to find a new bond tensor $B_{12}$,
\begin{equation}
  \parbox[c]{10em}{
  \centering
  \includegraphics[scale=0.5]{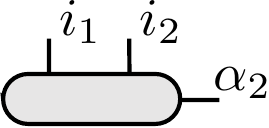}
  },
  \end{equation}
which is defined in the $(i_1,i_2,\alpha_2)$ basis and minimizes the effective Hamiltonian~\eqref{eq:effective_hamiltonian}.
If abstract space spanned by $\alpha_2$ well approximates the subspace on the ($i_3,\cdots,i_N$) space, in which the full eigenvector has support,
the full Hamiltonian can be minimized by optimizing the bond tensors. The DMRG algorithm tries to improve this abstract space by sequentially updating the bond tensors.

After obtaining the optimized bond tensor $B_{12}$, which is done by e.g., the Lanczos algorithms, we need to restore the bond tensor to the MPS form.
This can be done by using the singular value decomposition:
\begin{equation}
  \parbox[c]{10em}{
  \centering
  \includegraphics[scale=0.5]{figures/Bondtensor.pdf}
  }
  \simeq
  \parbox[c]{10em}{
  \centering
  \includegraphics[scale=0.5]{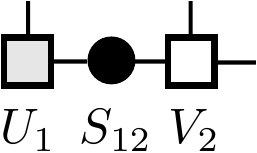}
  }
  =
  \parbox[c]{10em}{
  \centering
  \includegraphics[scale=0.5]{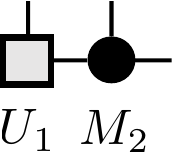}
  },
  \end{equation}
which is used as the update of eq.~\eqref{eq:MPS12}.
Here, we keep only the $m$ largest singular values of $B_{12}$ and corresponding columns (rows) of $U_1$ ($V_2$).
Alternatively, we can retrieve the MPS tensor by diagonalizing the density matrix:
\begin{equation}
  \parbox[c]{10em}{
  \centering
  \includegraphics[scale=0.5]{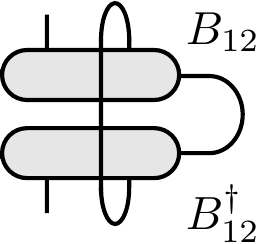}
  }
  =
  \parbox[c]{10em}{
  \centering
  \includegraphics[scale=0.5]{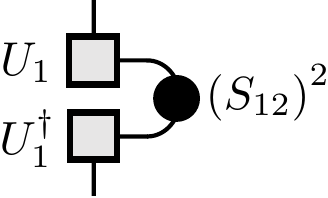}
  },
  \end{equation}
where we truncate $U_1$ according to the eigenvalue $\left(S_{12}\right)^2$, and $M_2$ is given as $U_1^\dagger\cdot B_{12}$.
Although these methods are formally equivalent, the latter has several advantages. 
One of them is the noise term perturbation~\cite{PhysRevB.72.180403}, which is used to accelerate the convergence of DMRG.
The MPS tensor $U_1$ obtained from these methods is left orthogonal, i.e., 
 \begin{equation}
  \parbox[c]{5em}{
  \centering
  \includegraphics[scale=0.5]{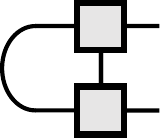}
  }
=
  \parbox[c]{5em}{
  \centering
  \includegraphics[scale=0.5]{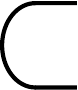}
  }.
 \end{equation}
We represent the left-orthogonal tensors by solid square shapes.

Next, we move on to the next site and update the MPS tensor with indices ($\alpha_1,i_2,i_3,\alpha_3$).
We define $L_1$ tensor from $U_1$ and the $i_1$ component of the MPO tensor of $\tilde{H}$ as
\begin{equation}
  \parbox[c]{5em}{
  \centering
  \includegraphics[scale=0.5]{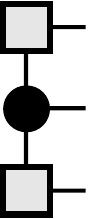}
  }
=
  \parbox[c]{5em}{
  \centering
  \includegraphics[scale=0.5]{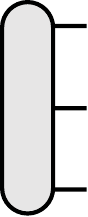}
  }
  =L_1.
 \end{equation}
Using this, $\tilde{H}$ is transformed as
\begin{equation}
  \parbox[c]{15em}{
  \centering
  \includegraphics[scale=0.5]{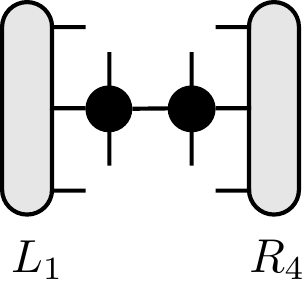}
  },
  \end{equation}
which is the effective Hamiltonian in the ($\alpha_1,i_2,i_3,\alpha_3$) basis.
By minimizing the transformed Hamiltonian with the Lanczos algorithms, we obtain the optimized bond tensor $B_{23}$.
Using the truncated singular value decomposition (or diagonalization of the density matrix), the bond tensor is restored to the MPS form:
  \begin{equation}
  \parbox[c]{10em}{
  \centering
  \includegraphics[scale=0.5]{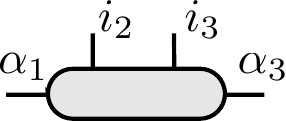}
  }
  \simeq
  \parbox[c]{10em}{
  \centering
  \includegraphics[scale=0.5]{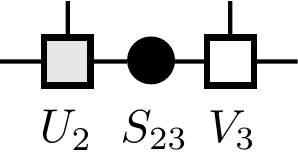}
  }
  =
  \parbox[c]{8em}{
  \centering
  \includegraphics[scale=0.5]{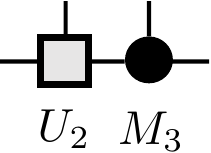}
  },
  \end{equation}
  which is used as the update of the MPS tensor.
  Similarly, by defining $L_2$ tensor from $L_1$ and $U_2$ as
\begin{equation}
  \parbox[c]{5em}{
  \centering
  \includegraphics[scale=0.5]{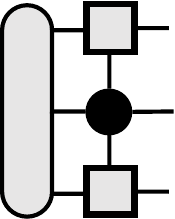}
  }
=
  \parbox[c]{5em}{
  \centering
  \includegraphics[scale=0.5]{figures/L1_right.pdf}
  }
  =L_2,
 \end{equation}
we next optimize the bond tensor $B_{34}$ defined on the ($\alpha_2,i_3,i_4,\alpha_4$) basis.
We continue this procedure until we reach the end of the chain.
This is half of a single DMRG sweep.
Next, starting from the bond located between $i=N-1$ and $N$, we continue the procedure in reverse order back to the first bond.
After reaching the first bond, one full sweep of DMRG is finished.
We iterate the sweep of DMRG with gradually increasing the accuracy of truncation of the singular value decomposition until the desired accuracy of the groundstate energy is reached.

\section{Numerical results}\label{sec:Numerical_results}

In this section, we show the numerical results using DMRG with the ITensor Library~\cite{itensor}.

\subsection{\texorpdfstring{$\mathrm{SU}(2)$}{SU(2)}}\label{sec:SU(2)}
DMRG was performed using iTensor~\cite{itensor} with maximum bond dimension $200$, truncation error cutoff $10^{-8}$, and noise strength $10^{-8}$ for the final DMRG sweeps. We typically need a few thousand DMRG sweeps for convergence (See appendix~\ref{sec:dmrg_convergence} for more details of the sweep dependence).

\subsubsection{Baryon size}
\begin{figure}[tb]
  \centering
  \includegraphics[width=0.49\linewidth]{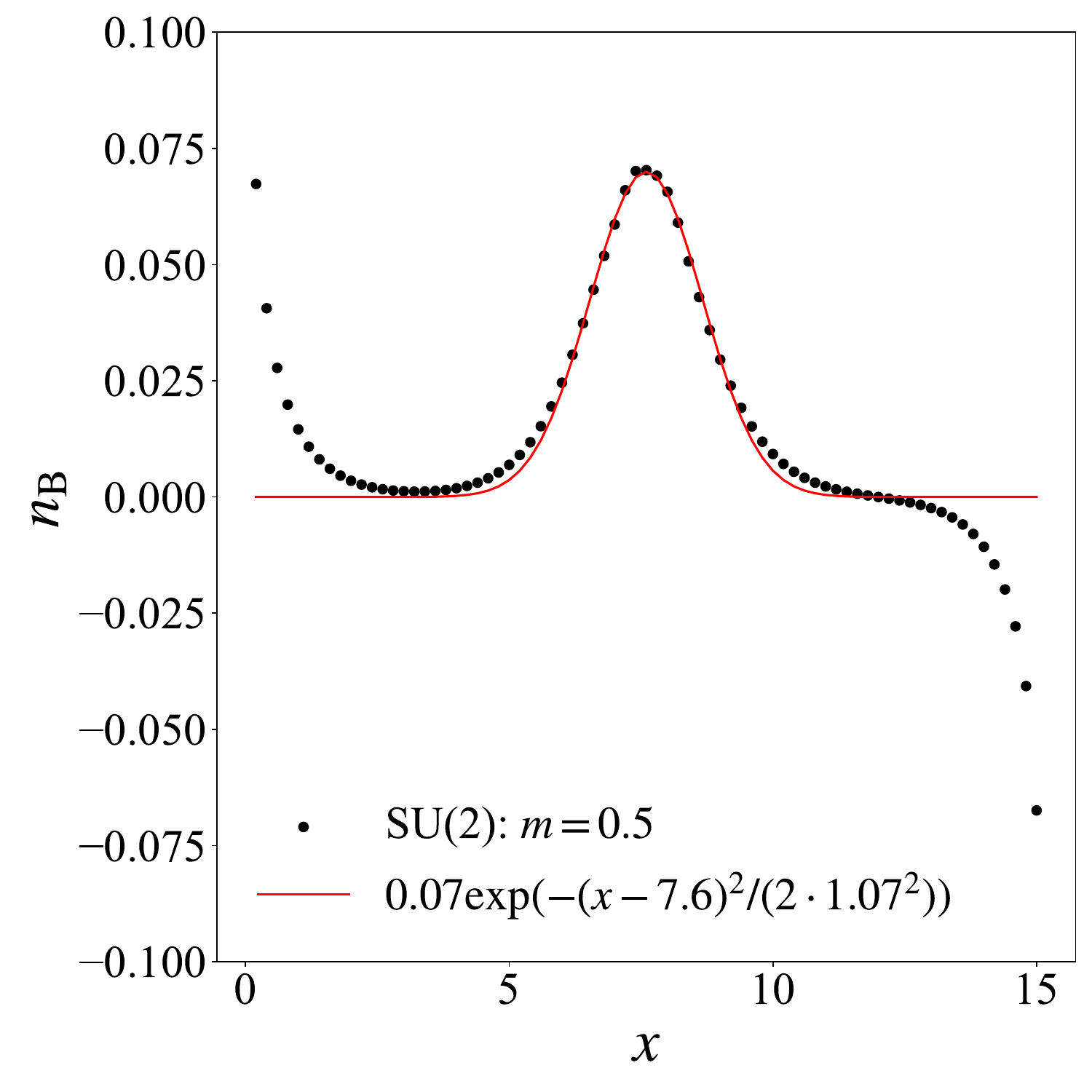}
  \includegraphics[width=0.49\linewidth]{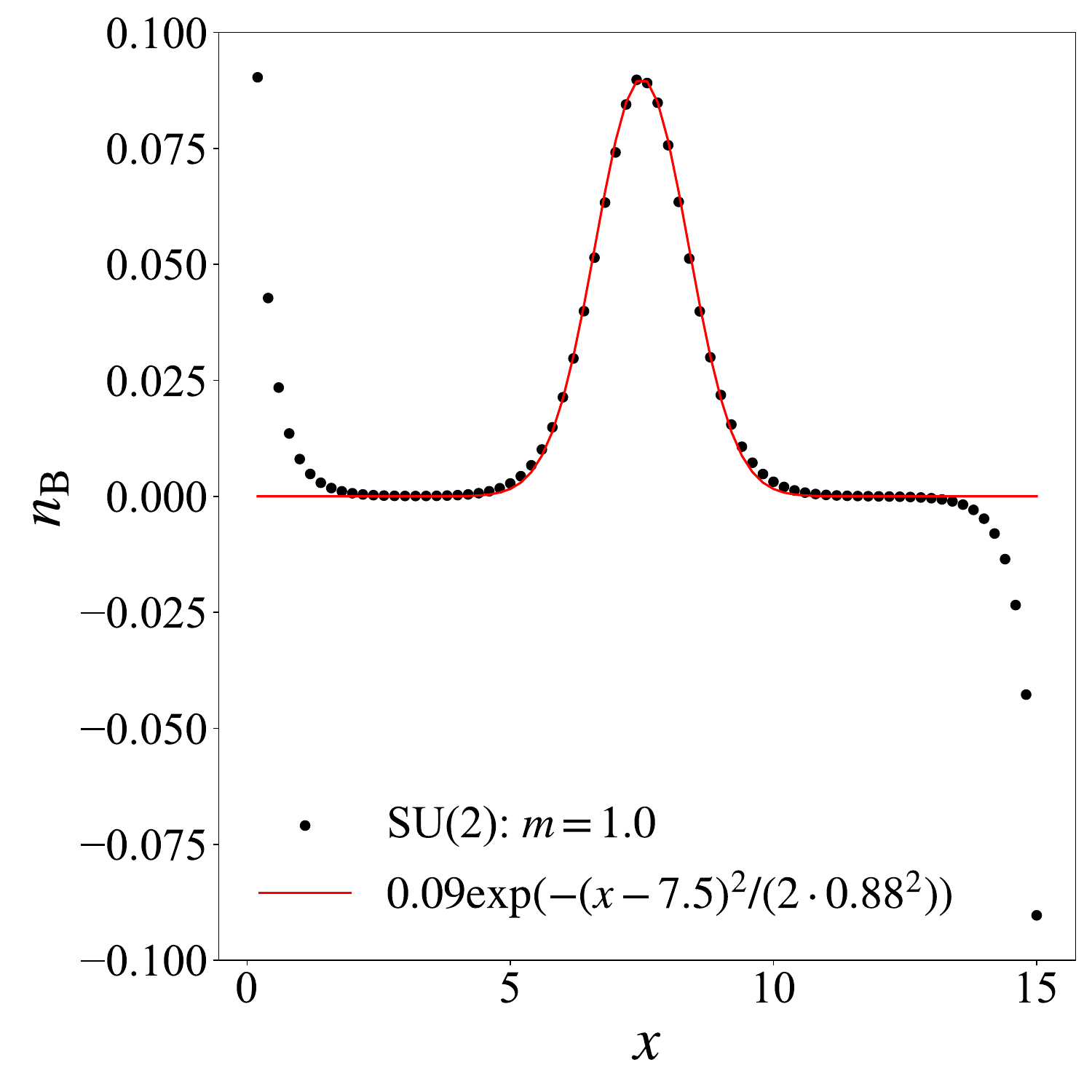}
  \caption{
  Spatial distribution of the baryon number $N_{\textrm{B}}=1$ for $m=0.5$ (left panel) and $m=1.0$ (right panel) with $N=150, \hop=5.0$.
  The black dots are numerical results, and the red lines are Gaussian fits of the black points.
  }
  \label{fig:single_baryon}
\end{figure}
\begin{figure}[tb]
  \centering
  \includegraphics[width=0.49\linewidth]{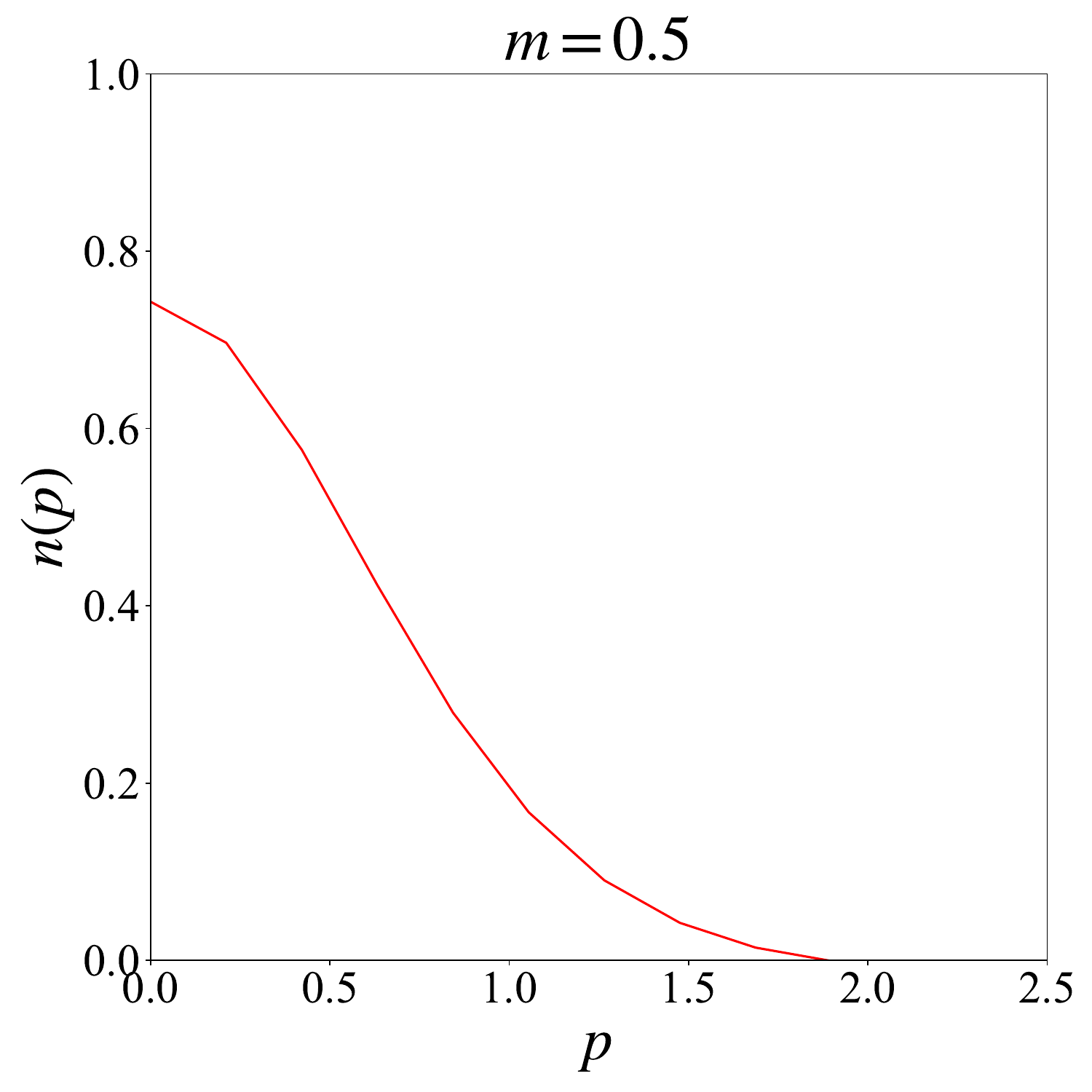}
  \includegraphics[width=0.49\linewidth]{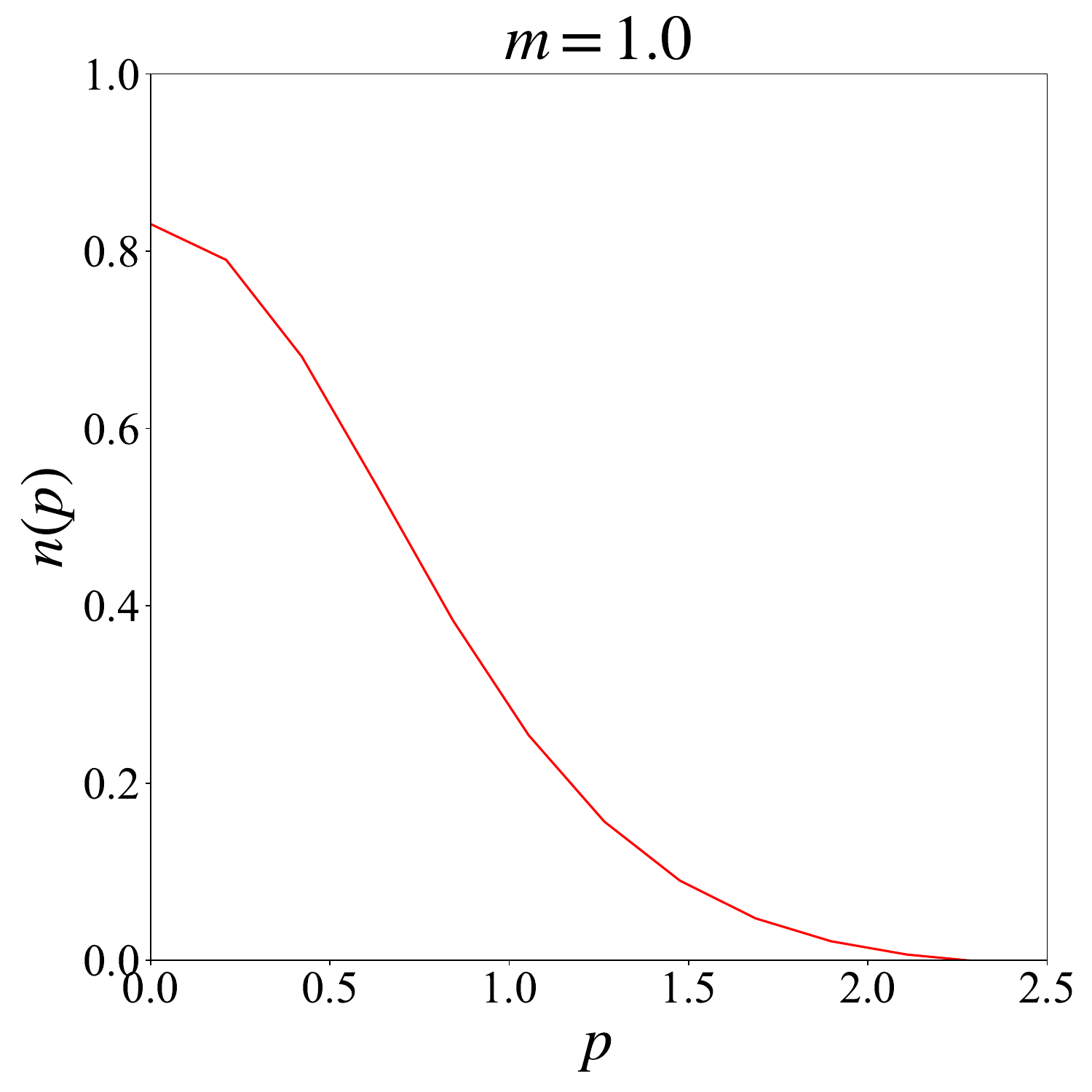}
  \caption{
  Quark distribution function for the single baryon state for $m=0.5$ (left panel) and $m=1.0$ (right panel) with $N=150, \hop=5.0$.
  }
  \label{fig:single_baryon_np}
\end{figure}
First, let us look at the single-baryon state to see the typical size of baryons.
Concretely, we prepare the vacuum state by minimizing the Hamiltonian,
by using the quantum-number conserving DMRG algorithm.
We employ parameters, $N=150$, $J=1/20$, and $w=5$.
The physical spatial volume is $V=N/(2w)=15$, which is sufficiently large to contain the single baryon, as will be seen numerically below.
Upon preparation of the vacuum state, we apply a baryon operator, $\psi_1^\dag(x) \psi_2^\dag(x)$ at the center position to create the $N_\mathrm{B}$=1 state. 
Then, the DMRG optimization is performed to minimize its energy with fixed baryon number $N_{\textrm{B}}=1$. This process is stopped at a finite number of sweeps, around 2000. 
The baryon remains localized at the center position in the resultant state that is not a true minimized energy state; however, the energy is low enough to see the distribution of the single baryon, where the energies are $1.59$, and $2.57$ for $m=0.5$ and $1.0$, respectively. Note that the baryon is localized at the boundary in the true minimum state.
The baryon size can be inferred from the expectation value of the baryon number density, shown in figure \ref{fig:single_baryon}. 
We can see that the baryon is localized at the center of the position. 
In addition, there are localized modes at the boundaries. 
This is due to the open boundary condition.
In calculations at finite density, particularly at low densities, boundary effects are expected to be unavoidable; thus, the volume should be as large as possible with keeping densities.
We fit the distribution by a Gaussian function $A \exp[-(x-x_0)^2/(2r^2)]$ with parameters $A$, $x_0$, and $r$, using the data points $3.4<x<11.8$.
The fitting results are also shown in figure \ref{fig:single_baryon}.
From these fitting results, the baryon sizes for $m=0.5$ and $1.0$ are estimated to be $r=1.07 \pm 1.15\times 10^{-2}$ and $0.88 \pm 5.40\times 10^{-3}$, respectively.
Here, the errors represent the standard deviation of the fit.

In figure~\ref{fig:single_baryon_np}, we show the quark-distribution function defined in eq.~\eqref{eq:n(p)} for the single baryon state. 
Both for $m=0.5$ and $m=1.0$, the distributions rapidly decay as $p$ increases.
The distribution function decays of order $1=g_0$, reflecting the fact that the baryon size is of order $1/g_0$.
The decay rate of the distribution function does not seem to depend significantly on mass.
Because the quarks are confined within the baryons, no Fermi surface or Fermi sea of quarks is formed in the single baryon state.

\subsubsection{Thermodynamic quantities}
\begin{figure}[tb]
  \centering
  \includegraphics[width=0.49\linewidth]{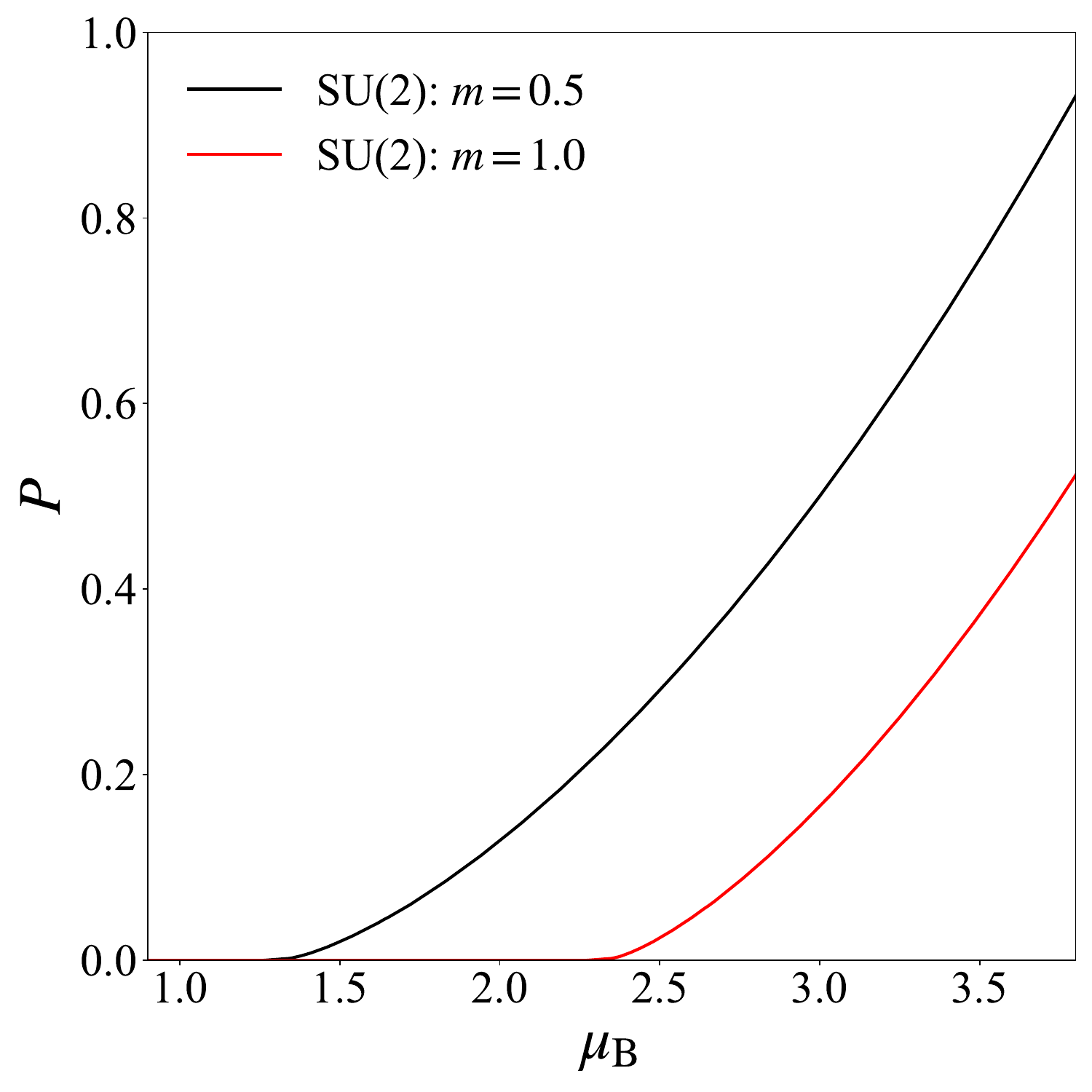}
  \includegraphics[width=0.49\linewidth]{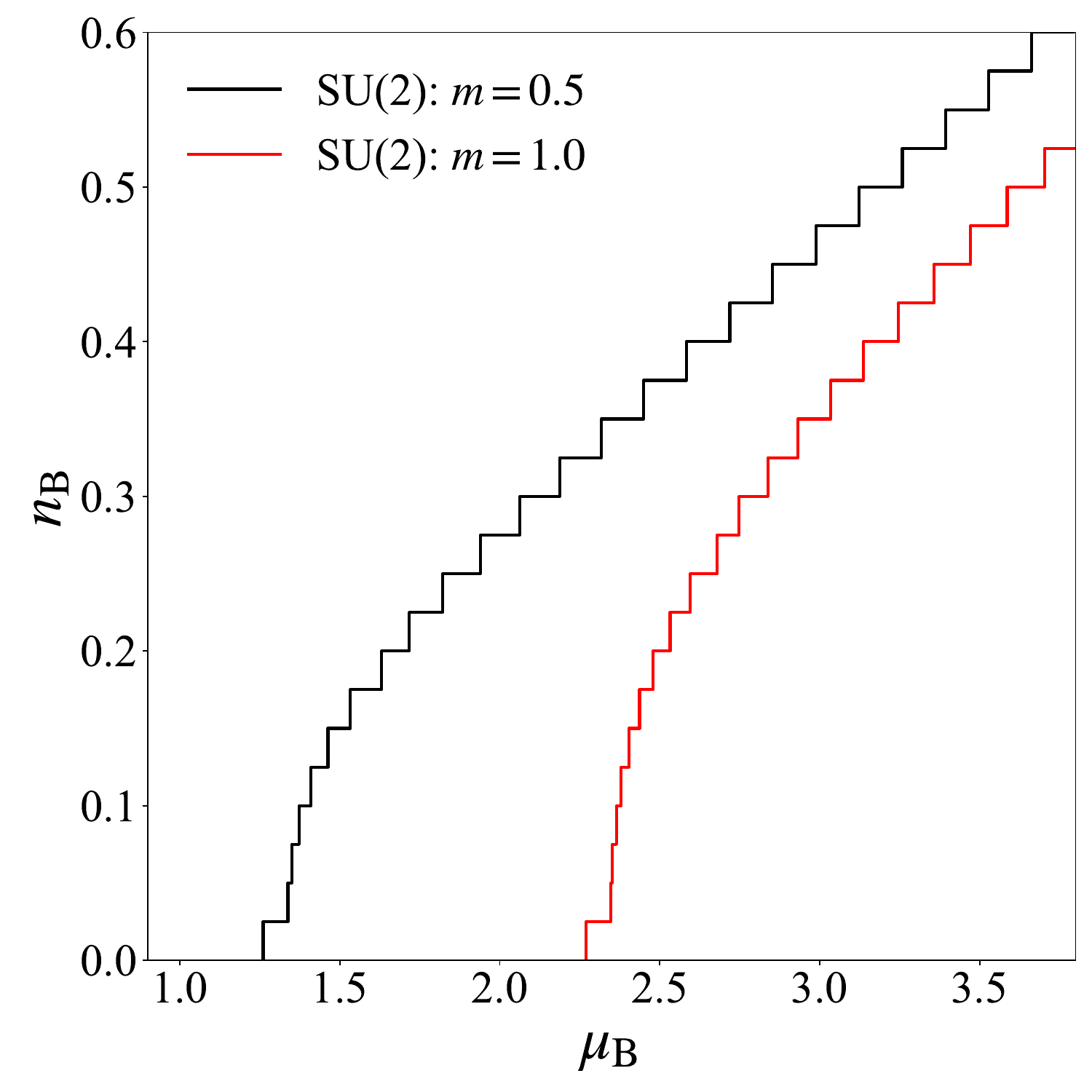}
   \caption{
    Pressure $P$ (left) and baryon number density $n_\mathrm{B}$ (right) for $\mathrm{SU}(2)$ as functions of $\mu_\mathrm{B}$ for $m=0.5$ and $1.0$ with $\hop=2$, and $N=160$.
    The pressure and number density begin to increase at $\mu_\mathrm{B}=1.26$ for $m=0.5$ and at $\mu_\mathrm{B}=2.27$ for $m=1.0$, respectively.
 }
  \label{fig:pressure_su2}
\end{figure}
Next, we examine the thermodynamic quantities such as the pressure and the number density.
At finite density, we minimize $H-\mu_\mathrm{B}N_\mathrm{B}$~\eqref{eq:H-muN} using the DMRG algorithm without quantum number conservation. 
We employ parameters $N=160$, $J=1/8$, and $w=2$, where the physical volume $V=40$ is much larger than the typical size of baryons $\sim1$.
The left panel of figure \ref{fig:pressure_su2} shows the pressure as a function of $\mu_\mathrm{B}$.
Due to a finite volume, the pressure is continuous but not smooth.
This fact can be seen from the baryon number density, which is the derivative of the pressure,
\begin{equation}
  n_\mathrm{B}=\frac{\dd P(\mu_\mathrm{B})}{\dd \mu_\mathrm{B}}=\frac{1}{V}\expval{N_\mathrm{B}},
\end{equation}
shown in the right panel of figure \ref{fig:pressure_su2}.
The baryon number density shows a step-like behavior.
Due to the confining energy, the pressure and baryon number density start to increase at a higher point than the threshold of the free theory $\mu_\mathrm{B}=N_\mathrm{c}m$.
The threshold values of the chemical potential are $1.26$ and $2.27$ for $m=0.5$ and $m=1.0$, respectively.
The baryon number density first rises sharply and then increases linearly.
The behavior can be understood as follows:
At high densities, $g_0/\mu\ll 1$, and $m/\mu\ll 1$, the contributions from the interaction and mass become negligible.
As a result, the system is approximated by free-quark fields, and then the baryon number density behaves as $n_\mathrm{B}=N_\mathrm{c}\mu/\pi=\mu_\mathrm{B}/\pi$, which is a linear function of $\mu_\mathrm{B}$.
The behavior at low density is more nontrivial, and we discuss the physical meaning along with other observables below.

When we compare physical quantities with different masses, it is better to consider them as functions of the baryon number density (or energy density) rather than the chemical potential.
For the energy, this can be determined unambiguously; however, subtleties arise when considering the pressure. This is caused by the ambiguity of the chemical potential; the baryon number density $n_\mathrm{B}$ is not a continuous function of $\mu_\mathrm{B}$, and therefore, the inverse function does not uniquely exist.
Given that the energy is uniquely determined as a function of the baryon number, the chemical potential can be defined as the change in the energy when one more baryon is added:
\begin{align}
  \mu^-_\mathrm{B}(n_\mathrm{B})&\coloneqq E(N_\mathrm{B})-E(N_\mathrm{B}-1),\\
  \mu^+_\mathrm{B}(n_\mathrm{B})&\coloneqq E(N_\mathrm{B}+1)-E(N_\mathrm{B}),
\end{align}
where $E(N_\mathrm{B})=\expval{H}$ is the total energy for a given baryon number.
The choice between a forward or backward difference causes ambiguity. 
Note that when we compute a physical quantity as a function of the chemical potential, the number and energy densities remain constant within the interval $\mu^-_\mathrm{B}(n_\mathrm{B})<\mu_\mathrm{B}<\mu^+_\mathrm{B}(n_\mathrm{B})$. Alternatively, we can introduce $\overline{\mu}_\mathrm{B}(n_\mathrm{B})$ as the average of $\mu_\mathrm{B}^\pm(n_\mathrm{B})$,
\begin{equation}
  \overline{\mu}_\mathrm{B}(n_\mathrm{B}) \coloneqq \frac{\mu^-_\mathrm{B}(n_\mathrm{B})+\mu^+_\mathrm{B}(n_\mathrm{B})}{2}\label{eq:average_mu}.
\end{equation}
In the large volume limit $V\to\infty$, all values of $\mu_\mathrm{B}^{\pm}(n_\mathrm{B})$ and $\overline{\mu}(n_\mathrm{B})$ converge to the same value.
Since the averaged chemical potential $\overline{\mu}_\mathrm{B}(n_\mathrm{B})$ has an improved volume dependence compared to $\mu_\mathrm{B}^{\pm}(n_\mathrm{B})$ at a large volume, we define the pressure as a function of $n_\mathrm{B}$ by using $\overline{\mu}_{\textrm{B}}$:
\begin{equation}
    {{P}}(n_\mathrm{B}) \coloneqq \overline{\mu}_\mathrm{B}(n_\mathrm{B})n_\mathrm{B}-\varepsilon(n_\mathrm{B}).
\end{equation}
\begin{figure}[tb]
  \centering
  \includegraphics[width=0.49\linewidth]{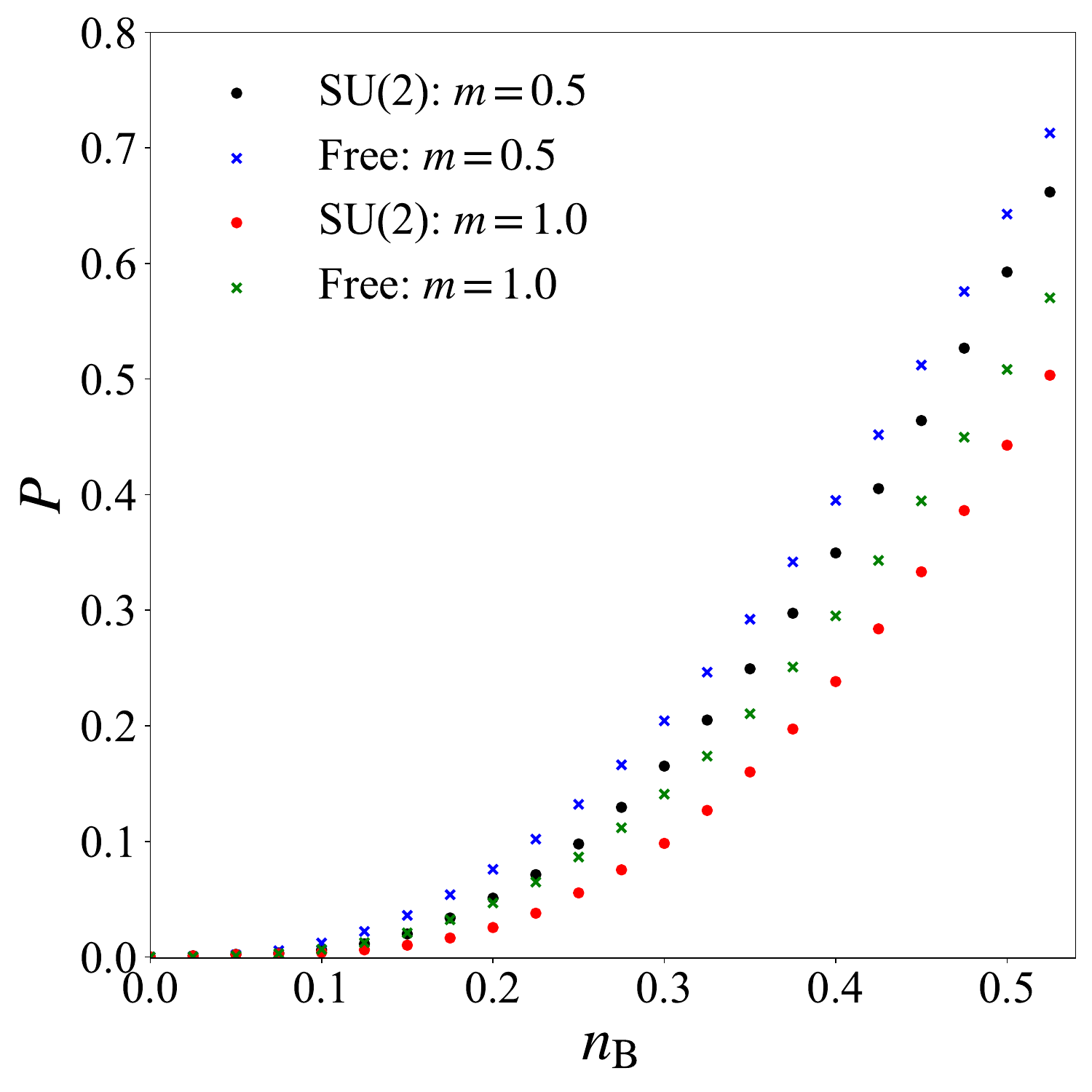}
  \includegraphics[width=0.49\linewidth]{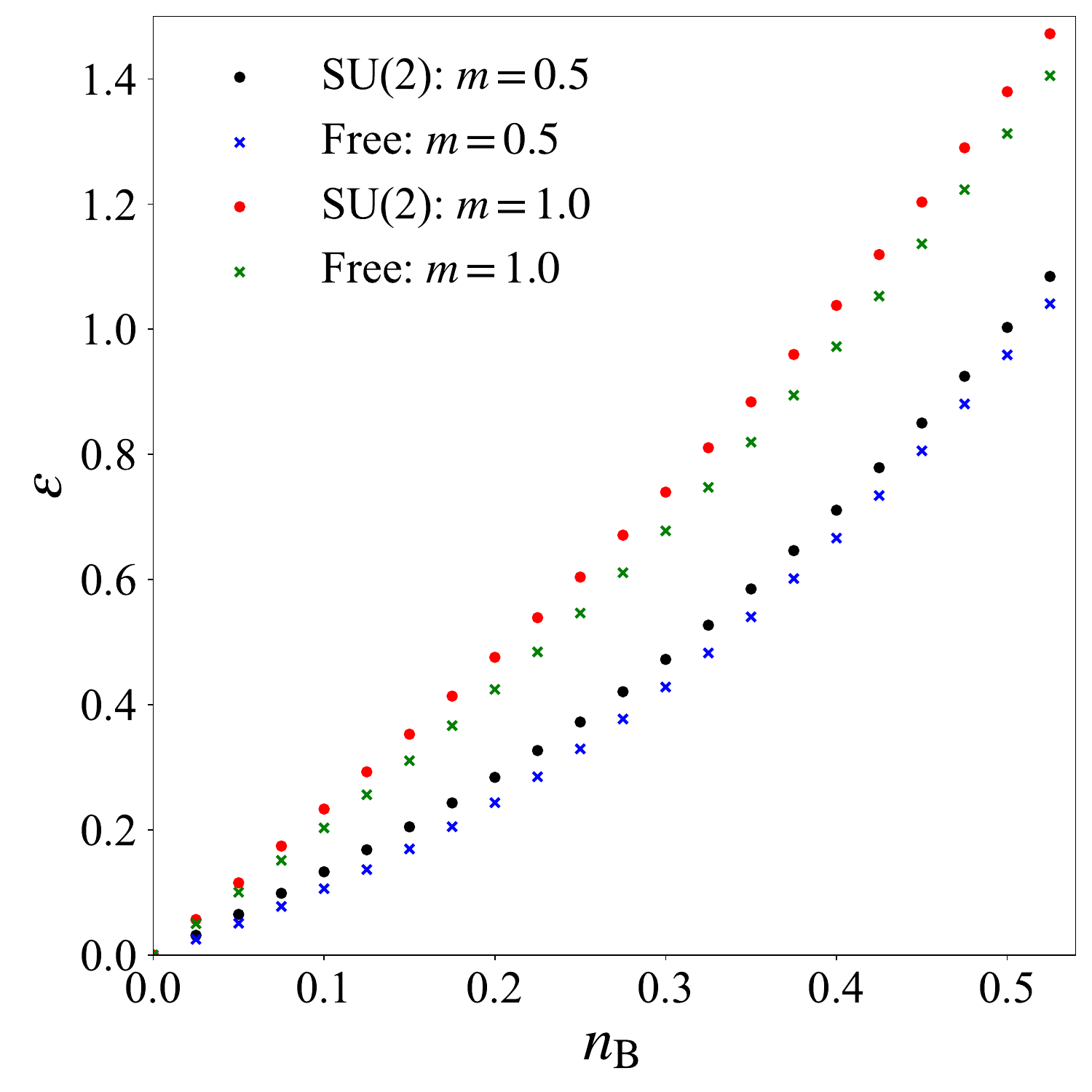}
  \caption{
  Pressure $P$ (left) and energy density $\varepsilon$ (right) for $\mathrm{SU}(2)$ as functions of $n_\mathrm{B}$  for $m=0.5$ and $m=1.0$ with $\hop=2.0$ and $N=160$.
  }
  \label{fig:nB_dependence_P_energy}
\end{figure}
The numerical results are presented in the left panel in figure \ref{fig:nB_dependence_P_energy}. We also plot the energy density as a function of $n_{\mathrm{B}}$ in the right panel in figure \ref{fig:nB_dependence_P_energy}.
For comparison, we plot the pressure and energy density of the free theory (See appendix~\ref{sec:Free_theory} for details of the free theory).
The overall behavior of $\textrm{QCD}_2$ is consistent with that of the free theory, although the pressure (energy density) in $\textrm{QCD}_2$ is slightly more suppressed (enhanced) than in the free theory.
\begin{figure}[tb]
  \centering
  \includegraphics[width=0.49\linewidth]{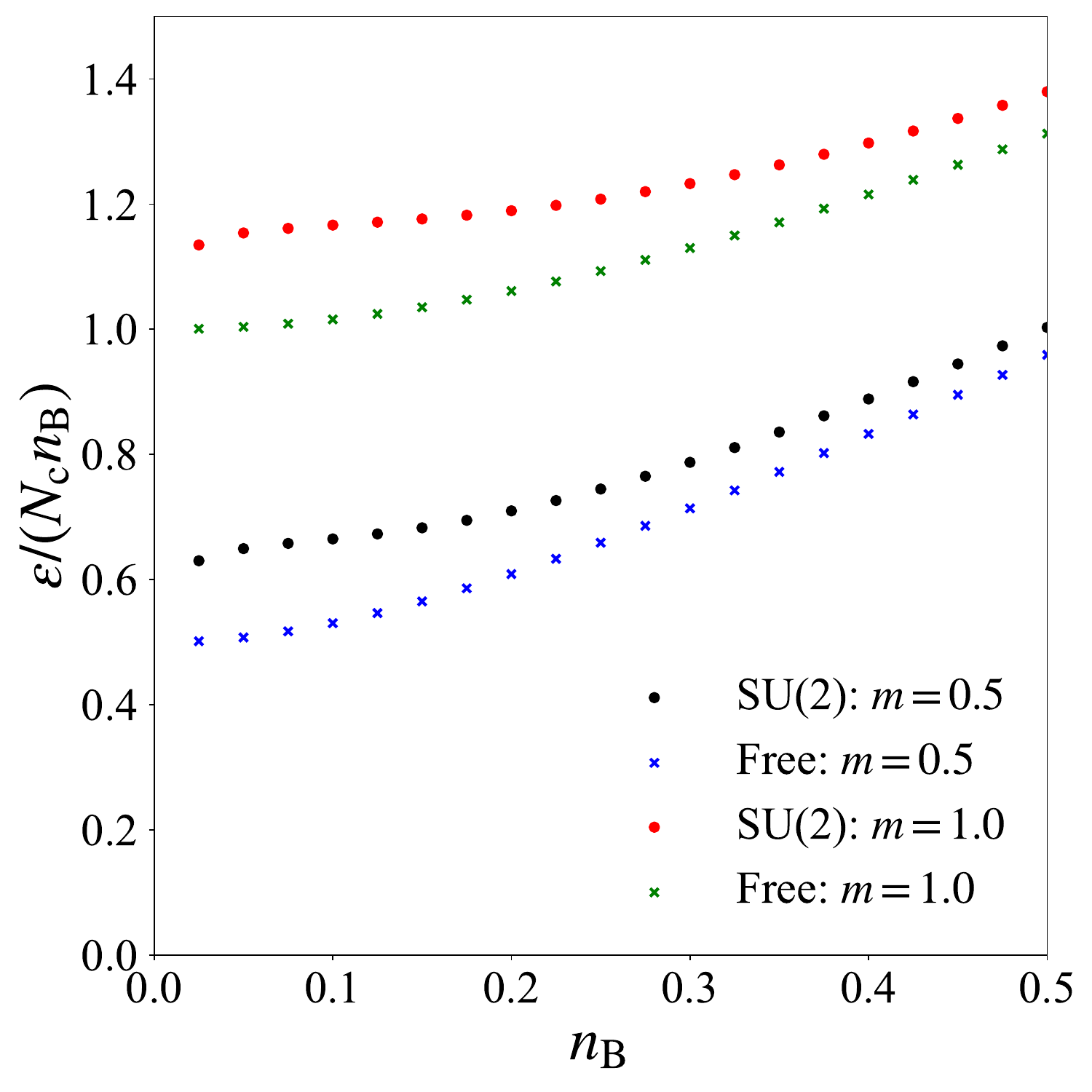}
  \includegraphics[width=0.49\linewidth]{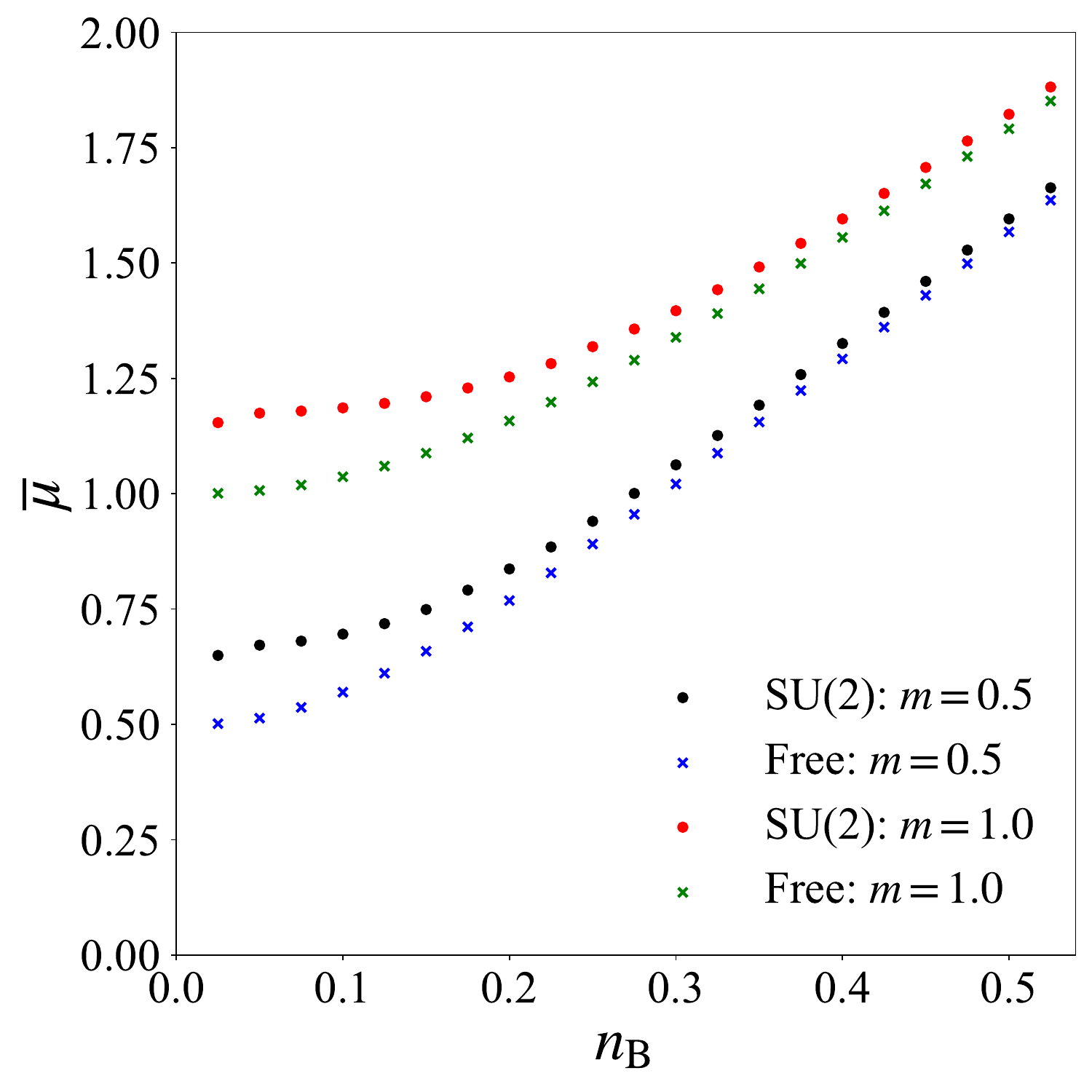}
  \caption{
  Ratio of $\varepsilon$ to $N_{\textrm{c}}n_{\textrm{B}}$ (left) and 
  averaged quark chemical potential $\overline{\mu}=\overline{\mu}_\mathrm{B}/N_\mathrm{c}$ (right) and for $\mathrm{SU}(2)$ as functions of $n_\mathrm{B}$ for $m=0.5$ and $m=1.0$ with $\hop=2.0$ and $N=160$.
  }
  \label{fig:energy_su2}
\end{figure}
This behavior is intriguing, given that the degrees of freedom in $\textrm{QCD}_2$ are baryonic, especially at lower densities.

To see the difference between $\textrm{QCD}_2$ and the free theory in more detail, let us focus on behaviors of the energy per quark and averaged chemical potential.
The left and right panels in figure \ref{fig:energy_su2} show the energy per unit quark number $\varepsilon/(N_{\textrm{c}}n_{\textrm{B}})$ and quark chemical potential $\overline{\mu}=\overline{\mu}_\mathrm{B}/N_\mathrm{c}$, respectively.
These quantities will help us understand whether the system behaves like quarks or like baryons, by comparing 
their behavior in the free theory.
In regions of a large baryon number density, $\varepsilon/(N_{\textrm{c}}n_{\textrm{B}})$ behaves like free quarks.
On the other hand, at low densities, the energy per quark and averaged chemical potential are higher than those of the free theory due to the effect of confinement energy.
The change in density is more gradual than that in the free theory, and the behavior seems to change to that of the free theory around $n_\mathrm{B}\approx 0.2$.

One of the important characteristics of two color $\textrm{QCD}_2$ is that the baryons are bosons due to the even number of colors.
Therefore, if we assume interactions are negligible at low density, the baryons will degenerate to the lowest energy state. 
Consequently, $\varepsilon/(N_{\textrm{c}}n_{\textrm{B}})$ and $\overline{\mu}$ are expected to be constant at low densities,
which is consistent with the behaviors in Fig.~\ref{fig:energy_su2}.
However, we must be careful with this consideration, since interactions are generally not negligible in $(1+1)$-dimensional systems. 

\begin{figure}[tb]
  \centering
  \includegraphics[width=0.49\linewidth]{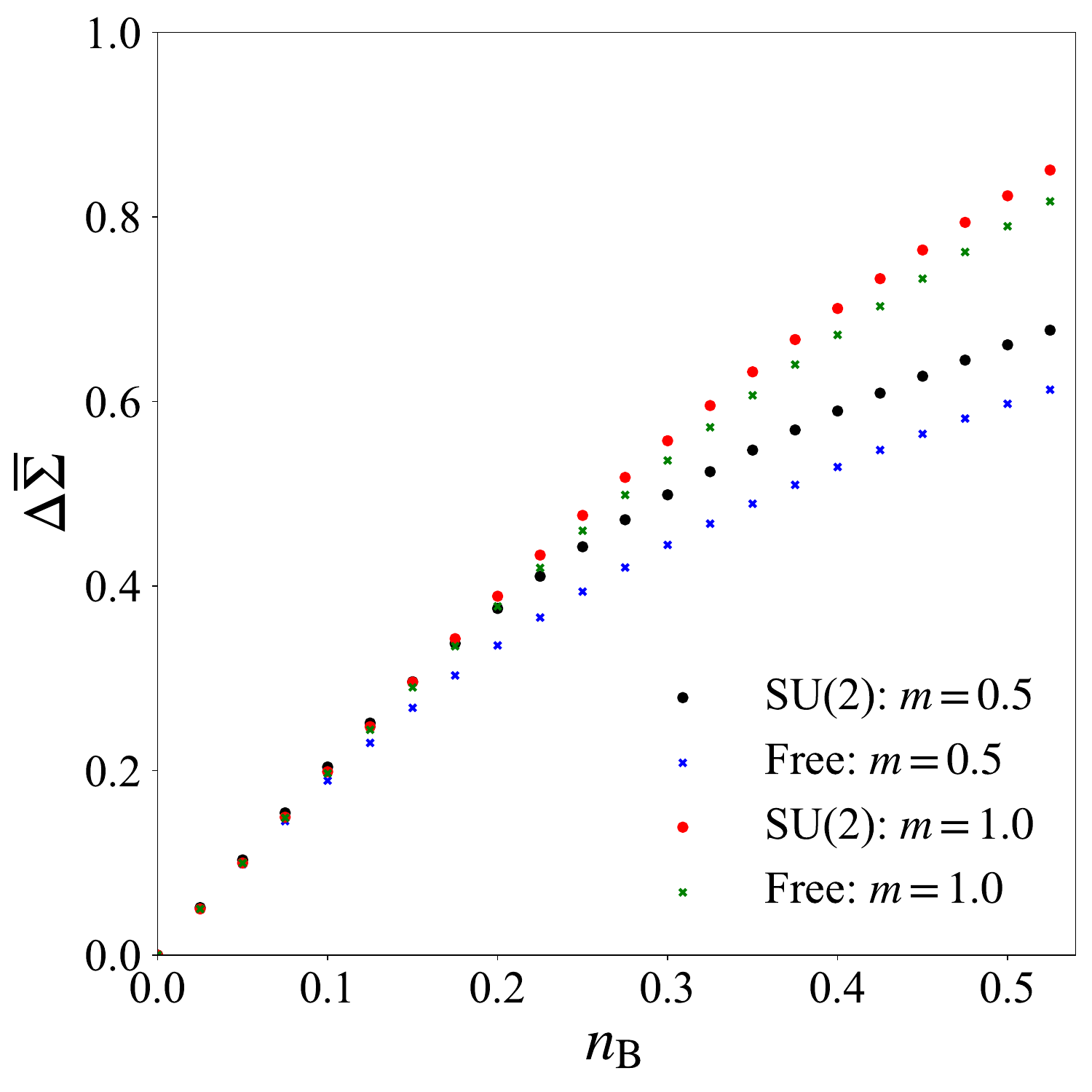}
  \includegraphics[width=0.49\linewidth]{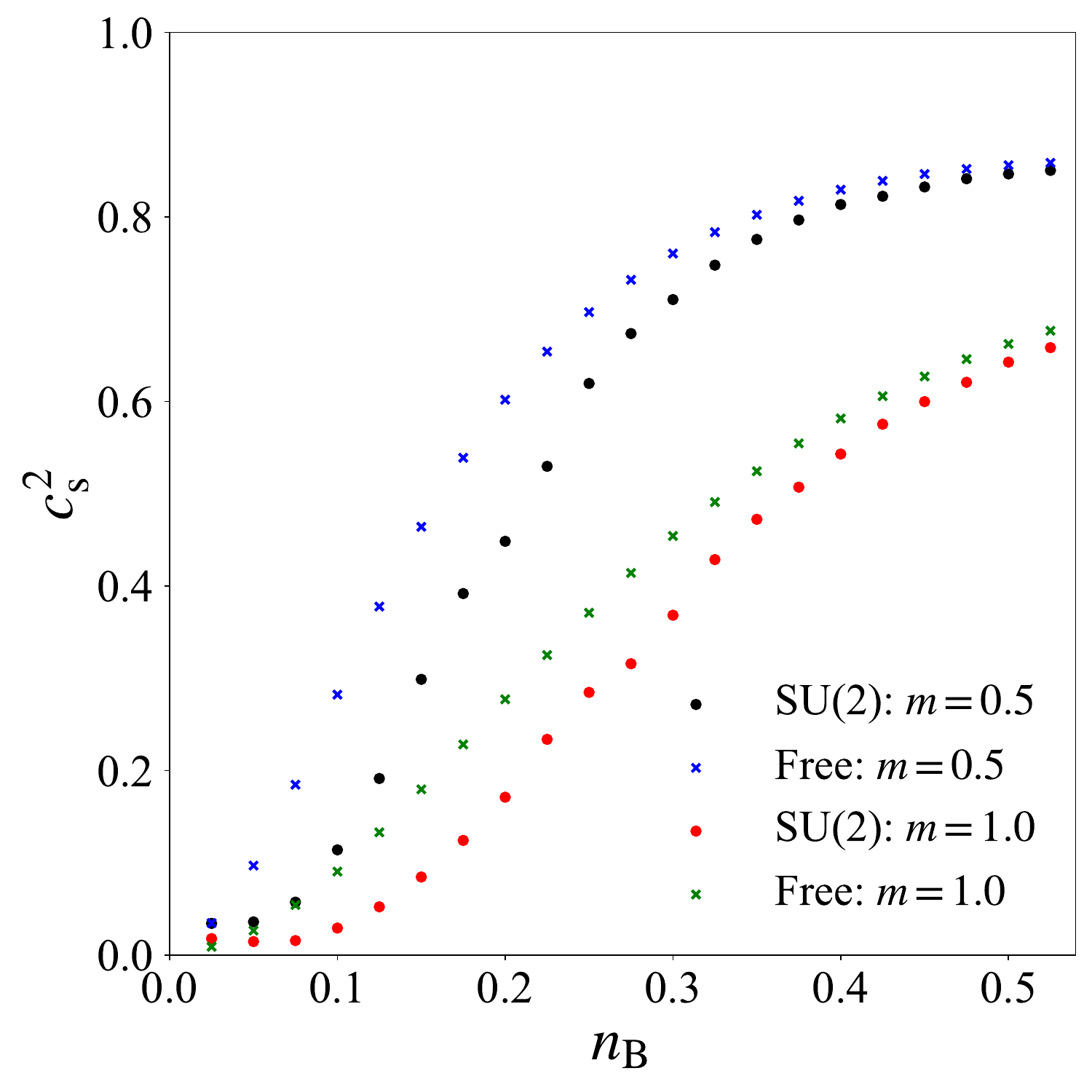}
  \caption{
    Chiral condensate subtracting the contribution from the vacuum
    $\Delta \overline{\Sigma}=\overline{\Sigma}-\overline{\Sigma}_\mathrm{vac}$ (left) and squared sound velocity (right) for $\mathrm{SU}(2)$ as functions of $n_\mathrm{B}$ for $m=0.5$ and $m=1.0$ with $\hop=2.0$ and $N=160$.
  }
  \label{fig:chiral_condensate_and_sound_velocity}
\end{figure}

Lastly, in this subsection, let us look at the averaged chiral condensate and sound velocity.
The averaged chiral condensate is given as
\begin{equation}
   \overline{\Sigma} =\frac{1}{V}\frac{\partial }{\partial m}\expval{H} =\frac{1}{V}\frac{1}{\hop}\sum_n\expval{\bar{\psi}\psi(n)}.
\end{equation}
We need a renormalization of $\Sigma$ because it diverges in the continuum limit. 
We are interested in the change of $\Sigma$ at finite density, so we introduce
\begin{equation}
  \Delta\overline{\Sigma} = \overline{\Sigma} - \overline{\Sigma}_\mathrm{vac}.
\end{equation}
Here, $\overline{\Sigma}_\mathrm{vac}$ is the unrenormalized chiral condensate in the vacuum.

The left panel of figure~\ref{fig:chiral_condensate_and_sound_velocity} shows $\Delta\overline{\Sigma}$ as a function of $n_\mathrm{B}$ together with the results of the free theory.
Both cases $m=1$ and $m=0.5$ behave similarly to the free theory, with the difference being more significant for the $m=0.5$ case. This may be because the lighter masses are more sensitive to chiral symmetry breaking.
However, we do not discuss the spontaneous breaking of chiral symmetry and its restoration for several reasons.
Firstly, the masses we employ are relatively heavy. Secondly, in the Hamiltonian formalism, staggered fermions do not possess continuous chiral symmetry. 
Lastly, the open boundary conditions explicitly break chiral symmetry.

In the continuum theory, the sound velocity $c_\mathrm{s}$ is given by
\begin{equation}
  c_\mathrm{s}^2\coloneqq\frac{\dd P}{\dd \varepsilon}
  \label{sound_velocity} ,
\end{equation}
which represents the response of the pressure to the change of energy.
At zero temperature, we can express it by using the chain rule as
\begin{equation}
  c_\mathrm{s}^2  = \frac{\dd P}{\dd n}\frac{\dd n}{\dd \varepsilon}
  \label{sound_velocity2} .
\end{equation}
We evaluate the sound velocity by replacing the derivative with the central difference, which can be expressed by $\overline{\mu}(N_{\textrm{B}})$ as
\begin{equation}
  \begin{split}
    c_\mathrm{s}^2(N_\mathrm{B}) &= V\frac{{P}(N_\mathrm{B}+1)-{P}(N_\mathrm{B}-1)}{E(N_\mathrm{B}+1)-E(N_\mathrm{B}-1)}
    \notag\\
    &=\frac{N_{\mathrm{B}}}{2}\frac{
    \overline{\mu}_\mathrm{B}(N_\mathrm{B}+1)
    -\overline{\mu}_\mathrm{B}(N_\mathrm{B}-1)
  }{\overline{\mu}_\mathrm{B}(N_\mathrm{B})}
  +\frac{1}{2}\frac{
    \overline{\mu}_\mathrm{B}(N_\mathrm{B}+1)
    +\overline{\mu}_\mathrm{B}(N_\mathrm{B}-1)
    -2\overline{\mu}_\mathrm{B}(N_\mathrm{B})
  }{\overline{\mu}_\mathrm{B}(N_\mathrm{B})}.
  \end{split}
\end{equation}
The right panel of figure \ref{fig:chiral_condensate_and_sound_velocity} shows the squared sound velocity as a function of $n_\mathrm{B}$ together with the results of the free theory.
The overall behavior is the same as before, behaving like a free theory. 
At low densities, the sound velocity is suppressed compared to the free theory. 
This is consistent with the free baryon picture, though interactions might play a significant role.

In the context of neutron star physics, the possibility of a peak in the sound velocity has been discussed~\cite{McLerran:2018hbz,Kojo:2021ugu}, and may exist in two color QCD~\cite{Kojo:2021hqh,Iida:2022hyy}.
This is due to the fact that in a $(d+1)$-dimensional system, $c_\mathrm{s}^2$ asymptotically approaches $1/d$ at high densities. Consequently, there may exist an intermediate density region where $c_\mathrm{s}^2 > 1/d$.
This is not the case in our $d=1$ situation. Assuming that the sound velocity does not exceed the speed of light, there is no peak satisfying $c_\mathrm{s}^2 >1/d$\footnote{This assumption might not necessarily hold. This is because excess of the sound velocity, defined in the limit of a small wavenumber over the speed of light does not necessarily violate causality. The condition for causality is that the front velocity, defined at an infinite wavenumber, does not exceed the speed of light~\cite{Brillouin,milonni2004fast}.}. 

\subsection{Size dependence}
\begin{figure}[tb]
  \centering
  \includegraphics[width=0.49\linewidth]{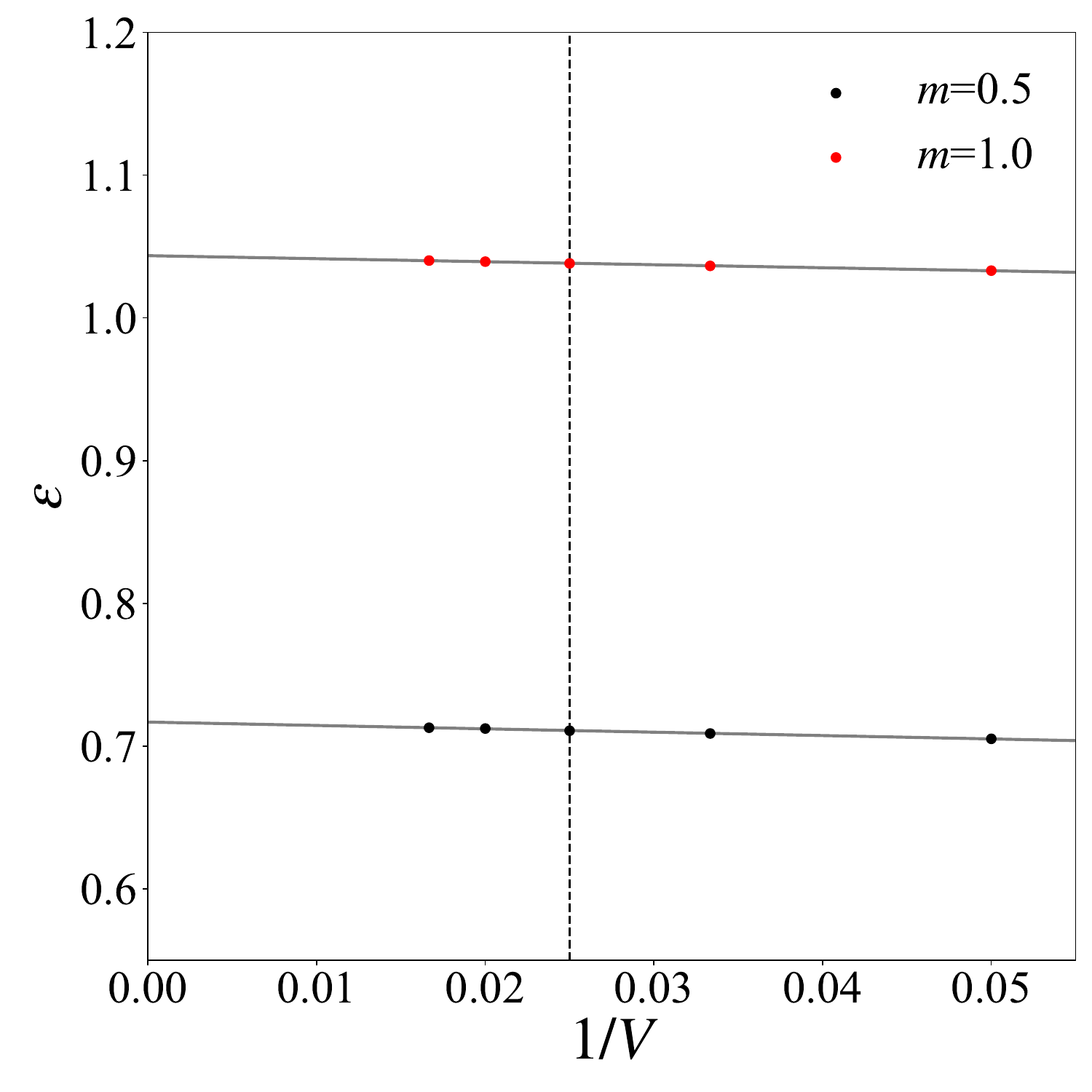}
  \caption{
Size dependence of the energy density fixed $w=2$ and $n_\mathrm{B}=0.4$ for $2w V=N=80,120,160,200,240$.
The dashed vertical line corresponds to the point of $N=160$.
The energy densities in the infinite volume limit, obtained by linear fitting, are $0.72$ for $m=0.5$ and $1.04$ for $m=1.0$, respectively.
  }
  \label{fig:size_dependence}
\end{figure}

While the purpose of this paper is to qualitatively understand $\textrm{QCD}_2$ at finite density, it is worthwhile to look at the lattice spacing and volume dependence.
For this purpose, we focus on the energy density at a fixed baryon number density with $w=2$.
Figure \ref{fig:size_dependence} shows the volume dependence of the energy density for $2w V=N=80,120,160,200,240$.
The volume dependence can be approximated by a linear function, with slope and intercept for $m=0.5$ being $-0.24$ and $0.72$, and for $m=1.0$ being $-0.21$ and $1.04$. 
Therefore, the energy densities in the infinite volume limit are 0.72 for $m=0.5$ and $1.04$ for $m=1.0$, respectively.
The results at $N=160$ corresponding to $V=40$ coincide with those in the infinite volume limit with an error of less than 
 $1\%$.
We also checked that the energy densities for $w=2$ $(a=0.25)$ and $w=4$ $(a=0.125)$ coincide with an accuracy of approximately $1\%$ in the range $0\leq n_\mathrm{B}<0.5$.

\subsubsection{Inhomogeneous phase in a finite volume}
\begin{figure}[tb]
    \centering
    \includegraphics[width=15.0cm]{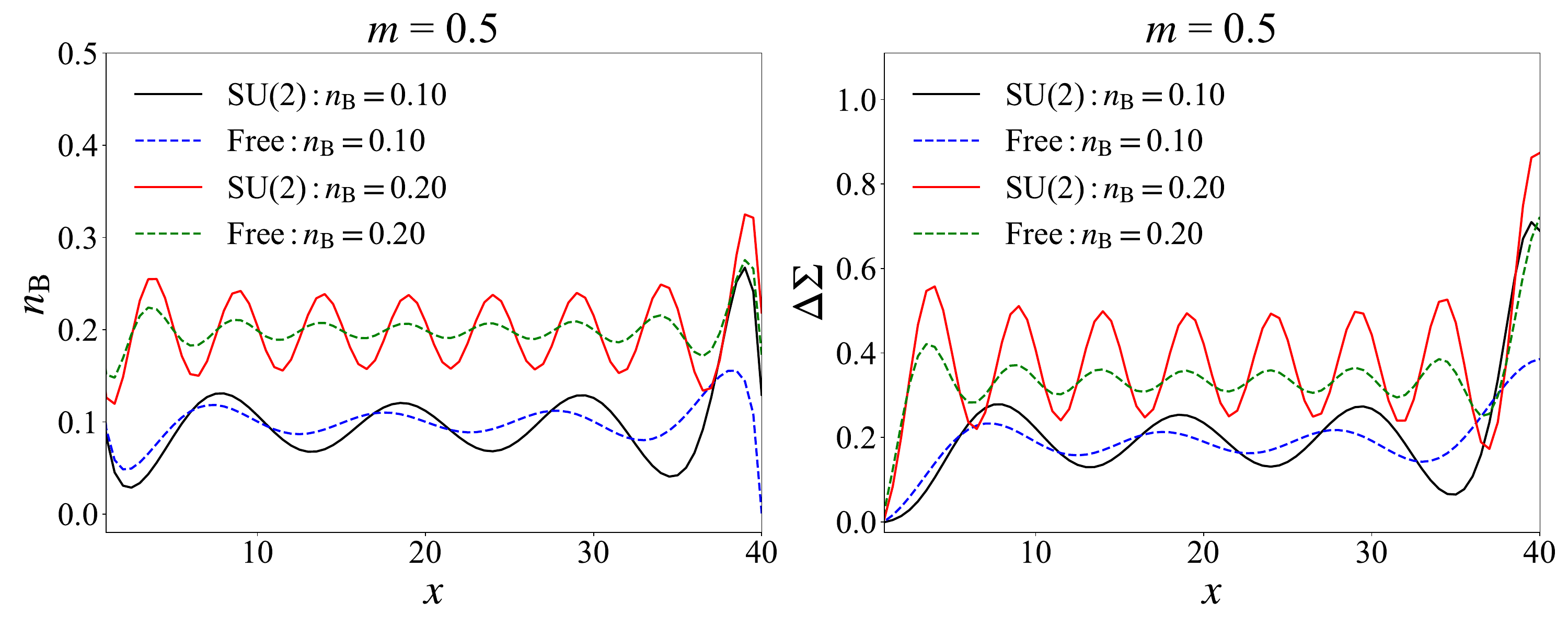}
    \includegraphics[width=15.0cm]{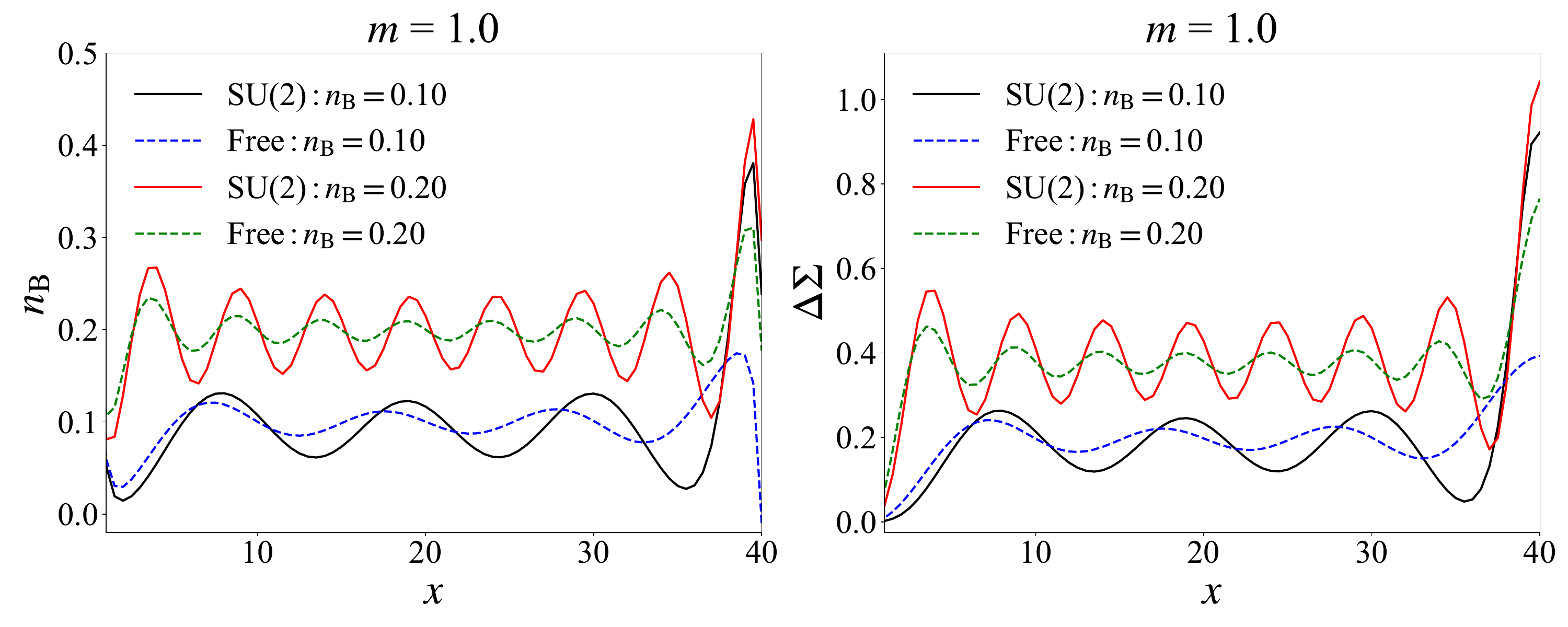}
    \caption{
    Spatial dependence of the baryon number density $n_\mathrm{B}(x)$ and the chiral condensate subtracting the contribution from the vacuum
    $\Delta {\Sigma}(x)={\Sigma}(x)-{\Sigma}_\mathrm{vac}(x)$  for $m=0.5$ (top panel) and $m=1.0$ (bottom panel)
    with $\hop=2.0$ and $N=160$.
    }
    \label{fig:inhomogeneous_phase}
\end{figure}
In Gross-Neveu, chiral Gross-Neveu models, and massless QCD in $(1+1)$ dimensions, inhomogeneous phases have been considered to exist within the mean-field approximation~\cite{Schon:2000he,Thies:2003kk,Kojo:2011fh}. It is an interesting question whether such an inhomogeneous phase can be realized in $\textrm{QCD}_2$ in the presence of quark mass. 
In this paper, we discuss the realization of the inhomogeneous phase in our method.
We, however, need to be careful about the possibility of an inhomogeneous phase.
In (1+1) dimensions, the spontaneous breaking of continuous symmetry is prohibited by the Hohenberg-Mermin-Wagner-Coleman theorem~\cite{Hohenberg:1967zz,Mermin:1966fe,Coleman:1973ci}.
Consequently, a translational symmetry cannot be spontaneously broken.
Nevertheless, the open boundary condition explicitly breaks the translational symmetry, so that it may have an inhomogeneous phase.

In figure \ref{fig:inhomogeneous_phase},
we show the spatial dependence of the baryon-number density (left) and chiral condensate (right) for $n_\mathrm{B}=0.1$ and $0.2$.
For comparison, we also plot those of the free theory.
In both the free and $\textrm{QCD}_2$ cases, the baryon-number density and chiral condensate are spatially modulated.
In the free theory, this is a finite volume artifact; we can see that the modulations vanish proportionally to $1/L$ in the infinite volume limit from eqs.~\eqref{eq:number_density_free} and \eqref{eq:chiral_condensate_free}.
Unfortunately, our current numerical calculations of $\textrm{QCD}_2$ cannot take a large size to make the volume dependence clear.
In the following, we will see that it is reasonable to believe that the inhomogeneous phase is also realized in the case of $\textrm{QCD}_2$.

\begin{figure}[tb]
    \centering
    \includegraphics[width=0.49\linewidth]{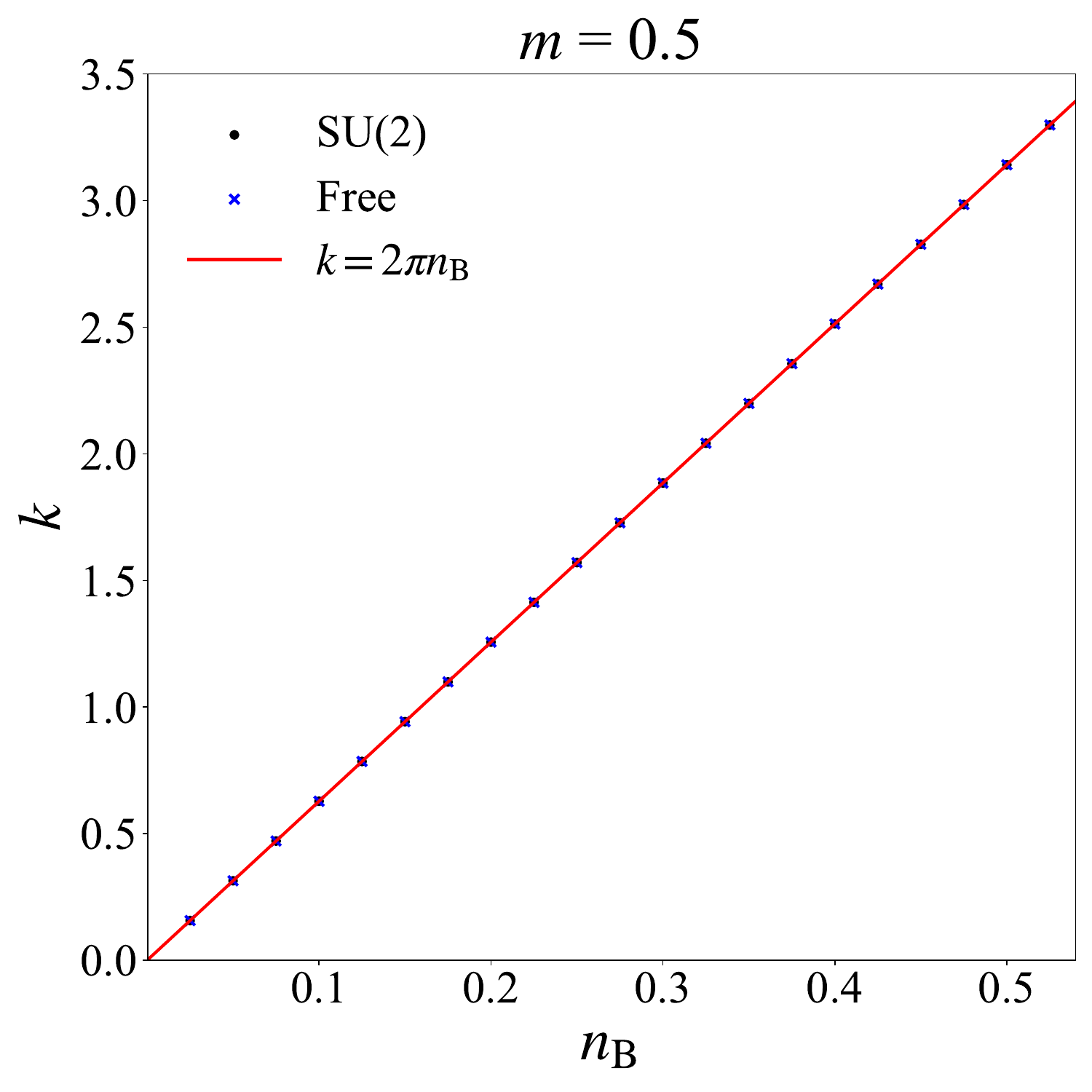}
    \includegraphics[width=0.49\linewidth]{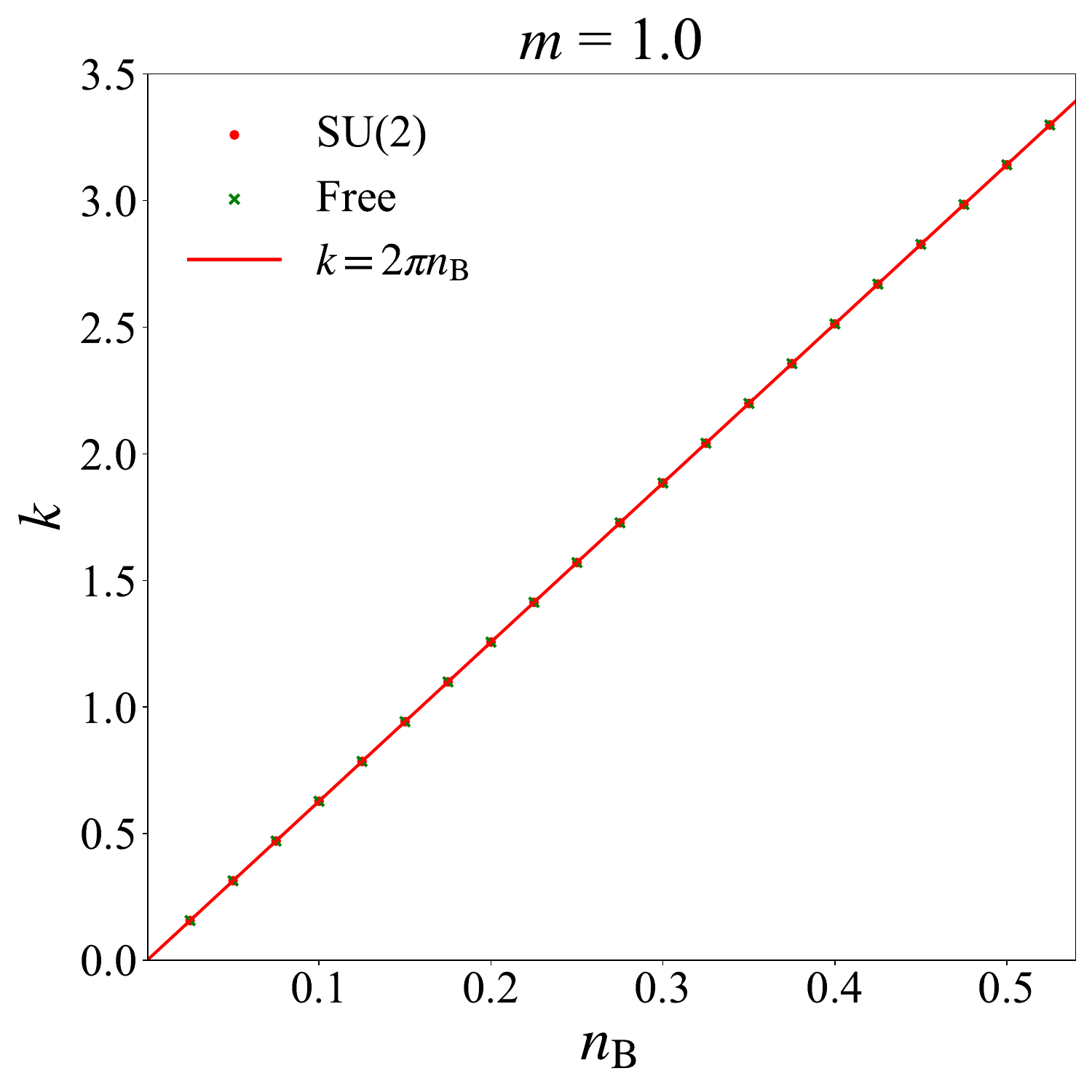}
    \caption{
    Wave number of $n_{\textrm{B}}(x)$ with the largest amplitude as a function of $n_\mathrm{B}$
    for $m=0.5$ (left) and $m=1.0$ (right) 
    with $\hop=2.0$ and $N=160$.
    }
    \label{fig:wave_number}
\end{figure}
\begin{figure}[tb]
  \centering
  \includegraphics[width=0.49\linewidth]{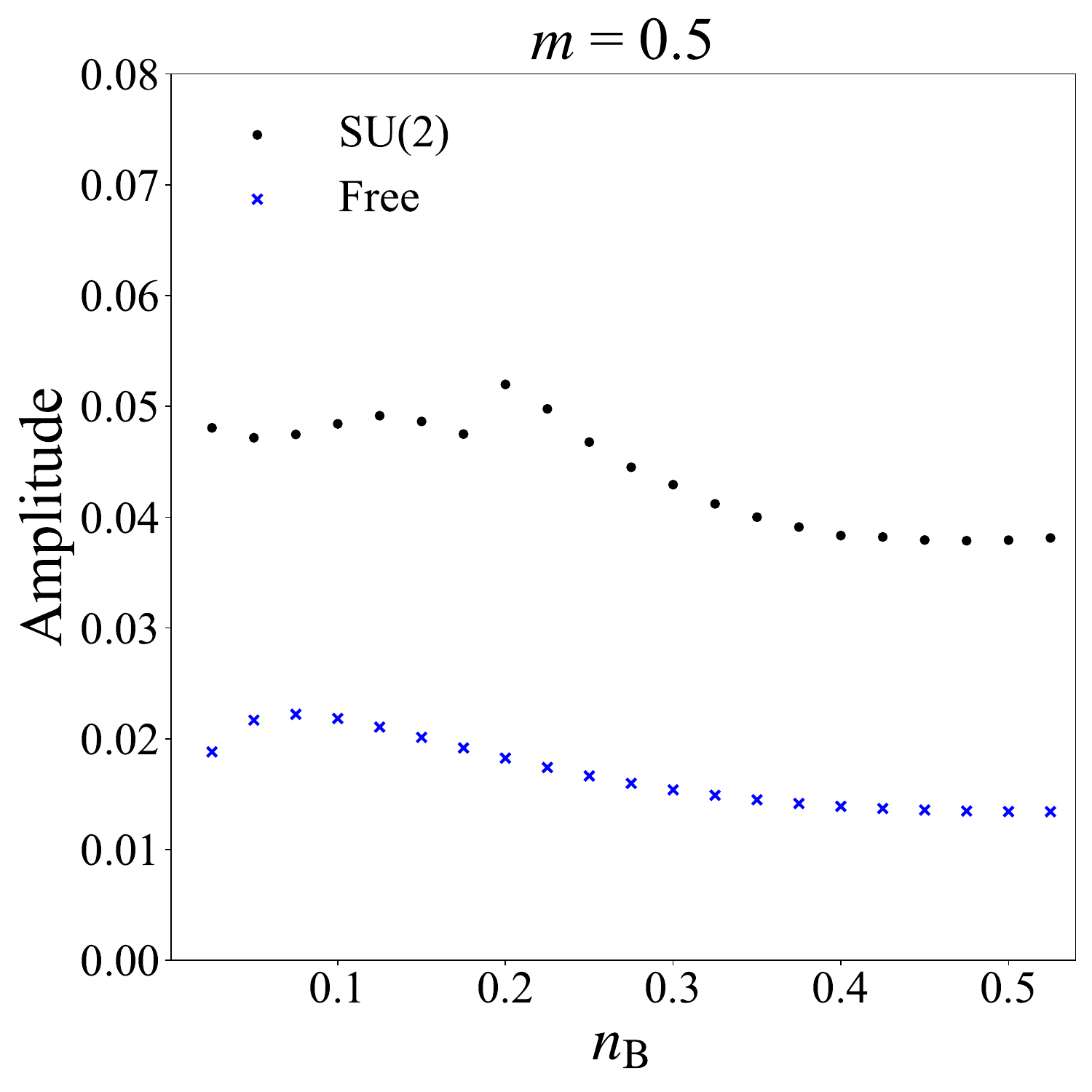}
  \includegraphics[width=0.49\linewidth]{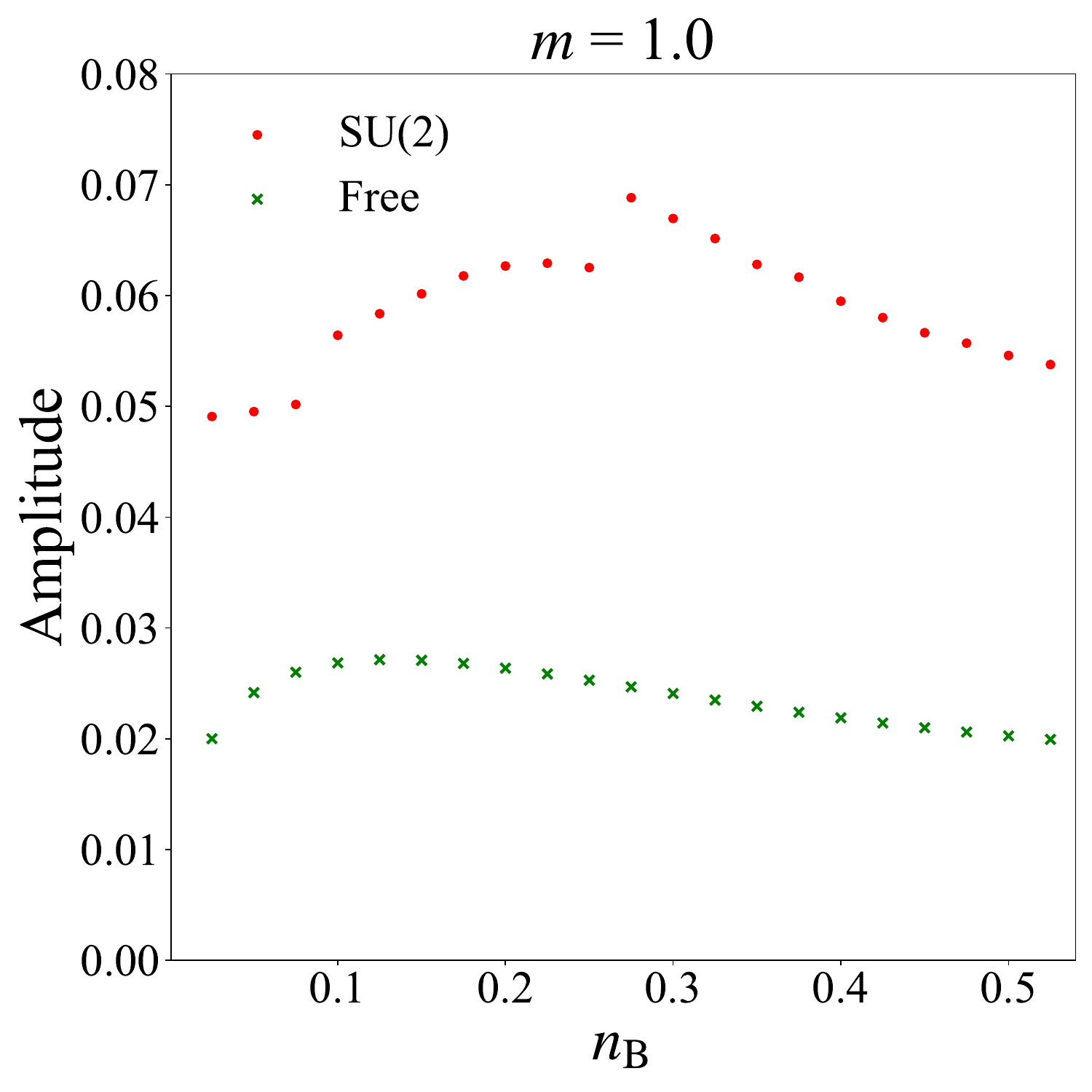}
  \caption{
    Largest amplitude in the Fourier expansion of $n_{\textrm{B}}(x)$ as a function of $n_\mathrm{B}$ for $m=0.5$ (left) and $m=1.0$ (right) with $\hop=2.0$ and $N=160$.
  }
  \label{fig:amplitude}
\end{figure}

We use the Fourier decomposition to analyze the modulation, thereby determining at which wavenumber the modulation occurs.
Figure~\ref{fig:wave_number} shows the wavenumber with the largest amplitude for $\textrm{QCD}_2$ and free theory.
Interestingly, the density dependence of the wave number almost coincides with each other. The behavior can be fitted by
\begin{equation}
  k = 2\pi n_\mathrm{B}.
\end{equation}
There are two ways of looking at this periodic structure.
One is based on the baryonic picture. Assuming repulsive interactions between baryons, if a box of length $V$ is packed with equally spaced baryons, the spacing is $V/N_\mathrm{B}=1/n_\mathrm{B}$, and thus, the wave number is $k=2\pi n_\mathrm{B}$.
The other is based on the quark picture.
If there is no interaction, the quarks form a Fermi surface at $p=p_\mathrm{F}=\pi n_\mathrm{B}$.
If one turns on an attractive interaction between quarks, the Fermi surface becomes unstable due to the Peierls instability and condensation of particle and hole pair occurs~\cite{RevModPhys.60.1129}.
In this condensation, the relative momentum of the particle and hole is zero, but the total momentum is $2p_\mathrm{F}$.
Therefore, the wave number is $k=2p_\mathrm{F}= 2\pi n_\mathrm{B}$.
The same modulation is obtained regardless of whether the quark or baryon picture holds.
This observation implies that the modulation is independent of the number of colors.
As will be discussed in section~\ref{sec:SU(3)}, the same behavior will be found in the $\mathrm{SU}(3)$ case.

We shall see that the amplitude of this modulation is significantly larger than the free case at any density.
Figure~\ref{fig:amplitude} shows the largest amplitude in the Fourier decomposition of the baryon number density.
The figure shows that the amplitude in $\textrm{QCD}_2$ is roughly twice that of the free theory.
In the case of $\textrm{QCD}_2$, there is a point where the amplitude behaves discontinuously, but this is presumably due to the finite volume and not a phase transition.
From this numerical calculation for $\textrm{QCD}_2$, it is challenging to ascertain whether the amplitude with modulation persists in larger volumes.
Instead, let us look at the distribution function of quarks and argue that the occurrence of condensation is plausible, given in the absence of Fermi surfaces.
\subsubsection{Quark distribution function}
\begin{figure}[tb]
  \centering
  \includegraphics[width=0.495\linewidth]{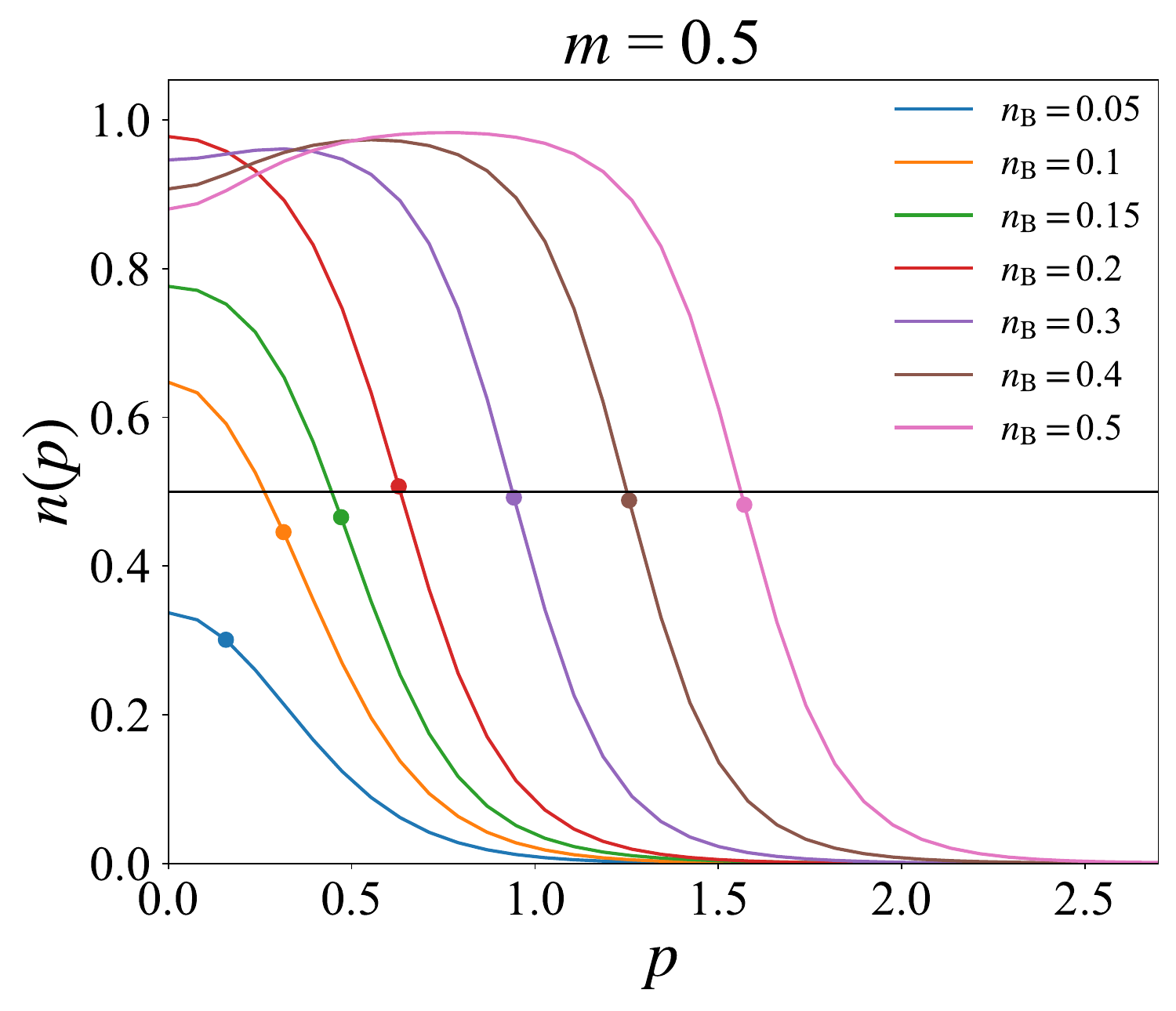}
  \includegraphics[width=0.495\linewidth]{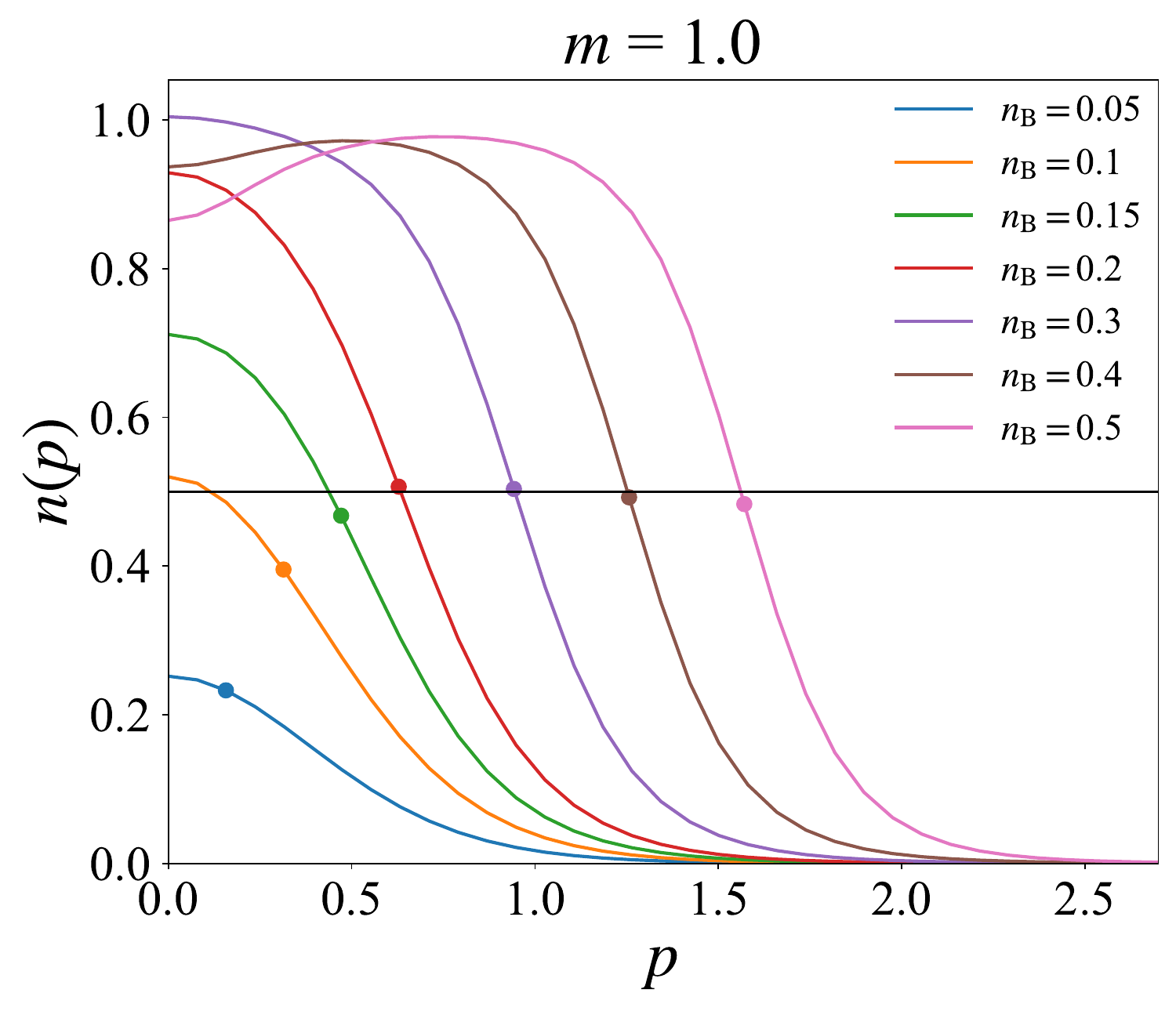}
  \caption{
  Quark distribution function $n(p)$ for $m=0.5$ (left) and $m=1.0$ (right) with $N=240$, $\hop=2.0$, and $n_{\textrm{B}}=0.1,0.2,0.3,0.4,0.5$.
  Colored dots represent the intersections of $n(p)$ and the Fermi momentum of the free theory $p_{\textrm{F}}=\pi n_{\textrm{B}}$.
  The solid black line shows $n(p)=0.5$.
  }
  \label{fig:quark_distribution}
\end{figure}
It is interesting to see how quarks are distributed in the inhomogeneous phases.
We defined the gauge-invariant quark distribution function $n(p)$ in eq.~\eqref{eq:n(p)}.
In order to observe the transition from hadronic matter to quark matter,
we calculate $n(p)$ at $N=240$ and $w=2.0$ varying the baryon density ($n_{\textrm{B}}=0.05,0.1,0.15,0.2,0.3,0.4,0.5$).
We employ the larger site number compared with $N=160$ used in the previous sections, to reduce the finite volume effect.
This is because $n(p)$ around $p=0$ is sensitive to the volume size.
The results for $m=0.5$ and $m=1.0$ are shown in the left and right panels of 
figure~\ref{fig:quark_distribution}, respectively.
As will be discussed below, these figures indicate that condensation is caused by the instability of the Fermi surfaces.

First, the two figures show that the difference in mass does not seem to affect the qualitative behavior of the distribution function.
Next, let us look at the density dependence of the distribution function.
At low densities, the structure has a peak at $p=0$ and decays at large $p$.
This shape is similar to the quark distribution in the single baryon shown in figure~\ref{fig:single_baryon_np}.
This would mean that the degrees of freedom of the system are baryonic.
As the baryon number density increases, the maximum of $n(p)$ at $p=0$ increases.
After the maximum reaches $1$,
$n(p)$ forms the Fermi sea.
This behavior is consistent with the argument based on the quarkyonic picture in $(3+1)$ dimensions discussed in ref.~\cite{Kojo:2021ugu}.
Numerical results show a depression at $p=0$ when the density is high.
We expect that the depression near $p=0$ is a lattice artifact because such depression also appears in the free theory.
Both panels of the figure~\ref{fig:quark_distribution} show that the Fermi sea begins to form near $n_{\textrm{B}}=0.2$.
We expect that the behavior of the system changes from a baryonic matter to a quark matter.
This is consistent with the behavior of the energy per unit baryon number and the averaged chemical potential shown in figure~\ref{fig:energy_su2},
which changes to the free-quark behavior around $n_{\textrm{B}}=0.2$.

We note that such a behavior can also be observed in ultracold atomic gases,
which exhibit the crossover from Bardeen-Cooper-Schrieffer (BCS) superfluids to molecular Bose-Einstein condensates (BEC),
tuning the interaction using the Feshbach resonance.
Although the distribution function of the ultracold atom fermion does not reach the maximum value, $1$, at the origin in the BEC (bosonic) region,
it forms the Fermi surface as it changes to the BCS (fermionic) region \cite{2005PhRvL..95w0405A,2005PhRvL..95y0404R}.

In BCS-type condensation, the half-density point coincides with that of the free theory because the degrees of freedom near the Fermi surface contribute symmetrically to the condensation. 
Our results in figure~\ref{fig:quark_distribution} show 
that this is realized for $n_\mathrm{B}>0.2$.
This is again consistent with the transition from baryon to quark degrees of freedom around $n_\mathrm{B}=0.2$.
Precisely speaking, it is an inhomogeneous phase with density waves, so it is different from BCS-type condensation, but the mechanism of condensation due to the instability of the Fermi surface itself is the same.

\subsection{\texorpdfstring{$\mathrm{SU}(3)$}{SU(3)}}\label{sec:SU(3)}
\begin{figure}[tb]
  \centering
  \includegraphics[width=0.49\linewidth]{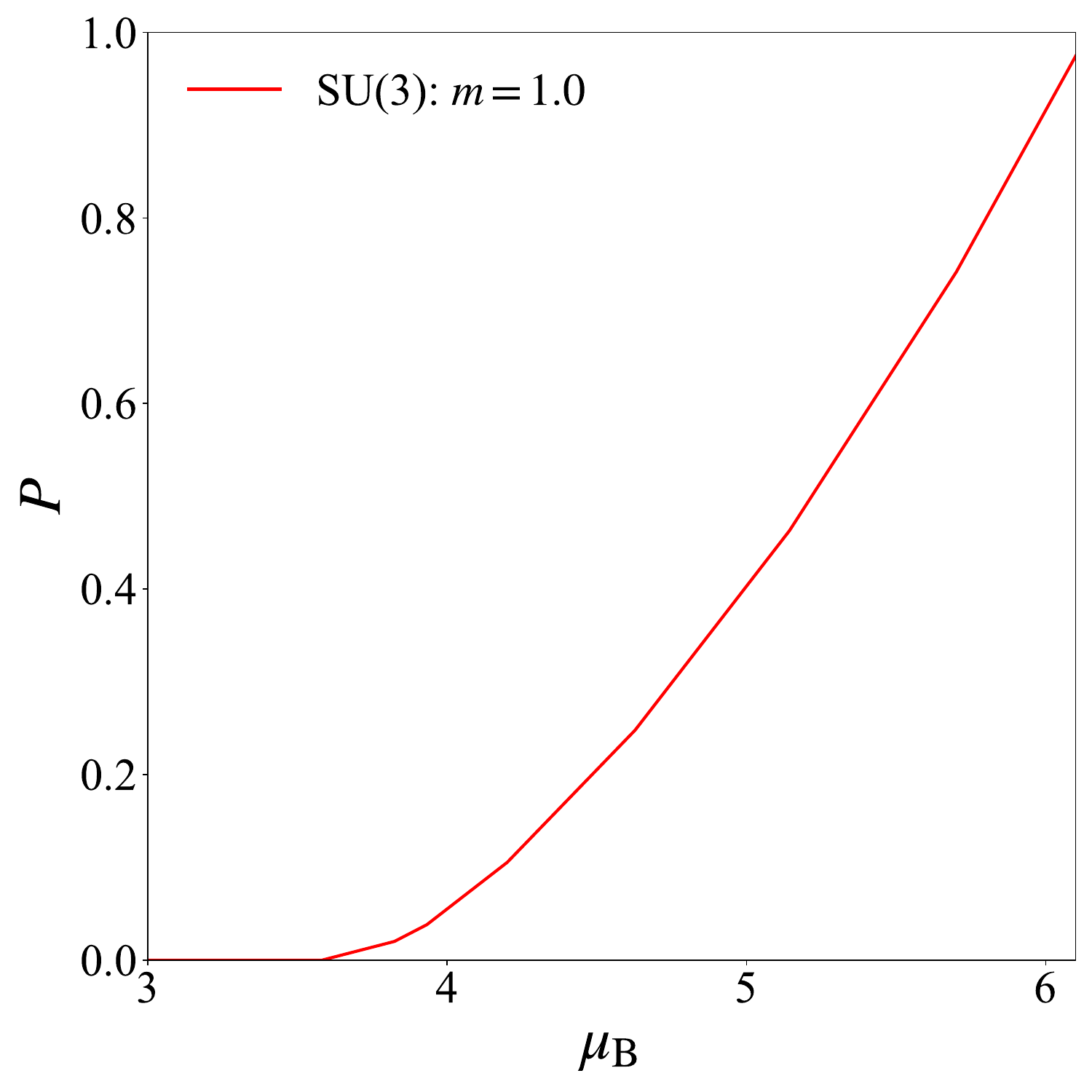}
  \includegraphics[width=0.49\linewidth]{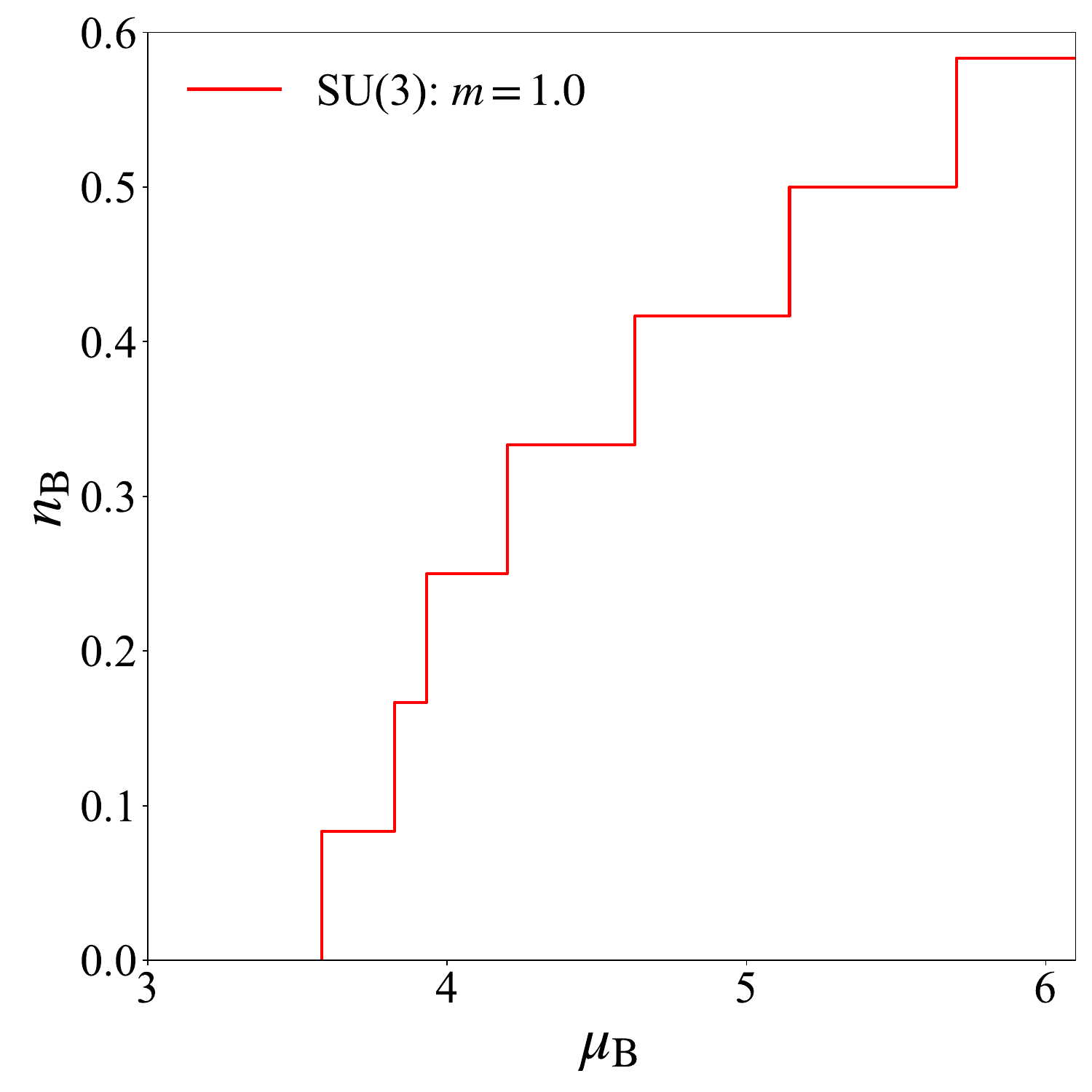}
   \caption{
  Pressure $P$ (left) and baryon number density $n_\mathrm{B}$ (right) for $\mathrm{SU}(3)$ as functions of $\mu_\mathrm{B}$ for $m=1.0$ with $\hop=2$ and $N=48$.
  The pressure and the number density begin to increase at $\mu_\mathrm{B}=3.58$.
  }
  \label{fig:pressure_su3}
\end{figure}
\begin{figure}[tb]
  \centering
  \includegraphics[width=0.49\linewidth]{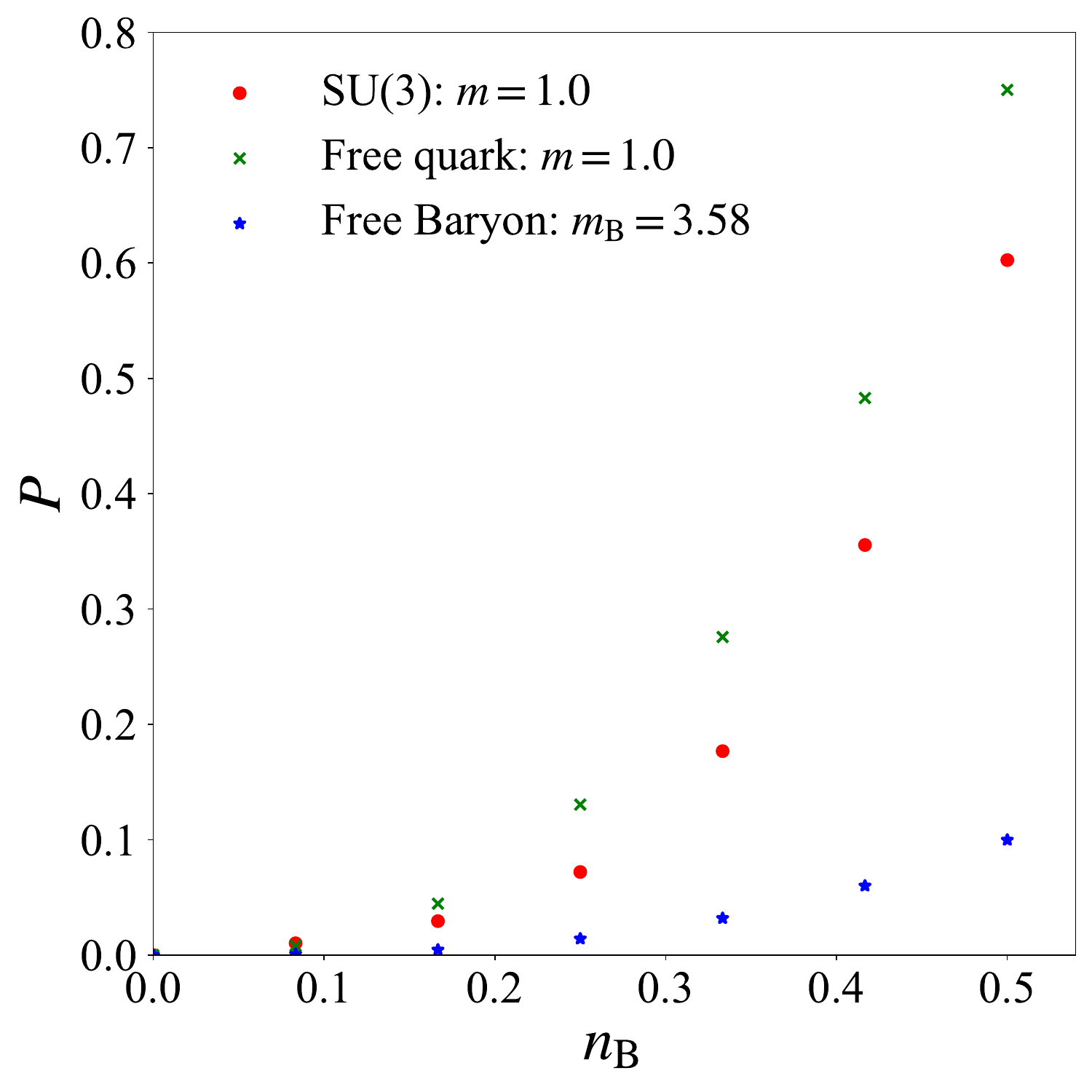}
  \includegraphics[width=0.49\linewidth]{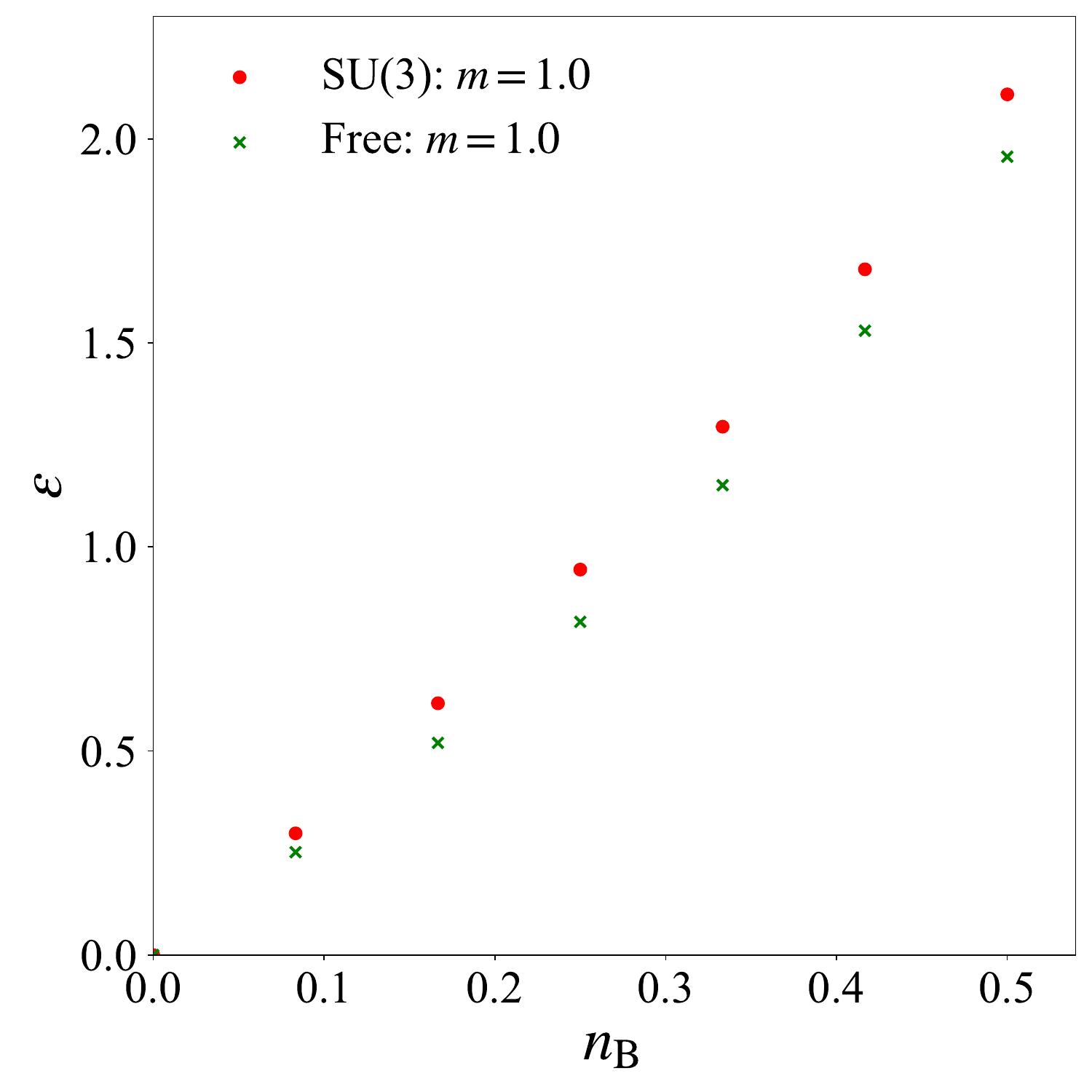}
  \caption{
  Pressure $P$ (left) and energy density $\varepsilon$ (right) for $\mathrm{SU}(3)$ as functions of $n_\mathrm{B}$  for  $m=1.0$ with $\hop=2.0$ and $N=48$.
  As a reference, the pressure of a free baryon with the mass $m_\mathrm{B}=3.58$ is also plotted in the left panel.
  }
  \label{fig:nB_dependence_P_energy_su3}
\end{figure}
In the previous sections, we have numerically studied $\mathrm{QCD}_{2}$
for $N_\mathrm{c}=2$ at finite density.
In this section, we generalize it to the case of $N_{\textrm{c}}=3$.
The numerical calculation for $N_{\textrm{c}}=3$ is more computationally demanding than that for $N_{\textrm{c}}=2$. This is due to the increased color degrees of freedom as well as the enhanced nonlocality of the Hamiltonian represented by the spin system. 
Therefore, we present simulation results for a relatively smaller size with parameters $N=48$, $w=2$, $J=1/8$, which corresponds to the physical volume size $V=12$ in $g_0=1$ unit.
DMRG was performed using iTensor~\cite{itensor} with maximum bond dimension $500$, truncation error cutoff $10^{-8}$, and noise strength $10^{-8}$ for the final DMRG sweeps. We need a few thousand DMRG sweeps for convergence.

\begin{figure}[tb]
  \centering
  \includegraphics[width=0.49\linewidth]{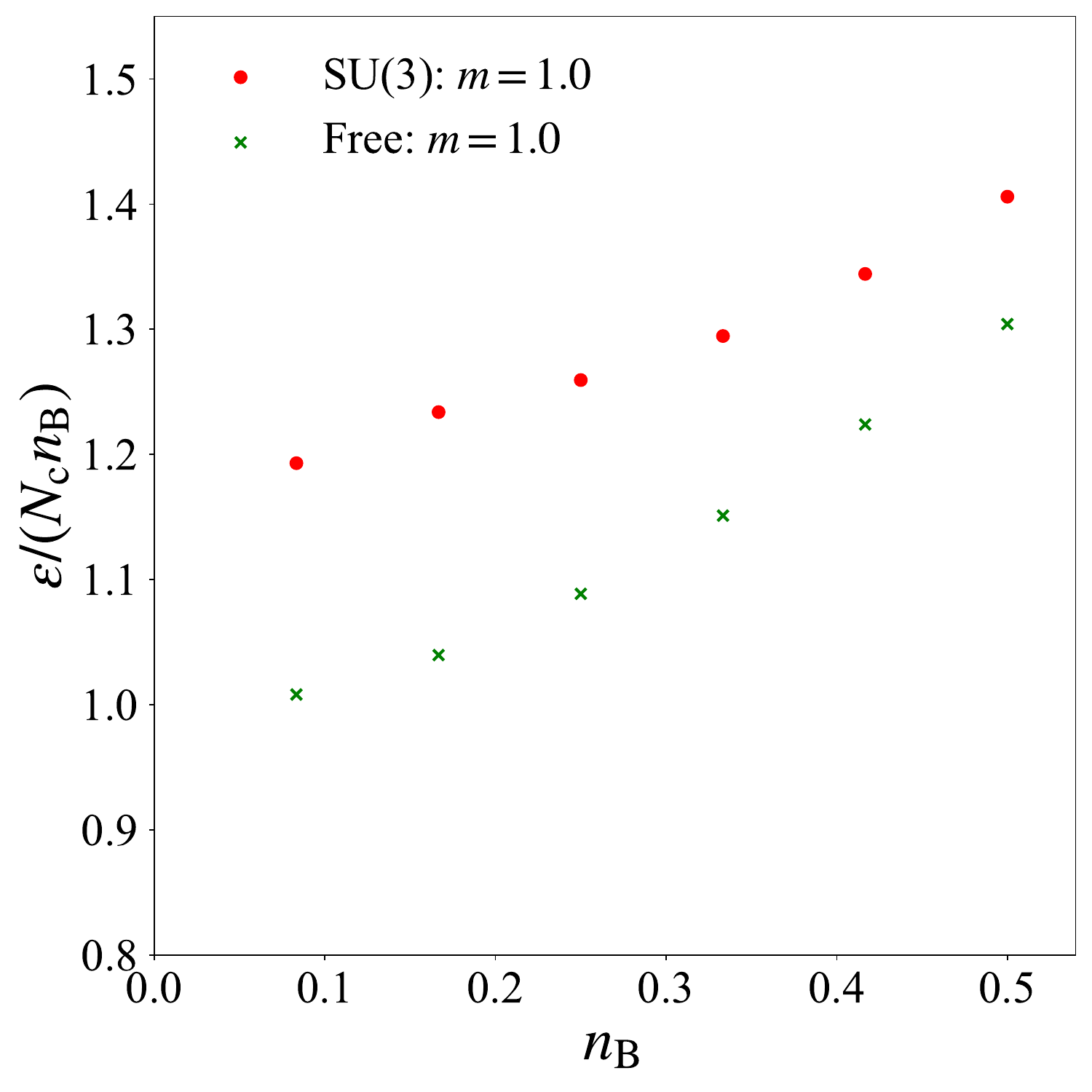}
  \includegraphics[width=0.49\linewidth]{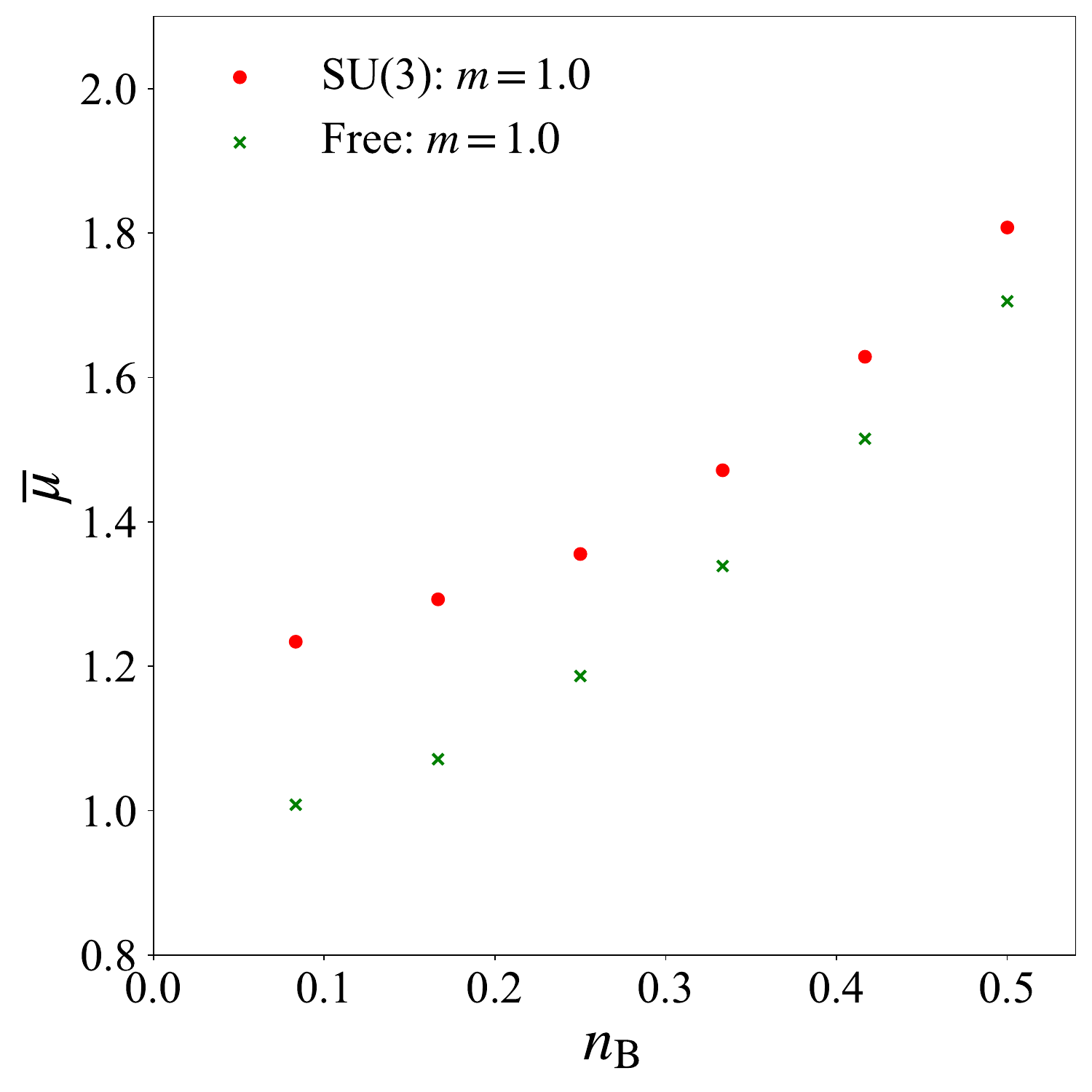}
  \caption{
  Averaged quark chemical potential $\overline{\mu}=\overline{\mu}_\mathrm{B}/N_\mathrm{c}$ (left) and ratio of $\varepsilon$ to $N_{\textrm{c}}n_{\textrm{B}}$ (right) for $\mathrm{SU}(3)$ as functions of $n_\mathrm{B}$ with $m=1.0$, $\hop=2.0$ and $N=48$.
  }
  \label{fig:energy_su3}
\end{figure}

\begin{figure}[tb]
  \centering
  \includegraphics[width=0.49\linewidth]{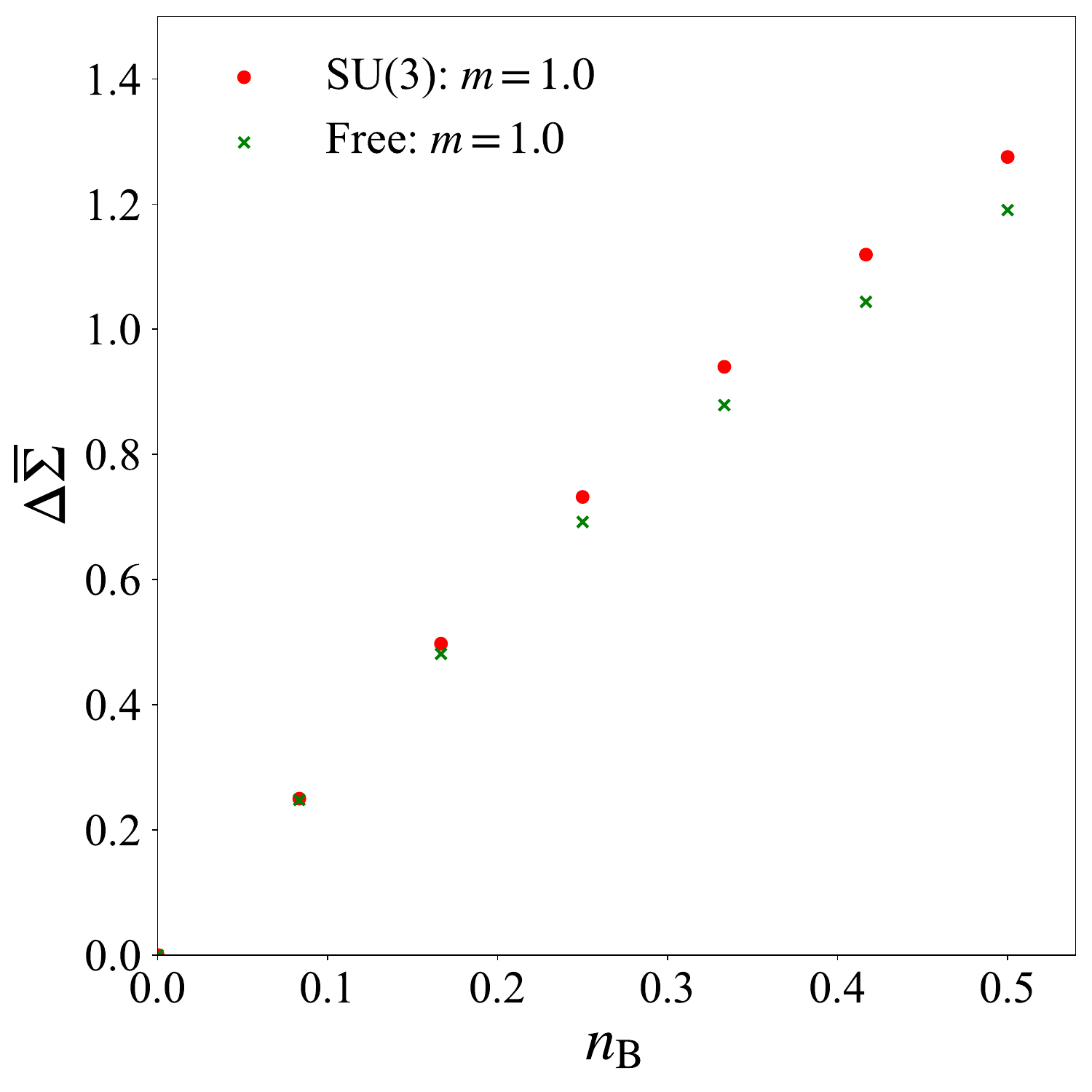}
  \includegraphics[width=0.49\linewidth]{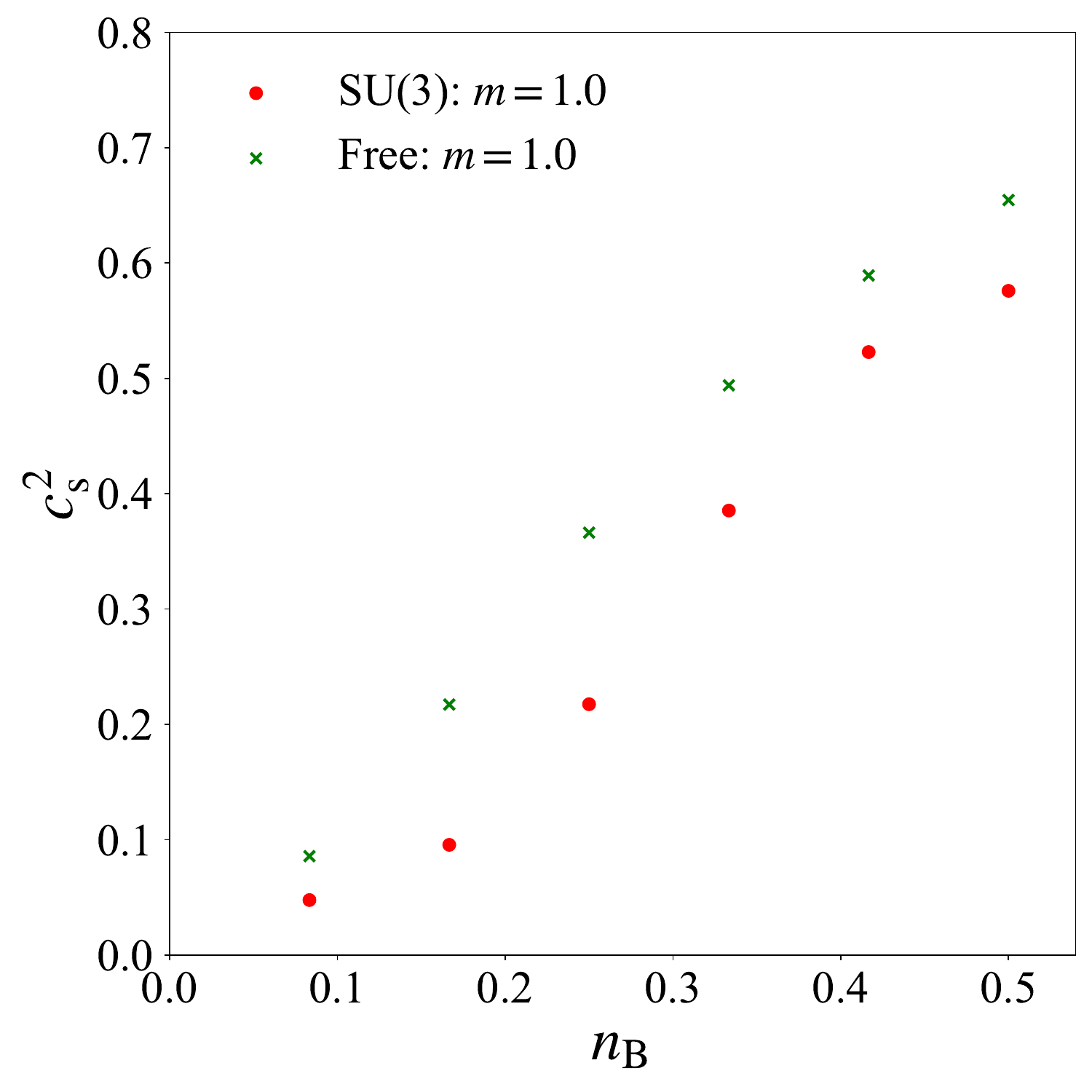}
  \caption{
    Chiral condensate subtracting the contribution from the vacuum
    $\Delta \overline{\Sigma}=\overline{\Sigma}-\overline{\Sigma}_\mathrm{vac}$ (left) and squared sound velocity (right) for $\mathrm{SU}(3)$ as functions of $n_\mathrm{B}$ with $m=1.0$, $\hop=2$ and $N=48$.
  }
  \label{fig}
\end{figure}

\begin{figure}[tb]
  \centering
  \includegraphics[width=15.0cm]{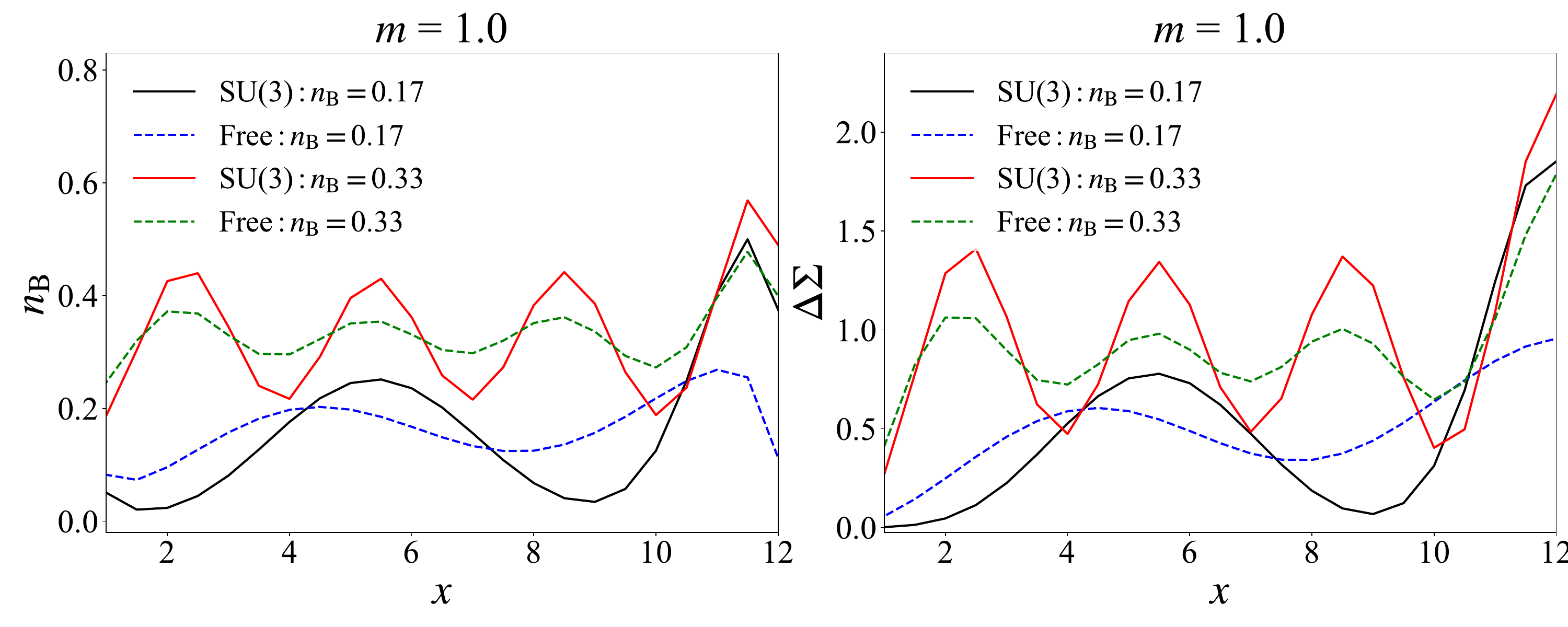}
  \caption{
    Spatial dependence of the baryon number density $n_\mathrm{B}(x)$ and the chiral condensate subtracting the contribution from the vacuum
    $\Delta {\Sigma}(x)={\Sigma}(x)-{\Sigma}_\mathrm{vac}(x)$ with $m=1.0$, $\hop=2$ and $N=48$.
  }
  \label{fig:inhomogeneous_phase_su3}
\end{figure}

\begin{figure}[tb]
  \centering
  \includegraphics[width=0.49\linewidth]{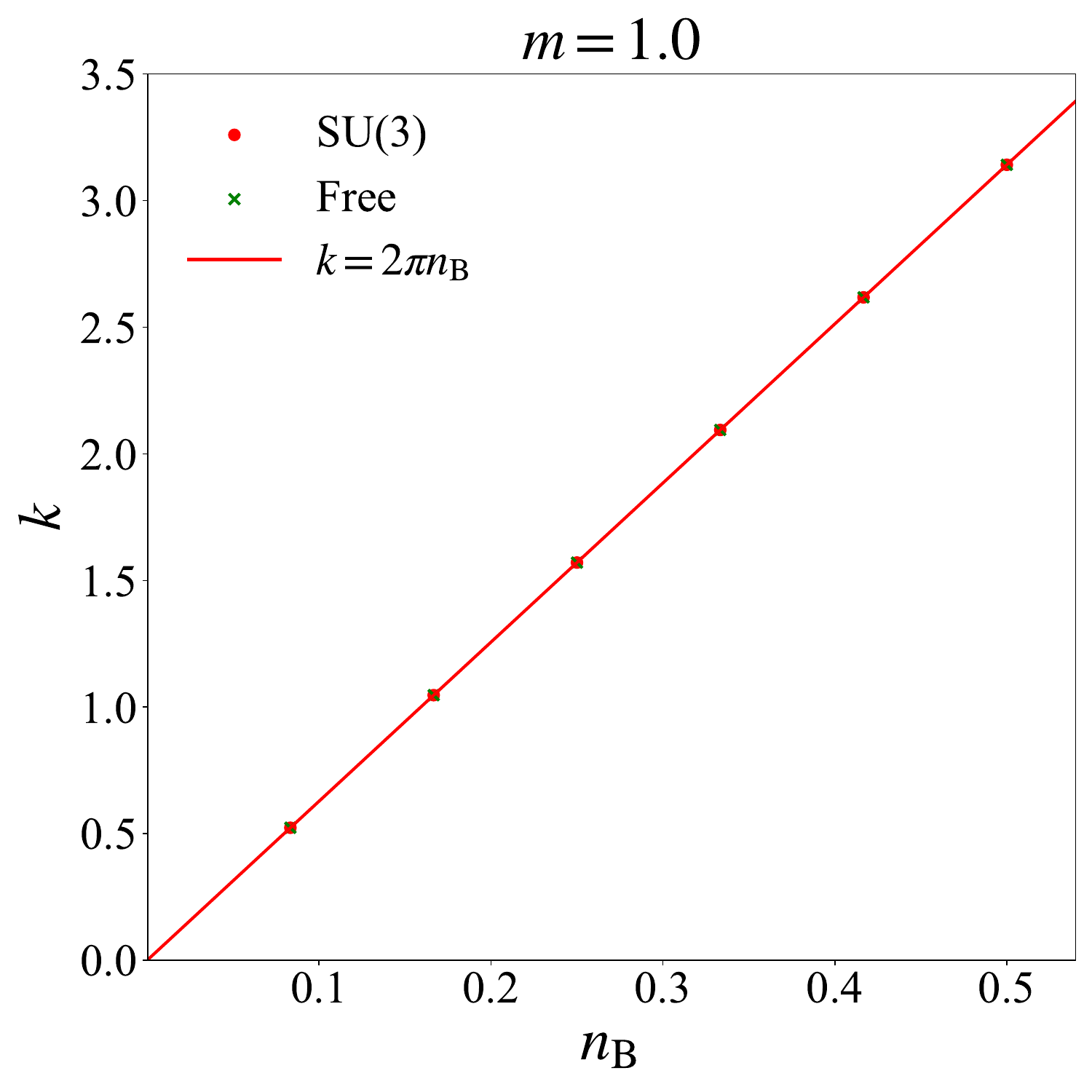}
  \includegraphics[width=0.49\linewidth]{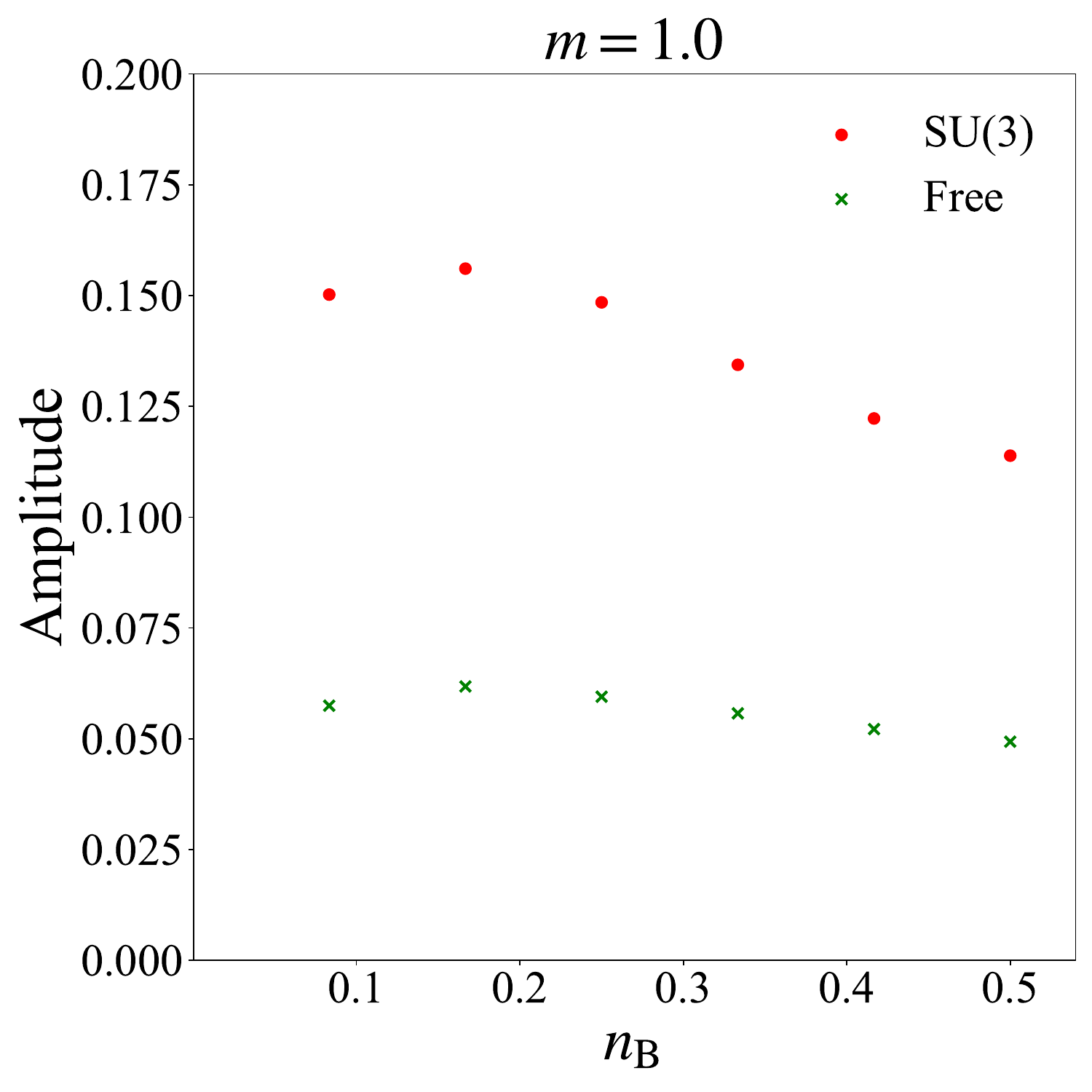}
 \caption{
  Wave number of $n_{\textrm{B}}(x)$ with the largest amplitude (left) and amplitude (right) as functions of $n_\mathrm{B}$ with $m=1.0$, $\hop=2$ and $N=48$.
  }
  \label{fig:wave_number_su3}
\end{figure}

\begin{figure}[tb]
  \centering
  \includegraphics[width=0.495\linewidth]{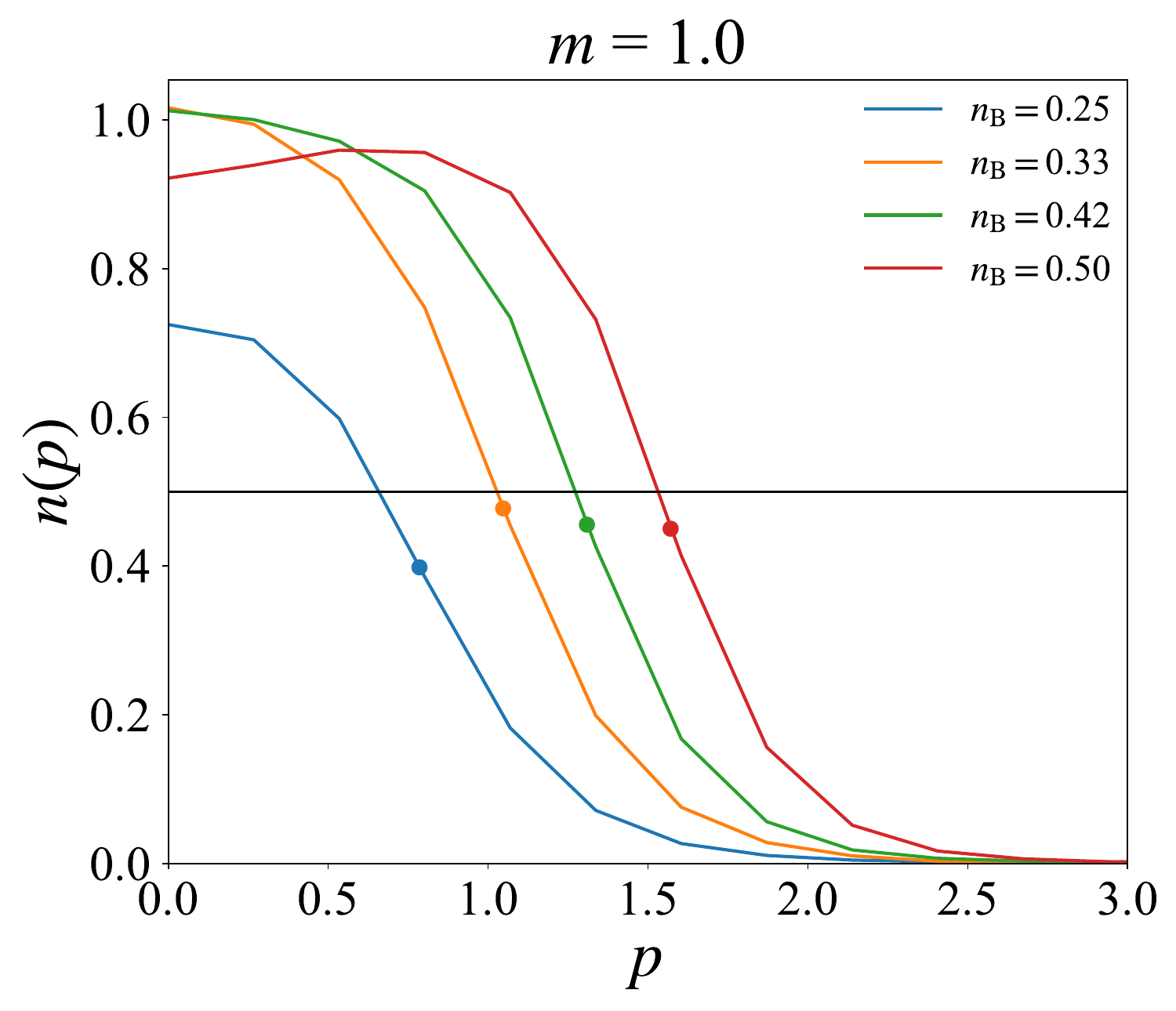}
  \caption{
  Quark distribution function $n(p)$ for $m=1.0$  with $N=48$, $\hop=2.0$, and $n_{\textrm{B}}=0.25,0.33,0.42,0.50$.
  Colored dots represent the intersections of $n(p)$ and the Fermi momentum of the free theory $p_{\textrm{F}}=\pi n_{\textrm{B}}$.
  The solid black line shows $n(p)=0.5$.
  }
  \label{fig:quark_distribution_su3}
\end{figure}

We show graphs corresponding to figures \ref{fig:pressure_su2}-\ref{fig:quark_distribution} (except figure~\ref{fig:size_dependence}) in figures \ref{fig:pressure_su3}-\ref{fig:quark_distribution_su3} for $m=1$.
As in the case of $N_{\textrm{c}}=2$, the pressure and baryon number density start to increase from $\mu_\mathrm{B}>N_\mathrm{c}m$ due to the confining energy (see figure \ref{fig:pressure_su3}).
The critical value of the baryon chemical potential is given as $\mu_\mathrm{B}=3.58$ ($>3m$).
The overall behavior is similar to the case of $N_{\mathrm{c}}=2$, even though the properties should be different since baryons are fermions and not bosons.
Thermodynamic quantities behave like free quarks at high densities. 
As in the case of $N_\mathrm{c}=2$, the inhomogeneous phases are realized, shown in figure~\ref{fig:inhomogeneous_phase_su3}, and the wavenumber of oscillations with the maximum amplitude is given by $k=2\pi n_\mathrm{B}$ (see figure \ref{fig:wave_number_su3}). The physical interpretation is the same as for $N_\mathrm{c}=2$, which is independent of the number of colors.

It is not easy to see the transition from baryons to quarks from behaviors of the energy per quark and the averaged chemical potential in figure~\ref{fig:energy_su3}. One of the reasons is that the volume is not large, so the number of points is not enough. 
On the other hand, the behavior of the distribution function shows the crossover transition from baryons to quarks around $n_\mathrm{B}=0.3$ in figure~\ref{fig:quark_distribution_su3}.

Finally, let us consider the low-density behavior from the point of view of large $N_\mathrm{c}$ counting.
When comparing free quarks with free baryons,
we find that both quarks and baryons have the same Fermi momentum $p_\mathrm{F}=\pi n_\mathrm{B}$. This is a unique property in $(1+1)$-dimensional theory.
The pressure of baryons and quarks for a given baryon number density $n_\mathrm{B}$ are 
\begin{align}
    P_\mathrm{B} &=\frac{p_\mathrm{F}^3}{3\pi m_\mathrm{B}}=\frac{\pi^2n_\mathrm{B}^3}{3 m_\mathrm{B}},\\
    P_\mathrm{q} &=N_\mathrm{c}\frac{p_\mathrm{F}^3}{3\pi m_\mathrm{q}}=N_\mathrm{c}\frac{\pi^2n_\mathrm{B}^3}{3 m_\mathrm{q}},
\end{align}
respectively. Here, $m_\mathrm{B}$ is the baryon mass of order $N_\mathrm{c}$, and $m_\mathrm{q}$ is the (constituent) quark mass of order $N_\mathrm{c}^0$.
Since $m_\mathrm{B} \sim N_\mathrm{c} m_\mathrm{q}$, the pressures for the free baryons and the free quarks are different by an order $N_\mathrm{c}^2$.
We plot the pressures of the free baryons in figure~\ref{fig:nB_dependence_P_energy_su3} for comparison. 
It can be seen that the pressure of free baryons is strongly suppressed compared to that of free quarks. 
The result from the numerical calculations (red dots) is closer to that for the free quarks rather than for the free baryons.
It implies that the contribution from the interactions of order $N_\mathrm{c}$ plays an essential role.
Note that the fact that the interaction between baryons is of order $N_\mathrm{c}$ is consistent with the large-$N_\mathrm{c}$ counting~\cite{Witten:1979kh}.


\section{Summary}\label{sec:summary}
In this paper, we numerically studied the one-flavor $\textrm{QCD}_2$ at finite baryon chemical potential using the DMRG method.
Using the Wigner-Jordan transformation, $\textrm{QCD}_2$ can be mapped to the spin chain system with non-local interaction.
Therefore, we can apply the DMRG method to $\textrm{QCD}_2$.
Not only is this computational method efficient, but also it is free of sign problems.
We have calculated thermodynamic quantities, such as the pressure, baryon number density, and energy density.
We have also calculated the quark distribution function from the Fourier transform of the Wigner function. 
Our study may pave the way toward ab-initio study of cold-dense QCD on the basis of tensor network methods.

We have several future directions, which make the analysis more realistic by introducing gauge fields, but may be reachable with current computational resources (although those are still far from real QCD).
First, if we employ periodic boundary conditions in $\textrm{QCD}_2$, one link corresponding to the spatial Wilson loop is no longer integrated out by gauge transformations.
This would be the most simple setup to simulate a nontrivial nonabelian gauge field, although it is not dynamical.
Furthermore, we may be able to study the effects of dynamical gauge fields by considering quasi-one-dimensional lattice such as a two-leg ladder geometry, and formulating gauge fields on the basis of the recently developed Hamiltonian Yang-Mills theory~\cite{Hayata:2021kcp,Yao:2023pht,Zache:2023dko,Hayata:2023puo,Hayata:2023bgh}.


\section*{Acknowledgements}
The numerical calculations were carried out on cluster computers at iTHEMS in RIKEN.
The source code for numerical calculations is available from \cite{dmrgqcd2}, which is implemented based on iTensor~\cite{itensor}. 
This work was supported by JSPS KAKENHI Grant Numbers~21H01007, and 21H01084.

\appendix

\section{Free theory on a lattice}\label{sec:Free_theory}
For comparison with $\textrm{QCD}_2$, we here present several results from the free theory with open boundary conditions.
For simplicity, we consider the case with $N_\mathrm{c}=1$ and $N_{\mathrm{f}}=1$
since the generalization to the case with arbitrary $N_\mathrm{c}$ and $N_{\mathrm{f}}$ is trivial.
The Hamiltonian of the free theory has the form:
\begin{align}
  {H}_\mathrm{free}&=\hop\sum_{n=1}^{N-1}\qty(\chi^\dag(n+1) \chi(n)+\chi^\dag(n)\chi(n+1))
  +m\sum_{n=1}^N(-1)^{n} \chi^\dag(n) \chi(n).
\end{align}
We choose the number of lattice sites $N$ to be even.
The equation of motion for the field reads
\begin{equation}
  \ri \partial_t \chi(n) = - [{H}_\mathrm{free}, \chi(n)]=\hop \qty(\chi(n+1)+\chi(n-1))
  + m (-1)^n\chi(n).
  \label{eq:eom fermion}
\end{equation}
Introducing two-component Dirac fields as
\begin{equation}
(-1)^n\begin{pmatrix}
\psi_\mathrm{L}(n)\\
\ri\psi_\mathrm{S}(n)
\end{pmatrix}
\coloneqq{\hop^{\frac{1}{2}}}
\begin{pmatrix}
  \chi(2n)\\
\chi(2n-1)
\end{pmatrix},
\end{equation}
we can express eq.~\eqref{eq:eom fermion} as
\begin{align}
  \ri\partial_t\psi_\mathrm{L}(n)&= -\ri\hop\qty(\psi_\mathrm{S}(n+1) -\psi_\mathrm{S}(n))+m\psi_\mathrm{L}(n),\\
  \ri\partial_t\psi_\mathrm{S}(n)&= -\ri\hop\qty(\psi_\mathrm{L}(n) -\psi_\mathrm{L}(n-1))-m\psi_\mathrm{S}(n).
\end{align}
This serves as the discrete analog of the Dirac equation in $(1+1)$ dimensions.
Here, $\mathrm{L}$ and $\mathrm{S}$ represent the large and small components of the wave function, respectively.
The open boundary conditions we impose are $\psi_\mathrm{L}(0)=0$ and $\psi_\mathrm{S}(L+1)=0$, where
$L$ is the number of physical sites defined by $L = N/2$.
The time-independent Dirac equation reads
\begin{align}
  (E-m)\psi_\mathrm{L}(n)&= -\ri\hop\qty(\psi_\mathrm{S}(n+1) -\psi_\mathrm{S}(n)), \label{eq:large_Eq}\\
  (E+m)\psi_\mathrm{S}(n)&= -\ri\hop\qty(\psi_\mathrm{L}(n) -\psi_\mathrm{L}(n-1)),
  \label{eq:small_Eq}
\end{align}
where $E$ represents an eigenvalue of the Hamiltonian.
Multiplying both sides of eq.~\eqref{eq:large_Eq} by $(E+m)$, and using eq.~\eqref{eq:small_Eq}, we obtain
\begin{align}
 (E^2-m^2)\psi_\mathrm{L}(n)&= 
 -\hop^2\qty(\psi_\mathrm{L}(n+1) -2\psi_\mathrm{L}(n) +\psi_\mathrm{L}(n-1)).
\end{align}
Given that this is a linear second-order difference equation, the solution can be expressed as
\begin{equation}
  \psi_\mathrm{L}(n)= A e^{\ri {2}pn}+B e^{-\ri {2}pn},
\end{equation}
with $E= \pm E_p=\pm\sqrt{4\hop^2\sin^2{p}+m^2}$,
where $A$, $B$, and $p$ are constants determined by the boundary conditions and the normalization of wave functions.
For a small $p$, $E_p^2\simeq {4}\hop^2 p^2 +m^2$,
which implies that $2\hop p$ is the momentum in the physical unit.
From one boundary condition $\psi_\mathrm{L}(0)=0$, we obtain, $A=-B=\mathcal{N}/(2\ri)$, and thus,
\begin{equation}
  \psi_\mathrm{L}(n)= \frac{\mathcal{N}}{2\ri}\qty( e^{\ri {2}pn}- e^{-\ri {2}pn})= \mathcal{N}\sin({2}pn),
\end{equation}
where $\mathcal{N}$ is the normalization constant.
$\psi_\mathrm{S}(n)$ for the positive energy solution, $E=+E_p$, is obtained from eq.~\eqref{eq:small_Eq} as
\begin{equation}
  \begin{split}
  \psi_\mathrm{S}(n)&=\frac{-\ri\hop}{E_p+m}(\psi_\mathrm{L}(n) -\psi_\mathrm{L}(n-1))\notag\\
  &=\frac{-\ri\hop}{E_p+m}\qty(
    \sin({2}pn)
    - \sin({2}p(n-1)))\mathcal{N}\notag\\
  &=\frac{-\ri2\hop\sin(p)}{E_p+m}\cos(p(2n-1))\mathcal{N}
      .
  \end{split}
\end{equation}
Here, we employed the sum-to-product identity,
\begin{equation}
  \sin({2}pn)-\sin({2}p(n-1))=2\sin(p)
    \cos(p(2n-1)),
\end{equation}
to obtain the last equality.
Therefore, the wave function with the positive energy solution $u(p,n)$ is expressed as
\begin{align}
  u(p,n) 
  =\begin{pmatrix}
    \psi_\mathrm{L}(n)\\
    \psi_\mathrm{S}(n)
  \end{pmatrix}
  & =\mathcal{N}
  \begin{pmatrix}
    \sin({2}pn)\\
    \frac{-\ri 2\hop\sin(p)}{E_p+m}\cos \qty(p\qty(2n-1))
  \end{pmatrix}.
  \label{eq:wave_function_positive_energy}
\end{align}
Similarly, we also obtain the wave function with the negative energy solution $v(p,n)$,
\begin{equation}
  v(p,n) = 
    \mathcal{N}
  \begin{pmatrix}
    \frac{-\ri 2\hop\sin(p)}{E_p+m}\sin({2}pn)\\
    \cos\qty(p(2n-1))
  \end{pmatrix}.
  \label{eq:wave_function_negative_energy}
\end{equation}
The other boundary condition $\psi_{\mathrm{S}}(L+1)=0$ or equivalently,
\begin{equation}
  \cos(p(2L+1))=0,
\end{equation}
quantize the momentum
\begin{equation}
  p=\frac{2\pi k-\pi}{{2}(2L+1)},
  \label{eq:momentum_free}
\end{equation}
where $k$ is an integer, and we choose $1\leq k\leq L$.

Let us determine the normalization constant $\mathcal{N}$. Notice that the sine and cosine satisfy orthogonal relations,
\begin{align}
  \sum_{n=1}^{L} \sin({2}pn)\sin( {2}p'n)&=\delta_{p,p'}\frac{1}{4}(2L+1), \label{eq:orthogonal_relation_sin}\\
  \sum_{n=1}^{L} \cos\qty(p\qty(2n-1))\cos\qty(p'\qty(2n-1))
  &=\delta_{p,p'}\frac{1}{4}(2L+1), \label{eq:orthogonal_relation_cos}
\end{align}
and
\begin{align}
  \sum_{p} \sin({2}pn)\sin({2}pn')
  &=\frac{1}{4}(2L+1)\delta_{n,n'}, \label{eq:orthognal1}\\
  \sum_{p} \cos\qty(p\qty(2n-1))\cos\qty(p\qty(2n'-1))
  &=\frac{1}{4}(2L+1)\delta_{n,n'}.\label{eq:orthognal2}
\end{align}
Equations~\eqref{eq:orthogonal_relation_sin} and \eqref{eq:orthogonal_relation_cos} lead to the orthogonality of the wave functions,
\begin{equation}
  \sum_{n=1}^Lu^\dag(p,n)u(p',n)=\sum_{n=1}^Lv^\dag(p,n)v(p',n)\propto \delta_{p,p'},
\end{equation}
and
\begin{equation}
  \sum_{n=1}^Lv^\dag(p,n)u(p',n)=\sum_{n=1}^Lu^\dag(p,n)v(p',n)=0.
\end{equation}
The normalization factor is determined by 
\begin{equation}
  \begin{split}
    1&={\frac{1}{\hop}}\sum_{n=1}^{L}u^\dag(p,n)u(p,n)\\
    &=|\mathcal{N}|^2{\frac{1}{\hop}}\sum_{n=1}^{L}\qty(
      \sin^2({2}pn)
      +\frac{4\hop^2\sin^2(p)}{(E_p+m)^2}\cos^2(p(2n-1))
    )\\
    &=|\mathcal{N}|^2\frac{1}{4\hop}(2L+1)\frac{2E_p}{E_p+m},
  \end{split}
\end{equation}
and thus, the normalization constant is
\begin{equation}
  \mathcal{N} = \sqrt{\frac{4\hop}{2L+1}}\sqrt{\frac{E_p+m}{2E_p}},
\end{equation}
where we choose $\mathcal{N}$ real.
Similarly, we can obtain the same normalization factor for $v$.

From the explicit form of the wave functions in eqs.~\eqref{eq:wave_function_positive_energy} and \eqref{eq:wave_function_negative_energy},
$u_{s}(p,n)u_{s'}^\dag(p,n')$ and  $v_{s}(p,n)v_{s'}^\dag(p,n')$ are given as
\begin{equation}
  \begin{split}
    &u_{s}(p,n)u_{s'}^\dag(p,n')\\
     &= 
    \frac{|\mathcal{N}|^2}{E_p+m}\begin{pmatrix}
      (E_p+m)\sin({2}pn) \sin({2}pn') &  
      \ri 2\hop\sin(p)\sin({2}pn)\cos \qty(p\qty(2n'-1))
      \\
      -\ri 2\hop\sin(p)\cos \qty(p\qty(2n-1))\sin({2}pn')&
      (E_p-m)
      \cos \qty(p\qty(2n-1))
      \cos \qty(p\qty(2n'-1))
    \end{pmatrix},
  \end{split}
\end{equation}
and
\begin{equation}
  \begin{split}
    &v_{s}(p,n)v_{s'}^\dag(p,n')\\
     &= 
    \frac{|\mathcal{N}|^2}{E_p+m}
    \begin{pmatrix}
    (E_p-m)\sin({2}pn)\sin({2}pn')& -\ri2\hop\sin(p)\sin({2}pn)\cos\qty(p\qty(2n'-1)) \\
    \ri2\hop\sin(p)\cos\qty(p\qty(2n-1))\sin({2}pn') & (E_p +m)\cos\qty(p\qty(2n-1))\cos\qty(p\qty(2n'-1))
    \end{pmatrix},
  \end{split}
\end{equation}
respectively.
The sum is diagonal:
\begin{equation}
  \begin{split}
  &u_{s}(p,n)u_{s'}^\dag(p,n')
  +v_{s}(p,n)v_{s'}^\dag(p,n')\\
  &=\frac{4\hop}{2L+1}
  \begin{pmatrix}
    \sin({2}pn)\sin({2}pn')&0 \\
    0& \cos\qty(p\qty(2n-1))\cos\qty(p\qty(2n'-1))
    \end{pmatrix}.
    \label{eq:completeness_p}
  \end{split}
\end{equation}
Summing eq.~\eqref{eq:completeness_p} over $p$ and using eqs.~\eqref{eq:orthognal1} and \eqref{eq:orthognal2}, we obtain the completeness relation,
\begin{equation}
  \begin{split}
    \sum_{p}\qty(
      u_{s}(p,n)u_{s'}^\dag(p,n')
      +v_{s}(p,n)v_{s'}^\dag(p,n')
    )=\hop\delta_{s,s'}\delta_{n,n'}.
  \end{split}
  \label{eq:completeness}
\end{equation}

The quantized field can be expanded as
\begin{equation}
  \begin{split}
    \psi_s(n) &=\sum_{p}\qty(u_s(p,n)a_p + v_s(p,n)b^\dag_p
    ).
  \end{split}
\end{equation}
Here, $a_p^\dag$, $a_p$, $b_p^\dag$, and $b_p$ are creation-annihilation operators for particles and antiparticles, which obey the anticommutation relations:
\begin{equation}
  \{a_p,b^\dag_{p'}\}=\delta_{p,p'},\quad
  \{b_p,b^\dag_{p'}\}=\delta_{p,p'}. \label{eq:commutation_relation_b}
\end{equation}
One can check this is consistent with the anticommutation relation of fields in the coordinate space,
\begin{equation}
  \begin{split}
    \{\psi_s(n),\psi_{s'}^\dag(n')\}
    =\sum_{p}\qty(
      u_{s}(p,n)u_{s'}^\dag(p,n')
      +v_{s}(p,n)v_{s'}^\dag(p,n')
    )=\hop\delta_{s,s'}\delta_{n,n'},
  \end{split}
\end{equation}
where we used eq.~\eqref{eq:completeness}.

Let us now evaluate observables.
We first consider the thermodynamic observables.
The Hamiltonian is written as
\begin{equation}
  H_\mathrm{free} = \sum_{p}\qty(E_p a^\dag_p a_p  -E_p b_p b^\dag_{p})
  =\sum_{p}\qty(E_p a^\dag_p a_p  +E_p b^\dag_{p} b_p ) - E_0,
  \label{eq:Hamiltonian_free}
\end{equation}
where $E_0$ is the vacuum energy,
\begin{equation}
  E_0 = -\sum_p E_p.
\end{equation}
In the last equality in eq.~\eqref{eq:Hamiltonian_free}, we used eq.~\eqref{eq:commutation_relation_b}.
We normalize the vacuum energy to $0$ and drop $E_0$ in the following.

The vacuum state $\ket{\Omega}$ at $\mu=0$ satisfies
\begin{equation}
  a_p\ket{\Omega} = 0 \quad  b_p\ket{\Omega} = 0
\end{equation}
for any $p$.
On the other hand, for $\mu\neq0$,
the state minimizes
\begin{equation}
  H_\mathrm{free}-\mu N_\mathrm{q} = \sum_{p}\qty((E_p-\mu) a^\dag_p a_p  +(E_p+\mu) b^\dag_{p} b_p ).
  \label{eq:H-mu_N}
\end{equation}
Here, $N_\mathrm{q}$ is the quark number operator:
\begin{equation}
  N_\mathrm{q} =\sum_p\qty( a^\dag_p a_p  - b^\dag_{p} b_p ).
\end{equation}
The solution that minimizes eq.~\eqref{eq:H-mu_N} is
\begin{equation}
  \ket{\Omega(\mu)} = \prod_{p\leq {\overline{p_\mathrm{F}}}}a_p^\dag \ket{\Omega},
\end{equation}
where ${\overline{p_\mathrm{F}}}$ corresponds to the Fermi momentum, i.e., the maximum $p$ satisfying $E_p<\mu$.
$\ket{\Omega(\mu)}$ satisfies $ b_p\ket{\Omega(\mu)} = 0$ for any $p$ and 
$a_p\ket{\Omega(\mu)} = 0$ for $p>{\overline{p_\mathrm{F}}}$.
The ground state energy and the quark number are given as
\begin{equation}
  E(\mu) =\sum_{p\leq {\overline{p_\mathrm{F}}}} E_p,
\end{equation}
and
\begin{equation}
  N_\mathrm{q} (\mu) = \sum_{p\leq {\overline{p_\mathrm{F}}}}=
  \frac{1}{2\pi}{2}(2L+1){\overline{p_\mathrm{F}}}+\frac{1}{2}
  .
\end{equation}
Noting the physical volume and the Fermi momentum
are given by $V=L/\hop$ and $p_\mathrm{F}=2\hop\overline{p_\mathrm{F}}$, respectively,
we obtain the quark number density as
\begin{equation}
  n_\mathrm{q} = \frac{1}{V}N_\mathrm{q} =
  \frac{p_\mathrm{F}}{\pi}\qty(1+\frac{1}{2L})+\frac{1}{2V}.
\end{equation}
In the large volume and continuum limit, $n_\mathrm{q}$ reduces to the continuum one, $n_\mathrm{q}=p_\mathrm{F}/\pi$.
Similarly, the pressure is expressed as 
\begin{equation}
  P(\mu) = \frac{1}{V}(\mu N_\mathrm{q} (\mu)- E(\mu))
  =\frac{1}{V}\sum_{p\leq {\overline{p_\mathrm{F}}}}(\mu-E_p).
\end{equation}

Next, we consider correlation functions and local observables.
The two-point correlation functions are calculated as
\begin{align}
  \begin{split}
  S_{s,s'}^>(n,n')&=\expval{\psi_{s}(n)\psi_{s'}^\dag(n')}\\
&  =\sum_{p}
      u_{s}(p,n)u_{s'}^\dag(p,n')(1-\theta(\mu-E_p) ),
  \end{split}\\
  \begin{split}
S_{s,s'}^<(n,n')
  &= \expval{\psi_{s'}^\dag(n')\psi_{s}(n)}\\
&=\sum_{p}
\qty(
      u_{s'}^\dag(p,n') u_{s}(p,n)\theta(\mu-E_p)
      + v^\dag_{s'}(p,n') v_{s}(p,n)
)\\
&=-S_{s,s'}^>(n,n') +\hop\delta_{s,s'}\delta_{n,n'}.
  \end{split}
\end{align}
From the two-point functions, the quark-number density is expressed as
\begin{equation}
  \begin{split}
    \expval{\bar{\psi}\gamma^0\psi(n)}
    &=S_{\mathrm{L},\mathrm{L}}^<(n,n)+S_{\mathrm{S},\mathrm{S}}^<(n,n)\\
    &=\sum_{p}
    (u_\mathrm{L}(p,n)u_\mathrm{L}^\dag(p,n)+u_\mathrm{S}(p,n)u_\mathrm{S}^\dag(p,n))
    (\theta(\mu-E_p)-1)
     +2\hop\\
     &=\frac{4\hop}{2L+1}\sum_{p\leq {\overline{p_\mathrm{F}}}}\frac{1}{2E_p}
     ((E_p+m)\sin^2({2}pn) +(E_p-m)\cos^2(p\qty(2n-1))
     )\\
      &-\frac{4\hop}{2L+1}\sum_{p}\frac{1}{2E_p}
      ((E_p+m)\sin^2({2}pn) +(E_p-m)\cos^2(p\qty(2n-1)))
      +2\hop.
    \end{split}
\end{equation}
Using
\begin{align}
  \sin^2({2}pn)+    \cos^2(p\qty(2n-1))
  &= 1+\sin(p)\sin(p\qty(4n-1)),\\
  \sin^2({2}pn)-
  \cos^2(p\qty(2n-1))&=
  -\cos(p)\cos(p\qty(4n-1)),
\end{align}
$\expval{\bar{\psi}\gamma^0\psi(n)}$ reduces to
\begin{equation}
  \begin{split}
    \expval{\bar{\psi}\gamma^0\psi(n)}
    &=\hop+n_\mathrm{q}
    \\
    &+\frac{4\hop}{2L+1}\sum_{p\leq {\overline{p_\mathrm{F}}}}
    \frac{1}{2}\qty(\sin(p)\sin(p\qty(4n-1))-\frac{1}{2L})\\
      &+\frac{4\hop}{2L+1}\sum_{p>{\overline{p_\mathrm{F}}}}\frac{m}{2E_p}
        \cos(p)\cos(p\qty(4n-1)).
        \label{eq:number_density_free}
    \end{split}
\end{equation}
Here, we used eqs.~\eqref{eq:orthognal1} and \eqref{eq:orthognal2} for the vacuum part.
$\hop$ is the vacuum contribution that can be eliminated through the renormalization of the charge.
The remaining terms are the oscillation terms due to the open boundary conditions, which will vanish in the large volume limit $L\to\infty$.
On the other hand, the spatial component of the current vanishes
\begin{equation}
  \begin{split}
    \expval{\bar{\psi}\gamma^1\psi(n)}
    &=S_{\mathrm{S},\mathrm{L}}^<(n,n)+S_{\mathrm{L},\mathrm{S}}^<(n,n)\\
&=0.
  \end{split}
\end{equation}

Similarly, $\expval{\bar{\psi}\psi(n)}$ and $\expval{\bar{\psi}\ri \gamma_5\psi(n)}$ can be evaluated as
\begin{equation}
  \begin{split}
    \expval{\bar{\psi}\psi(n)}
    &=S_{\mathrm{L},\mathrm{L}}^<(n,n)-S_{\mathrm{S},\mathrm{S}}^<(n,n)\\ 
&=-\frac{1}{V}\sum_{p>{\overline{p_\mathrm{F}}}}\frac{m}{E_p}\\
  &\quad-\frac{4\hop}{2L+1}\sum_{p>{\overline{p_\mathrm{F}}}}\frac{m}{2E_p}\qty(
    -\frac{1}{2L}+\sin(p)\sin(p\qty(4n-1)))\\
    &\quad-\frac{4\hop}{2L+1}\sum_{p\leq {\overline{p_\mathrm{F}}}}\frac{1}{2}\cos(p)\cos(p\qty(4n-1)),
  \end{split}\label{eq:chiral_condensate_free}
\end{equation}
and
\begin{equation}
  \begin{split}
    \expval{\bar{\psi}\ri\gamma^5\psi(n)}
    &=\ri S_{\mathrm{S},\mathrm{L}}^<(n,n)-\ri S_{\mathrm{L},\mathrm{S}}^<(n,n)\\
  &=-\frac{4\hop}{2L+1}\sum_{p>{\overline{p_\mathrm{F}}}}\frac{2\hop}{E_p}
  \qty(\frac{1}{2L}+\sin^2(p))\\
  &\quad-\frac{4\hop}{2L+1}\sum_{p>{\overline{p_\mathrm{F}}}}\frac{2\hop}{E_p}
  \qty(-\frac{1}{2L}+\sin(p)\sin(p(4n-1))),
  \end{split}
\end{equation}
respectively.
Here, we decompose them into static and oscillating parts.

Let us finally provide an expression for the two-point function that appeared in the distribution function:
\begin{equation}
  \begin{split}
    S_{\mathrm{L},\mathrm{L}}^<(n,n')+S_{\mathrm{S},\mathrm{S}}^<(n,n')
        &=
        \frac{4\hop}{2L+1}\sum_{p\leq {\overline{p_\mathrm{F}}}}\frac{1}{2}
          \cos(2p(n-n'))\\
        &\quad+\frac{4\hop}{2L+1}\sum_{p\leq {\overline{p_\mathrm{F}}}}\frac{1}{2}
        \sin(p)\sin(p\qty(2n+2n'-1))\\
        &\quad+\frac{4\hop}{2L+1}\sum_{p>{\overline{p_\mathrm{F}}}}\frac{m}{2E_p}
        \cos(p)\cos(p\qty(2n+2n'-1))
            +\hop\delta_{n,n'}.
  \end{split}
\end{equation}

Here, we used
\begin{align}
  &\sin({2}pn) \sin({2}pn')+    \cos \qty(p\qty(2n-1))
  \cos \qty(p\qty(2n'-1))\notag\\
  &= \cos(2p(n-n'))
  +\sin(p)\sin(p\qty(2n+2n'-1)),\\
  &\sin({2}pn) \sin({2}pn')-
  \cos \qty(p\qty(2n-1))
  \cos \qty(p\qty(2n'-1))\notag\\
  &=
  -\cos(p)\cos(p\qty(2n+2n'-1)).
\end{align}

\section{Sweep dependence}
\label{sec:dmrg_convergence}
In this appendix, we study the convergence of energy and entanglement entropy as a function of sweeps.
We have calculated the energy and entanglement entropy (EE) for $m=1.0$, $w=1.0$, $N=50$, $N_{\textrm{c}}=2$ and $N_{\textrm{B}}=5.0, 6.0, 8.0, 10.0, 11.0, 13.0$.
The entanglement entropy is calculated by dividing the system at its center.
We fix the maximum bond dimension to $40$ and perform DMRG algorithm in the following sequence: $1000$ sweeps with truncation error cutoff and noise strength of $10^{-6}$ and $10^{-5}$, $1000$ sweeps with $10^{-6}$ and $10^{-8}$, $6000$ sweeps with $10^{-8}$ and $10^{-5}$, $3000$ sweeps with $10^{-8}$ and $10^{-7}$, and $5000$ sweeps with $10^{-8}$ and $0$.

\begin{figure}[tp]
  \begin{center}
   \includegraphics[width=13cm]{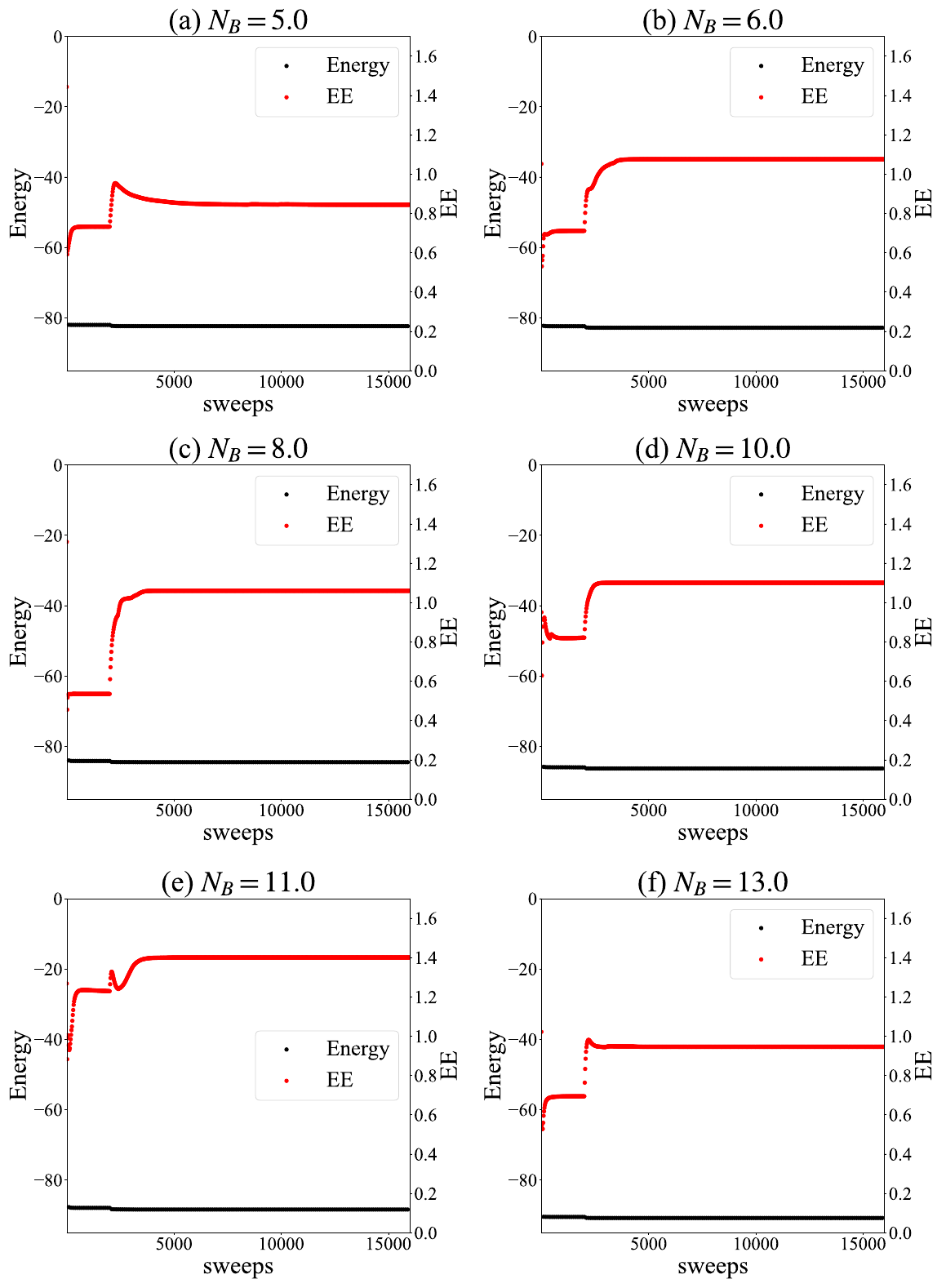}
  \end{center}
  \caption{
  The sweep dependence of the energy and entanglement entropy at the center of the system with fixed $m=1.0$, $w=1.0$, $N=50$, $N_{\textrm{c}}=2$ for (a) $N_{\textrm{B}}=5.0$, (b) $N_{\textrm{B}}=6.0$, (c) $N_{\textrm{B}}=8.0$, (d) $N_{\textrm{B}}=10.0$, (e) $N_{\textrm{B}}=11.0$, and (f) $N_{\textrm{B}}=13.0$.
  }
  \label{fig:sweep-dep1}
\end{figure}

\begin{figure}[tp]
  \begin{center}
   \includegraphics[width=10.0cm]{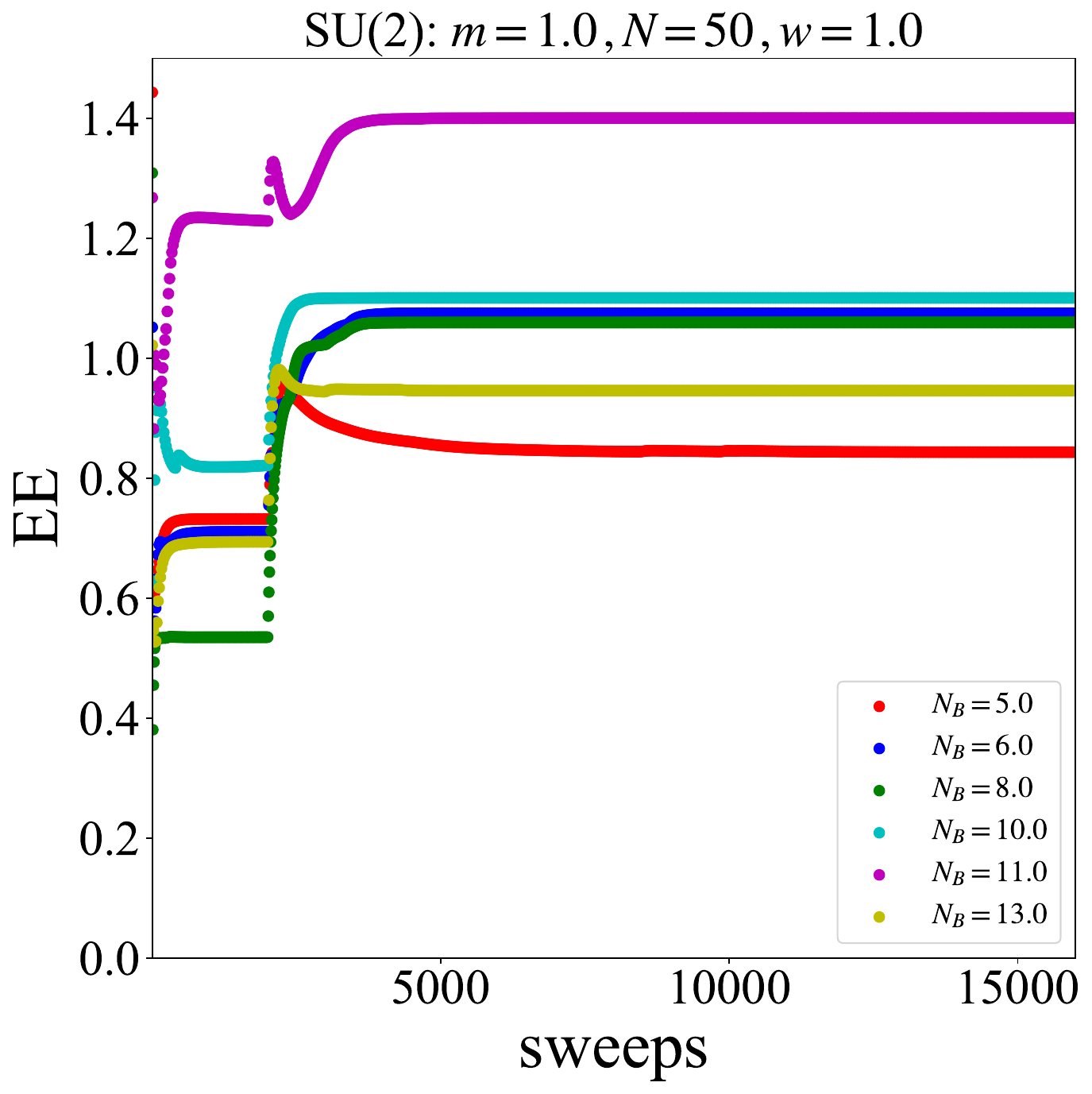}
  \end{center}
  \caption{
  The sweep dependences of the above entanglement entropies are summarized into one graph.
  }
  \label{fig:sweep-dep2}
\end{figure}

The numerical results are shown in figures \ref{fig:sweep-dep1} and \ref{fig:sweep-dep2}.
We observed that both the energy and the EE converge to constant values when sufficient optimization is performed. We also numerically demonstrated that the EE converges more slowly than the energy, although the energy almost converges with a few hundred sweeps.

\bibliographystyle{utphys}
\bibliography{HPbib}

\end{document}